\documentclass[12pt,a4paper]{article}
 
\usepackage{jheppub}
\usepackage[dvipsnames]{xcolor}
\usepackage{amssymb} 
\usepackage{amsmath}
\usepackage{mathtools}
\usepackage{amsfonts}    
\usepackage{dsfont}
\usepackage{pdfpages}
\usepackage{verbatim}
\usepackage{stmaryrd}
\usepackage{graphicx}

\usepackage{tensor}
\usepackage{mathrsfs}
\usepackage{textgreek} 
\usepackage{euscript}
\usepackage[smalltableaux]{ytableau}
\ytableausetup{centertableaux}
\usepackage{slashed}
\usepackage{datetime}
\usepackage{hyperref}
\usepackage{braket}
\hypersetup{
    pdfencoding=unicode,
	colorlinks=true,
	urlcolor=Maroon,
	linkcolor=RoyalBlue,
	citecolor=Maroon,
	pdftitle={A microscopic realization of dS3},
	pdfauthor={Scott Collier, Lorenz Eberhardt, Beatrix M\"uhlmann},
	pdfdisplaydoctitle=true,
	pdfstartview=FitH,
	linktocpage=true
}

\usepackage{tikz}
\usetikzlibrary{decorations,calc,shapes,patterns, math, shadings, fadings, arrows}
\tikzset{snake it/.style={decorate, decoration=snake}}
\tikzset{cross/.style={cross out, draw=black, minimum size=2*(#1-\pgflinewidth), inner sep=0pt, outer sep=0pt},
cross/.default={1pt}}

\usepackage{subcaption}

\newcommand{\ba}{\begin{align}}

\newcommand{\be}{\begin{equation}}
\newcommand{\ee}{\end{equation}}
\def\bd{\begin{tikzpicture}}
\def\ed{\end{tikzpicture}}

\DeclareMathOperator\tr{tr}

\renewcommand\Im{\mathop{\text{Im}}}

\renewcommand\Re{\mathop{\text{Re}}}
\newcommand\Res{\mathop{\text{Res}}}

\allowdisplaybreaks

\definecolor{light-gray}{gray}{0.75}

\newcommand\PSL{\text{PSL}}

\newcommand\SL{\text{SL}}
\newcommand\SU{\text{SU}}
\newcommand\SO{\text{SO}}

\newcommand\Diff{\text{Diff}}

\renewcommand\sl{\mathfrak{sl}}

\newcommand\CC{\mathbb{C}}

\newcommand\ZZ{\mathbb{Z}}
\newcommand\RR{\mathbb{R}}

\newcommand\CP{\mathbb{CP}}

\newcommand\EE{\mathbb{E}}

\renewcommand\d{\text{d}}

\newcommand{\e}{\mathrm{e}}

\renewcommand{\S}{\mathrm{S}}

\newcommand{\id}{\mathds{1}}

\renewcommand{\le}{\leqslant}
\renewcommand{\ge}{\geqslant}
\renewcommand{\leq}{\leqslant}
\renewcommand{\geq}{\geqslant}

\newcommand{\xx}{\mathsf{x}}
\newcommand{\yy}{\mathsf{y}}

\DeclareMathOperator*\Disc{Disc}

\DeclareMathOperator{\re}{Re}

\definecolor{bleudefrance}{rgb}{0.19, 0.55, 0.91}

\definecolor{vert}{rgb}{0.1367 0.543 0.1367}

\definecolor{pink}{rgb}{1.0, 0.13, 0.32}

\title{A microscopic realization of $\text{dS}_3$}

\author[1]{Scott Collier}\emailAdd{sac@mit.edu}
\author[2]{\!\!, Lorenz Eberhardt}\emailAdd{l.eberhardt@uva.nl}
\author[3]{\!\!, Beatrix M\"uhlmann}\emailAdd{beatrix@ias.edu}
\affiliation[1]{Center for Theoretical Physics, Massachusetts Institute of Technology, Cambridge, MA 02139, USA}
\affiliation[2]{Institute for Theoretical Physics, University of Amsterdam, PO Box 94485, 1090 GL Amsterdam, The Netherlands}
\affiliation[3]{School of Natural Sciences, Institute for Advanced Study, Princeton, NJ 08540, USA}

\abstract{
We propose a precise duality between pure de Sitter quantum gravity in $2{+}1$ dimensions and a double-scaled matrix integral. This duality unfolds in two distinct aspects. First, by carefully quantizing the gravitational phase space, we arrive at a novel proposal for the quantum state of the universe at future infinity. We compute cosmological correlators of massive particles in the universe specified by this wavefunction. Integrating these correlators over the metric at future infinity yields gauge-invariant observables, which are identified with the string amplitudes of the complex Liouville string \cite{Collier:2024kmo}. This establishes a direct connection between integrated cosmological correlators and the resolvents of the matrix integral dual to the complex Liouville string, thereby demonstrating one aspect of the dS$_3$/matrix integral duality.
The second aspect concerns the cosmological horizon of the dS static patch and the Gibbons-Hawking entropy it is conjectured to encode. 
We show that this entropy can be reproduced exactly by counting the entries of the matrix.

}

\begin{document}

\maketitle

\makeatletter
\g@addto@macro\bfseries{\boldmath}
\makeatother
\section{Introduction} \label{sec:introduction}

Low-dimensional models often offer greater calculable control than more realistic models and at the same time retain some of the physical lessons that we hope to learn about the real world.
This holds particularly true for quantum gravity, where the computational complexity of semi-realistic string compactification is staggering and a direct quantization of Einstein gravity in $3{+}1$ dimensions and higher is out of reach. 
Quantum gravity in two- and three-dimensional spacetime \emph{is} partially tractable by direct quantization and has provided physicists with many lessons over the years.

Constructing top-down long-lived de Sitter (dS) vacua is a notoriously hard problem and a low-dimensional viewpoint can be particularly useful \cite{Anninos:2017hhn, Maldacena:2019cbz, Cotler:2019nbi, Anninos:2021ene, Anninos:2021eit, Anninos:2023exn, Anninos:2024iwf, Verlinde:2024zrh, Verlinde:2024znh,  Coleman:2021nor, Batra:2024kjl}. A powerful guiding principle is provided by \emph{holographic duality} \cite{tHooft:1993dmi, Susskind:1994vu}, which postulates that theories of quantum gravity secretly admit a microscopic description in terms of an ordinary quantum system. Of course, this has a precise realization in anti-de Sitter (AdS) spaces \cite{Maldacena:1997re}. Various proposals exist in de Sitter space \cite{Strominger:2001pn, Maldacena:2002vr,  Witten:2001kn, Anninos:2011ui, Anninos:2017eib, Dong:2018cuv, Shyam:2021ciy}, but none comes close to matching the computational control and insight provided by the AdS/CFT correspondence.
A useful and concrete approach to dS quantum gravity is to take a global perspective of de Sitter space and compute cosmological correlators, which encode imprints left at future infinity $\mathcal{I}^+$ at the end of inflation. This has recently been implemented as the cosmological bootstrap \cite{Anninos:2014lwa, Hogervorst:2021uvp, DiPietro:2021sjt, Baumann:2022jpr}. A somewhat orthogonal approach is the introduction of an observer in the static patch and the associated algebra of observables in de Sitter quantum gravity \cite{Chandrasekaran:2022cip}.

\medskip

 In this paper, we propose a very precise microscopic realization of pure Einstein de Sitter quantum gravity in $2{+}1$ dimensions. The microscopic theory is a double-scaled matrix integral. Let us emphasize that this is drastically different than a duality involving a two-dimensional CFT, since we consider completely different observables. In AdS/CFT, observables are correlation functions of a boundary conformal field theory which can be computed systematically by Witten diagrams on the bulk side. In dS, such correlators naturally live at $\mathcal{I}^+$ and are interpreted as defining the \emph{wavefunction} of the universe. As in any quantum mechanical system, the observables are matrix elements, which in the absence of gauge-invariant operators simply correspond to the norm of the wavefunction which can be viewed as (integrated) cosmological correlators \cite{Maldacena:2002vr}. The basic dictionary maps these integrated cosmological correlators to a correlator of resolvents in the matrix model. Both admit a genus expansion: in 3d gravity, the future boundary $\mathcal{I}^+$ can have any topology and we have to sum over the genus in the gravitational path integral, while it arises as a $1/N$ expansion in the matrix model. 

Physically, such integrated cosmological correlators measure the correlations of massive non-interacting particles travelling through $\mathrm{dS}_3$ to $\mathcal{I}^+$, as well as correlations in the moduli of the two-dimensional surface that specifies the spatial topology. This is of course a far cry of what one hopes to compute in $3{+}1$ dimensions, where one can in principle observe correlations of primordial gravitational waves, or when adding matter, correlations of the cosmic microwave background (CMB). There is good reason to believe however that the matrix integral description captures the complete physics of the bulk in this vastly simplified setting.

\begin{figure}[ht]
\centering
    \begin{tikzpicture}[scale=1.5]
        \begin{scope}[shift = {(5,0)}]
\coordinate (A) at (0.02, 0.02);
    \coordinate (B) at (0.02,2.98);
    \coordinate (C) at (1.493, 1.493);
\coordinate (D) at (2.98, 0.02);
    \coordinate (E) at (2.98,2.98);
    \coordinate (F) at (1.493, 1.493);    
\shade[right color=blue!10, left color=blue!50] (A) -- (B) -- (C) -- cycle;
\draw[very thick](0,0) -- (0,3.02);   
\draw[very thick, gray](0,3) -- (3.02,3.);       
\draw[very thick](3,0) -- (3,3);  
\draw[very thick](3.02,0) -- (-.02,0);   
\draw[very thick](3,3) -- (0,0);  
\draw[very thick](0,3) -- (3,0);  
\node[scale=.7, rotate = 315] at (.9,2.3) {event horizon}; 
\node[scale=.8, black] at (1.5,3.2) {$\mathcal{I}^+$};
\node[cross out, thick, draw=red, scale=.79/1.2] at (1.5,3.) {};
\node[cross out, thick, draw=red, scale=.79/1.2] at (.75,3.) {};
\node[cross out, thick, draw=red, scale=.79/1.2] at (2.25,3.) {};
        \end{scope}
    \end{tikzpicture}
\caption{Penrose diagram of de Sitter. The square denotes the global patch with future infinity $\mathcal{I}^+$. This is a two-dimensional manifold that we take to be hyperbolic, such that the global metric is that of a Milne type universe. An observer today can only see a piece of dS, called the static patch, and realized in blue in the Penrose diagram. They are surrounded by an event horizon marking the boundary of their visible universe. }
\label{fig:penrose}
\end{figure}
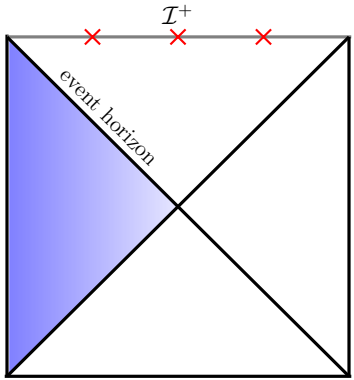

\begin{figure}
    \centering
    \includegraphics[width=0.5\textwidth]{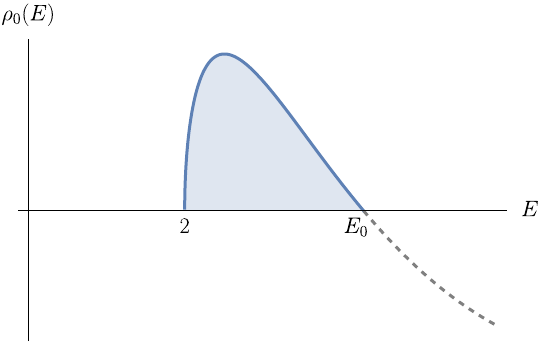}
    \caption{A plot of the leading density of eigenvalues of one of the matrices in the two-matrix integral dual of the complex Liouville string. The density exhibits the familiar square-root behavior near the edge of the spectrum, and oscillates on non-perturbative scales. We propose that the de Sitter microstates are enumerated by integrating the density of eigenvalues up to its first zero $E_0$. }
    \label{fig:eigenvalue density}
\end{figure}

The discussion of cosmological correlators takes a perspective of global de Sitter space. However, according to a conjecture of Gibbons and Hawking \cite{Gibbons:1976ue, Gibbons:1977mu} the cosmological horizon of the static patch of an observer encodes an entropy. The static patch, marked in blue in Figure~\ref{fig:penrose}, captures the part of the universe visible to an observer.
It can be computed by a Euclidean gravitational path integral on the three-sphere. We show that this entropy can be reproduced microscropically by the number of eigenvalues in the dual matrix model, 
\begin{equation}
    S_{\rm dS} = \log |\mathcal{Z}_{\mathrm{grav}}^{\mathrm{S}^3}|= \log N_\text{eff}^2 = S_{\text{dS}}^\text{micro} ~. \label{eq:intro entropy matching}
\end{equation}
In this equation $\mathcal{Z}_{\mathrm{grav}}^{\mathrm{S}^3}$ denotes the  Euclidean gravity partition function on $\mathrm{S}^3$. On the matrix model side, we declare the entropy $S_{\text{dS}}^\text{micro}$ in \eqref{eq:intro entropy matching} to be given by the total number of entries of the matrix. The matrix model is double scaled and thus $N=\infty$, but the leading density of eigenvalues is only positive from the edge $E \approx 2$ of the spectrum up to the first zero $E_0$, as displayed in Figure~\ref{fig:eigenvalue density}. At higher energies, the contour of the integral over eigenvalues has to be deformed.
We denote by $N_\text{eff}$ the eigenvalues in this first interval where the density of eigenvalues is positive. With these assumptions, we show that \eqref{eq:intro entropy matching} holds \emph{non-perturbatively} in $G_\text{N}$! 

\medskip

We now give some more technical details (and caveats) on the two central claims of this paper: the matching of the integrated cosmological correlators and of the de Sitter entropy. There are some assumptions that go into it and we have made reasonably optimistic guesses on how to continue based on the intuition and matching provided by the matrix model. Most of the work in this paper lays the ground work needed to establish this duality: we discuss the canonical quantization of $\mathrm{dS}_3$ gravity and the associated wavefunction in detail. After this, it is relatively straightforward to use previous results \cite{paper2} to establish the duality with the matrix model.

\paragraph{Canonical quantization of dS$_3$.} To talk about the wavefunction of $\mathrm{dS}_3$ gravity, we first need to discuss canonical quantization on a spatial slice of genus $g$ with $n$ punctures. The $n$ punctures are associated to worldlines of massive particles traversing the initial value surface $\Sigma_{g,n}$. Quantization proceeds similarly as for $\mathrm{AdS}_3$ gravity with some important differences.
The main result is that the Hilbert space is spanned by objects transforming as CFT correlation functions of central charge $c \in 13+i \RR$ and conformal weights $\Delta_i=h_i+\tilde{h}_i \in 1+i \RR$ under diffeomorphisms and Weyl transformations. This Hilbert space is endowed with an inner product taking the form 
\be 
\langle \Psi \, |\, \Psi' \rangle=g_\text{s}^{2g-2} \int_{\mathcal{M}_{g,n}} \Psi^* \Psi'~ , \label{eq:inner product intro}
\ee
which is structurally identical to a string theory path integral. Here, $\Im c=\frac{3 \ell_\text{dS}}{2 G_\text{N}}$ follows a similar dictionary as the Brown-Henneaux relation \cite{Brown:1986nw}. Related results have been obtained in \cite{Castro:2012gc, Verlinde:2024zrh, Gaiotto:2024osr,Godet:2024ich,Chakraborty:2023los}. An important input in the duality is the value of the `string coupling' in \eqref{eq:inner product intro} that sets the normalization of the inner product which by consistency of the three-dimensional theory is related to the three-sphere partition function as\footnote{The $\sim$ denotes equality up to a $G_\text{N}$-independent constant, which can be absorbed in the renormalization scheme of the $\mathfrak{b}\mathfrak{c}$ ghost partition function necessary to define \eqref{eq:inner product intro}. In particular, this constant hides an important factor of $i$ that is present in the three-sphere partition function.}
\be 
g_\text{s}^{-2} \sim \mathcal{Z}_\text{grav}^{\mathrm{S}^3}~. \label{eq:intro gstring ZS3}
\ee

\paragraph{The wavefunction of the universe.} The next link in our chain of reasoning is the determination of an appropriate wavefunction on $\mathcal{I}^+$.
There is very little guidance on how we should choose this wavefunction and this part is the most speculative part of our proposal. 
The no-boundary proposal \cite{Hartle:1983ai} would instruct us to sum over non-singular complex spacetimes with fixed boundary condition at $\mathcal{I}^+$, similarly to the computation of partition functions in AdS/CFT \cite{Witten:1998qj}. Such a sum is very hard to make sense of: it doesn't contain any on-shell topologies, but requires analytic continuation to metrics of $(-,-,-)$ signature \cite{Castro:2012gc, Cotler:2019nbi, Godet:2024ich} violating the Kontsevich-Segal criterion \cite{Kontsevich:2021dmb, Witten:2021nzp}, the topological expansion is ill-behaved for an imaginary central charge as there is no small parameter defining a hierarchy of spacetime topologies, and moreover such a wavefunction is anyway not normalizable with respect to the inner product (\ref{eq:inner product intro}). We take this as an indication that this is not particularly natural and instead propose to consider the gravitational path integral over an inflating de Sitter universe of the form
\be 
\d s^2=-\d t^2+ \sinh(t)^2 \d s_{\Sigma_{g,n}}^2~, \label{eq:intro inflating universe}
\ee
where $\d s_{\Sigma_{g,n}}^2$ is the hyperbolic metric on the spatial slice. 
The massive particles create deficit angles in the spatial surface of the appropriate strength.
Of course, \eqref{eq:intro inflating universe} has a big bang singularity at $t=0$. From the point of view of the gravitational path integral it is perhaps not clear how to deal with this. We circumvent this problem by relating $\mathrm{dS}_3$ gravity to a 3d TQFT as was done for $\mathrm{AdS}_3$ in \cite{Collier:2023fwi}. Even though this TQFT, which we call complex Virasoro TQFT, is much less well-understood than its AdS$_3$ counterpart, we can interpret the partition function on \eqref{eq:intro inflating universe} as the partition function on the interval $\Sigma_{g,n} \times I$, where we put topological boundary conditions on one side and dynamical boundary conditions on the other side. 
The result of this is perhaps not surprisingly that the cosmological wavefunction is given in terms of the correlation function of Liouville theory at central charge $c=1+6(b+b^{-1})^2 \in 13+i \RR$ and with spinless vertex operators $\Delta_i=2h_i=\frac{c-1}{6}-2p_i^2 \in 1+i \RR$ by
\be 
\Psi_{g,n}^{(b)}=\langle V_{p_1} \cdots V_{p_n} \rangle^{(b)} ~.\label{eq:intro wavefunction Liouville}
\ee
The Liouville momenta are fixed in terms of the particle masses by the usual dS/CFT relation (\ref{eq:conformal weight mass}).

One can also argue for \eqref{eq:intro wavefunction Liouville} by requiring physically desirable properties of $\Psi_{g,n}$, such as normalizability and factorizability under degenerations of the moduli of the surface. It turns out that these constraints essentially uniquely pin down the Liouville correlator \eqref{eq:intro wavefunction Liouville} as the cosmological wavefunction. 

\paragraph{Integrated cosmological correlators.}
In contrast to the no-boundary proposal discussed above, the Liouville correlator \emph{does} define a normalizable state and its norm in the Hilbert space defined by \eqref{eq:inner product intro} can be interpreted as an integrated cosmological correlator in the spirit of \cite{Maldacena:2002vr} of massive, non-dynamical particles in dS$_3$. The metric on $\mathcal{I}^+$ fluctuates, which necessitates the integral over moduli space $\mathcal{M}_{g,n}$ appearing in the 2d theory. We are then also led to considering a sum over topologies at $\mathcal{I}^+$ and take the full cosmological correlator to be, up to a suitable normalization of the vertex operators that we will discuss in the main text, given by
\be 
\sum_{g=0}^\infty g_\text{s}^{2g-2} \int_{\mathcal{M}_{g,n}} |\langle V_{p_1} \cdots V_{p_n} \rangle^{(b)} |^2~, \label{eq:intro sum of genera}
\ee
which has precisely the form of a string theory path integral.  In fact, the corresponding worldsheet theory consists of two Liouville theories of complex conjugated central charges $c \in 13+i \RR$. We studied this theory in detail in \cite{paper1, paper2, paper3, paper4} and in particular established a duality with a two-matrix model similar to other string theory/matrix model dualities known in the literature \cite{Gross:1989vs,Brezin:1990rb,Douglas:1989ve, Douglas1991, Collier:2023cyw}. Embedding this duality in the $\mathrm{dS}_3$ discussion, it becomes a duality between integrated cosmological correlators and resolvents in the matrix model. One conceptually important point is that the string coupling in \eqref{eq:intro sum of genera} is related to $G_\text{N}$ via \eqref{eq:intro gstring ZS3} and is not an independent parameter. 

\paragraph{Microstate counting?} We now discuss the matching \eqref{eq:intro entropy matching} further. The density of eigenvalues in the matrix model takes the form\footnote{We choose $-i b^2 \in \RR_{>0}$ so that $c \in 13+i \RR$. Here we write $|g_{\text{s}}|$ because, as we will see, the string coupling is imaginary. This density of eigenvalues is denoted by $\mathrm{e}^{S_0} \rho_0(E)$ in the rest of the paper.}
\begin{equation}\label{eq: intro eigenvalue distribution}
    \rho(E) = |g_{\text{s}}|^{-1}\frac{ (b^{-2}-b^2)\sin \left(- ib^2 \mathop{\text{arccosh}}\left(\frac{E}{2}\right)\right)}{2 \sin(\pi b^{-2})} ~,
\end{equation}
which has the qualitative form shown in figure~\ref{fig:eigenvalue density}. Here $g_{\text{s}}$ is the effective string coupling that controls the behavior of the matrix model resolvents at asymptotically large genera \cite{paper3}.

Checking \eqref{eq:intro entropy matching} boils down to computing $N_{\text{eff}} \equiv \int_2^{E_0} \d E \, \rho(E)$ and comparing it with the sphere partition function $\mathcal{Z}_\text{grav}^{\text{S}^3}$. Recall that the string coupling $g_{\text{s}}$ that appears in the density of eigenvalues above was itself related to the three-sphere partition function by consistency of the three-dimensional description in (\ref{eq:intro gstring ZS3}). The resulting match between the integrated density of eigenvalues $N_{\text{eff}}$ and the sphere partition function is non-trivial (and the argument is not circular), but a byproduct of (\ref{eq:intro gstring ZS3}) is that it works \emph{regardless} of the actual value of $\mathcal{Z}_\text{grav}^{\text{S}^3}$. 
This is perhaps a bit disappointing, since rather than giving us a clue about the nature of the holographic dual, the matrix model is in some sense insensitive to $\mathcal{Z}_\text{grav}^{\text{S}^3}$, which instead plays the role of a topological expansion parameter in the model via (\ref{eq:intro gstring ZS3}).
In particular, our discussion does not actually compute the specific value of $\mathcal{Z}_\text{grav}^{\text{S}^3}$. One may hope that it becomes computable in the future via TQFT techniques akin to those recently developed in $\text{AdS}_3$ \cite{Collier:2023fwi} and we make a speculative proposal along these lines in the discussion section~\ref{sec:discussion}.

Evaluating the gravitational path integral on $\mathrm{S}^3$ is in fact famously subtle because of the conformal mode problem \cite{Gibbons:1978ac}. The logarithm of the sphere path integral is known up to one-loop order and takes the form \cite{Anninos:2020hfj, Carlip:1992wg, Guadagnini:1994ahx, Castro:2012gc, Anninos:2022hqo, Anninos:2021ihe} 
\begin{equation}
    \log \mathcal{Z}_{\rm grav}^{\text{S}^3}  = S_{\rm GH} - 3\log S_{\rm GH} + 5 \log 2\pi \pm \frac{5\pi i}{2}+ \sum_{n\geq 1} c_n S_{\rm GH}^{-n}~,\quad S_{\rm GH} = \frac{\pi \ell_{\rm dS}}{2G_{\text{N}}}\gg 1~, \label{eq:sphere partition function}
\end{equation}
where $S_{\rm GH}=\frac{A_\text{h}}{4 G_{\text{N}}}$ denotes the leading de Sitter (Gibbons-Hawking) entropy in the semiclassical expansion and $A_\text{h}$ the area of the cosmological horizon. The logarithmic correction comes from the one-loop determinant with the prefactor $3=6 \times \frac{1}{2}$ originating from the number of isometries of the three-sphere. The imaginary part $\pm \frac{5\pi i}{2}$ is a result of the conformal mode problem which requires one to Wick rotate the integral over the Weyl factor in the gravitational path integral \cite{Polchinski:1988ua}. The arbitrary sign reflects the freedom in rotating the integration contour in any direction. 
It is not clear to us what the interpretation of the phase for the entropy should be and we took the pragmatic solution of including an absolute value in \eqref{eq:intro entropy matching}. 
We make however a speculative proposal for the exact three-sphere partition function \eqref{eq:sphere partition function} in the discussion~\ref{sec:discussion} (see also \cite{Anninos:2020hfj}), but this is independent from the rest of the paper.
We should also note that a recent proposal \cite{Maldacena:2024spf} gives a natural explanation of the phase in the sphere partition function in the presence of an observer.

\paragraph{Outline.}  We start in section \ref{sec:Hilbert space} with a discussion of the canonical quantization of $\text{dS}_3$ gravity. We keep the discussion in the main text somewhat general and provide some of the technical details in appendix~\ref{app: canonical quantization}. We then discuss in section \ref{sec:wavefunction} the physical choice of the wavefunction of the universe. In section \ref{sec:dual} we discuss observables in this universe and their dual descriptions. We discuss integrated cosmological correlators of massive non-dynamical particles and tie them to the string amplitudes of the complex Liouville string and its matrix integral dual. In section \ref{sec:dual} we also revisit the Gibbons-Hawking de Sitter entropy conjecture via the three-sphere partition function and establish a precise relation between this entropy and a count of the number of entries of the dual matrix integral.  We end with an extended discussion in section~\ref{sec:discussion}. In the appendices \ref{app:dS} and \ref{app:matrix model dual} we revisit some key features of the dS geometry and the two-matrix integral dual to the complex Liouville string, respectively.

\section{The gravitational Hilbert space}\label{sec:Hilbert space}
We consider three-dimensional gravity with a positive cosmological constant. It is classically described by the Einstein-Hilbert action with positive cosmological constant 
\begin{equation}
    S_{\mathrm{EH}} = \frac{1}{16 \pi G_{\text{N}} }\int \d^3 x \sqrt{-g} \left(\mathcal{R} -2\Lambda\right) ~,\quad \Lambda >0~.
\end{equation}
In the following we will discuss the canonical quantization of the theory. Some aspects of this are known \cite{Witten:1989ip, Cotler:2019nbi, Verlinde:2024zrh}, but we give a somewhat complete discussion with various new results. We have relegated various technical computations to appendix~\ref{app: canonical quantization}. This section can be read independently of the matrix model dual that we propose in section~\ref{sec:dual}.

\subsection{First order formalism} \label{subsec:relation CS}
In order to study the quantum theory, it is useful to relate the theory to Chern-Simons theory. We mostly use this as a bookkeeping device. We strongly emphasize that $\mathrm{dS}_3$ gravity is \emph{not} equivalent to Chern-Simons theory.
\paragraph{Rewriting the action.} We pass to a first-order formalism with the dreibein and the spin connection as the independent variables. We then form the linear combinations 
\be
\mathcal{A}^a=\omega^a + \frac{i}{\ell_\text{dS}} e^a~,\quad \bar{\mathcal{A}}^a=\omega^a  - \frac{i}{\ell_\text{dS}} e^a \label{eq:SL(2,C) gauge fields}
\ee
where $\ell_\text{dS} = 1/\sqrt{\Lambda}$ is the de Sitter length.
In terms of these variables, the (Lorentzian) Einstein-Hilbert action can be written in Chern-Simons form
\cite{Witten:1988hc}
\be 
S=\frac{k}{4\pi} \int \tr \left(\mathcal{A} \wedge \d \mathcal{A}+\frac{2}{3} \mathcal{A} \wedge \mathcal{A} \wedge \mathcal{A} \right)+\frac{\bar{k}}{4\pi} \int \tr \left(\bar{\mathcal{A}} \wedge \d \bar{\mathcal{A}}+\frac{2}{3} \bar{\mathcal{A}} \wedge \bar{\mathcal{A}} \wedge \bar{\mathcal{A}} \right)~, \label{eq:PSL(2,C) CS action}
\ee
where $\mathcal{A}$ is the complex gauge field and 
\be 
k=\frac{i\, \ell_\text{dS}}{4\, G_\text{N}} \in i \RR_+ \label{eq:level PSL(2,C) CS gravity}
\ee
is the level.\footnote{In general, the level of $\sl(2,\CC)$ Chern-Simons theory can also have a non-vanishing real part. It corresponds to a gravitational Chern-Simons term. 
We restrict ourselves to ordinary gravity for which the level is purely imaginary and hence set $k \in i \RR$ in the following. }
Under infinitesimal diffeomorphisms and local Lorentz transformations of the dreibein, $\mathcal{A}^a$ transforms like an $\mathfrak{sl}(2,\CC)$ gauge field. We will postpone the discussion of global issues for now and hence only use the Lie algebra.
We can take the trace in any faithful $\mathfrak{sl}(2,\CC)$ representation, but for normalization purposes the trace is conventionally taken in the fundamental representation.

Notice that this can be obtained by analytic continuation of the better known $\mathrm{AdS}_3$ relation to imaginary AdS length, $\ell_\text{AdS} \to i \, \ell_\text{dS}$. 

It is hence tempting to declare that $\mathrm{dS}_3$ gravity is equivalent to $\mathfrak{sl}(2,\CC)$ Chern-Simons theory. 
This is not quite right since there are several subtleties that we haven't addressed and that will all become important below. Let us discuss them first. In particular, even though we will use the Chern-Simons variables, the quantization of the gravitational phase space that we will discuss is \emph{not} equivalent to the quantization of Chern-Simons theory.

\paragraph{Invertibility.} The metric is a positive definite tensor which imposes some restrictions on the gauge field $\mathcal{A}^a$ corresponding to smooth gravitational backgrounds. For example, the flat gauge field $\mathcal{A}^a=0$ satisfies the Chern-Simons equations of motion, but clearly doesn't correspond to a good gravitational solution. For $\Lambda<0$, this condition can be implemented in a rather nice way since the phase space of $\mathrm{PSL}(2,\RR)$ Chern-Simons theory is disconnected and one can single out one component corresponding to smooth gravitational solutions \cite{Krasnov:2005dm} and consider its quantization \cite{Eberhardt:2022wlc, Collier:2023fwi}. For $\Lambda>0$, there is no analogous statement and regularity of the metric leads to some open subset of the Chern-Simons phase space.

\paragraph{Signature.} There is an enormous amount of confusion in the literature about the Wick rotation of the theory. This is not very important to our discussion since we will mostly work in Lorentzian signature. If one runs the same argument as above, one naively relates Euclidean $\mathrm{dS}_3$ to $\mathfrak{su}(2)$ Chern-Simons theory, but still with imaginary level. It is then tempting to use $\mathrm{SU}(2)$ or $\mathrm{SO}(3)$ Chern-Simons partition functions and try to relate them to Euclidean gravity partition functions. This theory is ill-defined since in any global form of $\mathfrak{su}(2)$ Chern-Simons theory the level $k$ has to be integer. Analytic continuation in $k$ is therefore not even unique \cite{Witten:2010cx}. 

More importantly, this is physically not the correct thing to do. The Hilbert space we get from canonical quantization of $\mathfrak{sl}(2,\CC)$ Chern-Simons theory differs drastically from the one of $\mathfrak{su}(2)$ Chern-Simons theory. The physically correct Hilbert space comes from the Lorentzian theory since the Hilbert space is by definition invariant under Wick rotation. 
The Wick rotation is then simply achieved by considering $\mathfrak{sl}(2,\CC)$ Chern-Simons theory on a Euclidean topological background manifold.\footnote{A familiar analogy comes from conformal field theory, where the spectrum of local operators is defined by the Hilbert space of the theory on the quantized on the circle in Lorentzian signature via the state-operator correspondence (and similarly, the unitarity constraints on the spectrum are those inherited from the Lorentzian conformal group). Despite its Lorentzian origin, this is the Hilbert space one uses when computing CFT correlation functions in Euclidean signature.} For attempts to study $\text{dS}_3$ gravity using $\mathrm{SU}(2)$ Chern-Simons theory, see e.g. \cite{Castro:2023dxp, Anninos:2020hfj, Hikida:2021ese}.

\paragraph{Global structure of the gauge group.} The Einstein-Hilbert action only tells us about the infinitesimal form of the gauge algebra. $\mathfrak{sl}(2,\CC)$ has several global forms given by the various covers of $\PSL(2,\CC)$. Let us discuss in particular the two-fold cover $\SL(2,\CC)$ and $\PSL(2,\CC)$. Since the theory doesn't contain fields in the fundamental representation, a better approximation of the theory is given by $\PSL(2,\CC)$ Chern-Simons theory. If we would add fermions as in $\mathrm{dS}_3$ supergravity, we would need to define a spin structure for which $\SL(2,\CC)$ becomes relevant. However, saying that the gauge group is $\PSL(2,\CC)$ is also not quite accurate since the phase space is further restricted to invertible dreibeins as we discussed above. 

\paragraph{Large diffeomorphisms.} Finally, the Chern-Simons description misses that also \emph{large} diffeomorphisms --- i.e.\ those not isotopic to the identity --- are to be gauged in gravity. Relatedly, we also need to sum over all topologies in gravity. Thus gravity is loosely obtained from Chern-Simons theory by gauging the mapping class group
\be 
\text{Map}(M)=\Diff(M)/\Diff_0(M)~.
\ee
Gravity computations on a single topology will typically not produce results that are invariant under this mapping class group, but gauge invariance is only achieved after summing over topologies consistent with given boundary conditions.

One therefore has two options: First quantize and then study the mapping class group action on the resulting Hilbert space and gauge it or first gauge the mapping class group and then quantize.\footnote{Gauging the mapping class group before quantization was studied in the $\mathrm{AdS}_3$ context in \cite{Maloney:2015ina, Eberhardt:2022wlc}.}
Both perspectives are useful and we therefore discuss them both. The Hilbert space obtained by canonical quantization before gauging the mapping class group will be denoted by a hat below. It carries a representation of the two-dimensional mapping class group of the chosen Cauchy slice.

\subsection{Wavefunction} \label{subsec:wavefunction}
After these preliminaries, we now discuss the wavefunctions of the theory. We give here an overview of the logic and refer to appendix~\ref{app: canonical quantization} for some more technical aspects.

\paragraph{Cauchy slice.} To talk about a Hilbert space we first have to fix a Cauchy slice, which will be a two-dimensional surface, which we denote by $\Sigma_{g,n}$.\footnote{We only consider orientable manifolds. One can presumably extend some results to the non-orientable case.}
Here, $g$ labels the genus and $n$ some number of punctures. These punctures can be thought of as massive particles passing through the initial-value surface $\Sigma_{g,n}$. Thus punctures will carry additional labels specifying the mass and spin of these particles.

\paragraph{Wheeler-DeWitt equation.} In the metric formalism, the phase space is formed by the metric $g_{ij}$ and the extrinsic curvature $K_{ij}$ of the initial value surface. To quantize, one has to pick a polarization, meaning a choice of `positions' and `momenta'. The wavefunction will then depend only on the positions, say. If we choose Dirichlet boundary conditions, the wavefunction would only depend on the boundary metric, $\Psi[g_{ij}]$, which gives a valid choice of polarization. Since diffeomorphisms are gauged in gravity, the wavefunctions also obey the Hamiltonian and momentum constraints
\be 
\mathcal{H} \Psi=0~, \qquad \mathcal{H}_i \Psi=0~, \label{eq:WdW equation}
\ee
where $\mathcal{H}$ generates time translations (i.e.\ it is the ADM Hamiltonian) and $\mathcal{H}_i$ generates infinitesimal diffeomorphisms along the Cauchy slice. The former equation is the Wheeler-DeWitt equation.

Roughly speaking, the momentum constraint tells us that the wavefunction is really only a function on the space of Riemannian manifolds with genus $g$ and $n$ punctures (with some boundary condition at the punctures that we will come to). The Hamiltonian constraint in turn tells us that the wavefunction descends to a function on the space of conformal structures on $\Sigma_{g,n}$, i.e.\ metrics up to rescaling by a positive function. This is intuitively clear since in a positive cosmological constant spacetime, time evolution will lead to an inflating universe which changes the scale factor of the spatial metric as usual in FRW cosmology. This statement can be directly derived from the Hamiltonian formulation of ($2{+}1$)-dimensional gravity \cite{Moncrief:1989dx, Krasnov:2005dm}. As usual when doing such Hamiltonian reductions, the wavefunction is actually not a function on the coset spaces, but rather a section of some hermitian line bundle over it.

So far, the discussion holds for any spacetime dimension. Restricting to $2{+}1$ dimensions makes life much easier since the space of conformal structures on the initial value surface $\Sigma_{g,n}$ is finite-dimensional (of complex dimension $3g-3+n$). Of course, this space is nothing other than the moduli space of Riemann surfaces $\mathcal{M}_{g,n}$! To be slightly more precise, we can, as we have mentioned above, gauge the mapping class group either before or after quantization. In the latter case, we want wavefunctions to be sections of a hermitian line bundle over the universal cover of $\mathcal{M}_{g,n}$ known as Teichm\"uller space $\mathcal{T}_{g,n}$, related by $\mathcal{M}_{g,n}=\mathcal{T}_{g,n}/\text{Map}(\Sigma_{g,n})$.

To summarize, depending on whether we gauge the mapping class group before or after quantization, we expect wavefunctions to be non-holomorphic sections of some (hermitian) line bundle over moduli space $\mathcal{M}_{g,n}$ or Teichm\"uller space $\mathcal{T}_{g,n}$.\footnote{For $\mathrm{AdS}_3$, it is convenient to make a different choice of polarization in which wavefunctions are holomorphic sections of a line bundle over the product $\mathcal{T}_{g,n} \times \mathcal{T}_{g,n}$. The Mess map \cite{Mess:2007} gives a symplectomorphism $\mathcal{T}_{g,n} \times \mathcal{T}_{g,n} \cong T^* \mathcal{T}_{g,n}$, which relates the two choices. In the $\mathrm{dS}_3$ case, we can however proceed with the more naive choice.} 

\paragraph{dS/CFT correspondence.} To continue our overview discussion, we can further motivate from the dS/CFT correspondence what such sections over $\mathcal{T}
_{g,n}$ should be. For this purpose we discuss the Hilbert space at late times on $\mathcal{I}^+$. Of course the Hilbert space at earlier times is isomorphic to that at late times, but the wavefunctions take a potentially more complicated form.\footnote{They are tentatively related by a type of timelike $T\bar{T}$ deformation in analogy with the situation in AdS/CFT \cite{McGough:2016lol, Araujo-Regado:2022gvw}.} The dS/CFT dictionary is essentially just an analytic continuation of AdS/CFT, where we think of the boundary partition function as a wavefunction. So the wavefunction $\Psi[g_{ij}]$ after imposing the constraints should behave like a CFT$_2$ partition function with central charge $c\sim 6k \sim \frac{3i \, \ell_{\text{dS}}}{2 \, G_\text{N}} \in i \RR$, which is the analytic continuation of the Brown-Hennaux formula \cite{Brown:1986nw}. We write $\sim$ since this relation receives loop-corrections, to be discussed below. For punctures, the conformal weight $(h,\tilde{h})$ associated to them is related to the mass and spin of the particle via the familiar conformal weight-mass relation \cite{Strominger:2001pn}:
\be 
\Delta=h+\tilde{h}=1 \pm \sqrt{1-m^2 \ell^2_\text{dS}}~, \label{eq:conformal weight mass}
\ee
while $h-\tilde{h}=s$ corresponds to the spin of the particle.
It will turn out from the quantization that only principal series representations of $\PSL(2,\CC)$ are allowed. These correspond to $\Delta \in 1+i \RR$ and hence to masses heavier than $\ell_\text{dS}^{-1}$. 

Thus the dS/CFT dictionary suggests that the wavefunctions transform like CFT correlation functions with an imaginary central charge 
\begin{equation}
    c \in 13+i \RR\, ,
\end{equation} 
and conformal weights 
\begin{equation}
    \Delta_i \in 1+i \RR\, .
\end{equation} 
We inserted a real part $\Re(c)=13$ for the central charge. It is a one-loop effect and is the analytic continuation of the corresponding shift in $\mathrm{AdS}_3$ computed in \cite{Giombi:2008vd,Cotler:2018zff}. Its dS incarnation was previously discussed in \cite{Cotler:2019nbi}.\footnote{An intuitive argument for this one-loop shift can be obtained by summing up the ground state energy of the Virasoro modes acting on the vacuum and comparing this to the Casimir energy $-\frac{c}{24}$. We obtain $-\frac{c}{24}\overset{!}{=}\frac{1}{2}\sum_{n=2}^\infty n=\frac{1}{2}(\zeta(-1)-1)=-\frac{13}{24}$ and hence $c=13$. The computation is completely analogous to that of a free boson except that the $n=1$ term is missing since $L_{-1} \ket{0}=0$.}
Contrary to the imaginary part of the central charge (which can receive scheme dependent renormalizations at higher loops), it follows from canonical quantization that the real part is one-loop exact. Such a correlation function picks up non-trivial factors under diffeomorphisms and Weyl rescalings due to the conformal weights of the vertex operators and the conformal anomaly. Thus the data $(c;\Delta_i,s_i)$ determines the line bundle over the polarized constrained phase space $\mathcal{T}_{g,n}$ (or $\mathcal{M}_{g,n}$) that arises from the Hamiltonian reduction discussed above.

Such a CFT correlation function is written as a linear combination of products of left- and right-moving Virasoro conformal blocks labelled by the external scaling dimensions and spins $(\Delta_i,s_i)_{i=1,\dots,n}$ and internal weights $(\Delta_a,s_a)_{a=1,\dots,3g-3+n}$, together with the central charge $c$.
A CFT correlation function must be crossing symmetric, which is akin to saying that it is invariant under the action of the mapping class group on the 2d surface $\Sigma_{g,n}$. This in particular implies that the internal spins are integer, $s_a=h_a-\tilde{h}_a \in \ZZ$. The individual products of left- and right-moving conformal blocks are obviously not crossing symmetric on their own and we could consider conformal blocks with real (not necessarily integer) internal spins $s_a\in\mathbb{R}$ and principal series internal dimensions $\Delta_a \in 1 + i\mathbb{R}$ as a basis of states for the Hilbert space $\hat{\mathcal{H}}_{g,n}$ defined by quantizing before gauging the mapping class group.\footnote{It turns out that ordinary Virasoro conformal blocks do not quite form an orthonormal basis to this Hilbert space under the inner product discussed in \eqref{eq:3d naive inner product}. This is perhaps expected since from the study of cosmological correlators \cite{Baumann:2022jpr}, we would expect such a basis to be spanned not by Virasoro conformal blocks but rather by a suitable notion of `Virasoro partial waves'. We have not found a convincing proposal for such Virasoro partial waves, but this will also not be needed for the following discussion. Such a basis will be important for a more systematic formulation of a complex version of Virasoro TQFT suited to dS$_3$ quantum gravity. We comment further on the issue in the discussion section \ref{sec:discussion}.}

To recapitulate, we have found the following two spaces of wavefunctions depending on whether we gauge the mapping class group after quantization or before:
\be 
\hat{\mathcal{H}}_{g,n}^{(b)}(\Delta_1,s_1,\dots,\Delta_n,s_n)~, \qquad
\mathcal{H}_{g,n}^{(b)}(\Delta_1,s_1,\dots,\Delta_n,s_n)~, 
\ee
which are spanned by linear combinations of products of left- and right-moving Virasoro conformal blocks. 
The combinations appearing in the Hilbert space $\mathcal{H}_{g,n}^{(b)}$ where we gauge the mapping class group before quantization are additionally crossing symmetric, and hence may be thought of as local CFT correlation functions. We will refine this statement once we have discussed the inner product and discuss normalizability of states in the Hilbert space. The Hilbert space carries a label $b$ which is a proxy for the central charge via the usual Liouville parameterization
\begin{equation}
    c = 13 + 6(b^{2}+b^{-2})\, .
\end{equation}
In the case of interest in this paper we hence have that $b^2$ is purely imaginary.

\subsection{Inner product} 
So far, everything is essentially just the analytic continuation of the quantization of $\mathrm{AdS}_3$ gravity. The story differs crucially in one aspect from the $\mathrm{AdS}_3$ setting. Technically, it arises because the relevant phase space $T^* \mathcal{T}_{g,n}$ (or $T^* \mathcal{M}_{g,n}$) is hyperk\"ahler and we may view the chosen polarization as either real or complex, depending on which complex structure we consider. This means that we can endow the Hilbert space with an inner product coming from the real polarization in which we naturally only integrate over the real slice of phase space, given by the zero section $\mathcal{T}_{g,n} \subset T^* \mathcal{T}_{g,n}$ and similarly for $\mathcal{M}_{g,n}$. We refer to appendix~\ref{app: canonical quantization} for more details. 
\paragraph{Inner product on $\hat{\mathcal{H}}_{g,n}^{(b)}$.} The inner product on $\hat{\mathcal{H}}_{g,n}^{(b)}(\Delta_1,s_1,\ldots,\Delta_n,s_n)$ is essentially trivial to write down, 
\be 
\langle \Psi' | \Psi \rangle = g_\text{s}^{2g-2}\int_{\mathcal{T}_{g,n}}  (\Psi')^* \Psi~, \qquad \ket{\Psi},\, \ket{\Psi'} \in \hat{\mathcal{H}}^{(b)}_{g,n}(\Delta_1,s_1,\ldots,\Delta_n,s_n)~. \label{eq:3d naive inner product}
\ee
The external operator labels of the states $\Psi'$ and $\Psi$ must agree because they live in the same Hilbert space. Together the mass-shell and level-matching conditions constrain this data as follows
\be 
h_i+\tilde{h}_i^*=1~,
\ee
which together with $h_i-\tilde{h}_i=s_i \in \RR$ give the reality conditions
\be 
h_i=\frac{1+s_i+i \lambda_i}{2}~, \qquad \tilde{h}_i=\frac{1-s_i+i \lambda_i}{2}~, \qquad \lambda_i \in \mathbb{R}~. \label{eq:h htilde reality conditions}
\ee
Thus external conformal dimensions naturally live in principal series representations of $\SL(2,\mathbb{C})$
\begin{equation}
    \Delta_i = h_i + \tilde h_i =  1 + i\lambda_i, \qquad \lambda_i\in\mathbb{R}.
\end{equation}

To define the integral \eqref{eq:3d naive inner product}, we coupled to the $\mathfrak{bc}$-ghosts as is familiar from string theory which are left implicit in the notation.  Writing down \eqref{eq:3d naive inner product} does not require us to choose a measure or metric on $\mathcal{T}_{g,n}$ and is hence canonical.
This makes the integral in \eqref{eq:3d naive inner product} well-defined (though not necessarily convergent).\footnote{Notice also that this integrand does not require the inclusion of a K\"ahler potential as is necessary for a complex polarization. In $\mathrm{AdS}_3$ gravity, the role of the K\"ahler potential is played by the partition function of timelike Liouville CFT that appears in the inner product \cite{Verlinde:1989ua, Collier:2023fwi}.} 

We also included in \eqref{eq:3d naive inner product} the possibility of a `string coupling' which simply accounts for the arbitrary inclusion of the relevant counterterm when defining the ghost path integral. $g_\text{s}$ is not arbitrary and we will fix it in section~\ref{subsec:3d consistency} by requiring consistency of the three-dimensional theory. We could in principle also include some leg factors $\mathcal{N}(\Delta_i,s_i)$ for the external punctures. We will eventually include this below, but suppress it for now from the notation, since this is clearly truly arbitrary.

\paragraph{Inner product on $\mathcal{H}_{g,n}^{(b)}$.} In analogy to \eqref{eq:3d naive inner product} we can also write down the inner product on the Hilbert space $\mathcal{H}^{(b)}_{g,n}$ where we gauge the mapping class group before quantization, which takes the form\footnote{We could in principle include the leg factors and the normalization of the path integral as one does in defining the string amplitudes $\mathsf{A}_{g,n}^{(b)}$ \cite{paper1} that will appear later in this norm. This will not play a conceptual role in the following and we will suppress it.} 
\be 
\langle \Psi' | \Psi \rangle = g_\text{s}^{2g-2}\int_{\mathcal{M}_{g,n}}  (\Psi')^* \Psi~, \qquad \ket{\Psi},\, \ket{\Psi'} \in {\mathcal{H}}^{(b)}_{g,n}(\Delta_1,s_1,\ldots,\Delta_n,s_n)~.\label{eq:3d naive inner product after Map}
\ee
The integral now runs over the moduli space $\mathcal{M}_{g,n}$ of the Cauchy surface, and the definition of this integral is locally equivalent to the integral over Teichm\"uller space $\mathcal{T}_{g,n}$ discussed above. Here the wavefunctions $\Psi$ and $\Psi'$ are assumed to be crossing-symmetric so restricting the integral to run only over moduli space is necessary and well-defined.
Evidently, this inner product precisely has the structure of a string theoretic moduli space integral, which lies at the root of the bridge to string theory and matrix models that we discuss below.

\paragraph{Higher 3d topology corrections.} We should note that \eqref{eq:3d naive inner product} and \eqref{eq:3d naive inner product after Map} correspond to the leading inner product for 3d quantum gravity. It can be viewed as the 3d gravity partition function on the interval $\Sigma_{g,n} \times I$ with one state associated to each boundary. In principle, it is possible that the inner product receives corrections from higher 3d topologies with two Riemann surfaces $\Sigma_{g,n}$ as boundaries. It was discussed in \cite{Marolf:2020xie, Iliesiu:2024cnh} that this quantum corrected inner product could lead to the emergence of many null states and consequently a dramatically smaller Hilbert space of quantum gravity. Nevertheless, we will consider the simple inner product \eqref{eq:3d naive inner product}, as it is computationally very useful.

\subsection{Consistency and the normalization of the inner product} \label{subsec:3d consistency}

The inner products \eqref{eq:3d naive inner product} and \eqref{eq:3d naive inner product after Map} depend on a `string coupling' that sets an overall normalization. As long as we only consider a single Hilbert space $\mathcal{H}_{g,n}$, it is completely arbitrary, but it is fixed from consistency of the three-dimensional theory.

To even talk about fixing $g_\text{s}$, we first have to pick a scheme in which we are computing the ghost partition function, since changing the scheme changes $g_\text{s}$. We will not attempt to completely fix this scheme, but rather just pick one scheme which does not depend on $G_\text{N}$. Thus we aim to determine $g_\text{s}$ as a function of $G_\text{N}$ up to an overall constant. The following simple argument is not very rigorous and it would be very desirable to improve upon it.

\paragraph{Splitting the three-sphere.} In order to determine the string coupling we consider the three-sphere. It constitutes a saddle of Euclidean $\mathrm{dS}_3$ Einstein gravity. From canonical quantization, we could compute its partition function in multiple ways by splitting it along surfaces of different topologies. This will provide an important consistency relation between the inner products on the Hilbert spaces associated with different topologies. Here we explain the computations in a simpler class of compact rational TQFTs such as those based on a modular tensor category whose Hilbert space on a given Cauchy surface is spanned by a discrete, finite set of conformal blocks. We will then discuss the generalization of these computations to the main case of interest relevant for dS$_3$ quantum gravity, which is a complex version of Virasoro TQFT whose Hilbert space is by contrast infinite dimensional and spanned by a continuum of Virasoro conformal blocks.

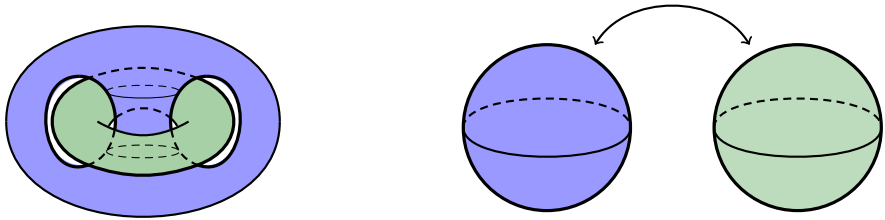
\begin{figure}[ht]
    \centering
    \begin{subfigure}[a]{.37\textwidth}
        \begin{tikzpicture}[scale=1.2]
              \draw[draw=none, fill=blue, opacity=.4] (-1.5,0) to[out = -90, in = -90,looseness=1.2] (1.5,0) to[out=90,in=90,looseness=1.2] (-1.5,0);
            \draw[draw=none, fill=white] (-1.05,0) to[out=90,in=90,looseness=2.25] (-.32,0) to[out=-90,in=-90,looseness=2.25] (-1.05,0);
            \draw[draw=white, fill=white] (1.05,0) to[out=90,in=90,looseness=2.25] (.32,0) to[out=-90,in=-90,looseness=2.25] (1.05,0);
            \draw[draw=none, fill=white] (-.31,-.09) to [out=-20,in=200] (.31,-.09) to[out=-80,in= 140,looseness=0.85] (.6,-.48) to[out=200,in=-20] (-.6,-.48) to[out=40,in=260,looseness=.8] (-.31,-.09);
            \draw[draw=none,fill=vert,opacity=.4] (-1,0) to[out=90,in=200] (-.6,.48) to[out=-20, in=100] (-.31,0.04) to (-3/8,-.05) to[out=-20,in=200,looseness=1.2] (3/8,-.05) to (.31,.04) to[out=85,in=200] (.6,.48) to[out=-20,in=90] (1,0) to[out=-90,in=-90] (-1,0);
    
            \draw[very thick] (-.31,-.09) to[out=80, in = 0] (-.7,1/2);
            \draw[very thick] (.31,-.09) to[out=100, in = 180] (.7,1/2);
            \draw[ thick, densely dashed] (-.31,-.09) to[out=-100, in =20] (-.6,-.48);
            \draw[ thick, densely dashed] (.31,-.09) to[out=-80, in = 160] (.6,-.48);
    
            \draw[thick] (1.5,0) to[out=90,in=90,looseness=1.2] (-1.5,0);
            \draw[thick] (1.5,0) to[out=-90,in=-90,looseness=1.2] (-1.5,0);
    
            \draw[very thick] (.7,1/2) to[out=0,in=-20,looseness=1.47] (.6,-.48);
            \draw[very thick] (-.7,1/2) to[out=180,in=200,looseness=1.47] (-.6,-.48);
    
            \draw[very thick] (-.6,.48) to[out=200, in = 160,looseness=1.5] (-.6,-.48);
            \draw[very thick] (-.6,-.48) to[out=-20,in=200,looseness=.95] (.6,-.48);
            \draw[very thick] (.6,-.48) to[out=20,in=-20, looseness=1.5] (.6,.48);
            \draw[ thick, densely dashed] (.6,.48) to[out=160,in=20,looseness=.95] (-.6,.48);
            \draw[thick] (-1/2,0) to[out=-30,in=160] (-.3,-.095);
            \draw[thick] (-.3,-.095) to[out=-20, in =200,looseness=.95] (.3,-.095);
            \draw[thick] (.3,-.095) to[out=20, in = 210] (1/2,0);
            \draw[thick] (-3/8,-.05) to[out=60,in=220] (-.3,0.04);
            \draw[thick, densely dashed] (-.3,0.04) to[out=40,in=140,looseness=.95] (.3,.04);
            \draw[thick] (.3,0.04) to[out=-40,in=120] (3/8,-.05);

            \draw (-.4,.33) to[out=-90, in=-90, looseness=0.3] (.4,.33);
            \draw[densely dashed] (-.4,.33) to[out=90, in=90, looseness=0.3] (.4,.33);
            \draw[densely dashed] (-.4,-.33) to[out=-90, in=-90, looseness=0.3] (.4,-.33);
            \draw[densely dashed] (-.4,-.33) to[out=90, in=90, looseness=0.3] (.4,-.33);
        \end{tikzpicture}
    \end{subfigure}
    ~
    \begin{subfigure}[b]{.37\textwidth}
        \begin{tikzpicture}[baseline={([yshift=-2ex]current bounding box.center)},scale=1.1]
            \draw[fill=blue,draw=blue,opacity=.4] (-3/2,0) circle (1);
            \draw[very thick] (-3/2,0) circle (1);
            \draw[thick] (-5/2,0) to[out=-90,in=-90,looseness=.6] (-1/2,0);
            \draw[thick,densely dashed] (-5/2,0) to[out=90,in=90,looseness=.6] (-1/2,0);
    
            \begin{scope}[shift=({3,0)})]
                \draw[fill=vert,draw=vert,opacity=.3] (-3/2,0) circle (1);
                \draw[very thick] (-3/2,0) circle (1);
                \draw[thick] (-5/2,0) to[out=-90,in=-90,looseness=.6] (-1/2,0);
                \draw[thick,densely dashed] (-5/2,0) to[out=90,in=90,looseness=.6] (-1/2,0);
            \end{scope}
    
            \node(a) at ({-3/2+cos(60)},{sin(60)}) {};
            \node(b) at ({3-3/2+cos(120)},{sin(120)}) {};
    
            \draw[thick, <->] (a) to[out=60, in = 120] (b);
        \end{tikzpicture}
    \end{subfigure}
    \caption{Two splittings of the three-sphere. On the left the sphere is split into two interlocked solid tori glued along their torus boundaries. On the right the sphere is split into equatorial three-balls glued along their two-sphere boundaries, with the interior of one three-ball identified with the exterior of the other.}\label{fig:S3 splittings}

\end{figure}

First we imagine cutting the three-sphere along the equator into the northern and southern hemisphere. We then first evaluate the TQFT path integral on the two hemispheres, which prepares a state on the equator.
The two hemispheres are topologically 3-balls. Thus as usual in topological theories, the path integral prepares the vacuum block associated with the boundary chiral algebra, which we denote by $\ket{\id_{0,0}}$. It is the unique state in the Hilbert space $\mathcal{H}_{0,0}^{(b)}$ associated with the two-sphere, and compactness of the TQFT means that its norm is simply determined by the normalization of the inner product itself, which for the zero-punctured sphere is given by $g_{\text{s}}^{-2}$. Thus we have
\begin{equation}\label{eq:gs ZS3 TQFT}
    \mathcal{Z}_{\text{TQFT}}^{\text{S}^3} = \braket{\id_{0,0}|\id_{0,0}} 
    = g_{\text{s}}^{-2}\, .
\end{equation}
However one may also compute the sphere partition function by splitting the three-sphere along different surfaces. For example, we may imagine splitting the sphere along two interlinked solid tori glued along their torus boundaries as shown in figure \ref{fig:S3 splittings}. The TQFT path integral on each solid torus computes a special state in the torus Hilbert space corresponding to the vacuum character of the boundary chiral algebra, but in the modular S-dual channels. Thus we can also express the TQFT three-sphere partition function in terms of the identity-identity component of the modular S-matrix of the chiral algebra
\begin{equation}
    \mathcal{Z}_{\text{TQFT}}^{\text{S}^3} = \braket{\id_{1,0}|\mathbb{S}|\id_{1,0}} = \mathbb{S}_{\id\id}\, . \label{eq:ZS3 S11}
\end{equation}
For rational compact TQFT this is just a number that is part of the data of the boundary chiral algebra. Notably the string coupling $g_{\text{s}}$ that determines the normalization of the inner product drops out in the case of the torus inner product so together with (\ref{eq:gs ZS3 TQFT}) this fixes it completely.

\paragraph{The case of $\mathrm{dS}_3$.} In the case of interest, computing $\mathbb{S}_{\id\id}$ is a very subtle business because the inner product in \eqref{eq:ZS3 S11} diverges. Thus, we will not perform this computation here since the rest of the paper does not depend on it. In the discussion section \ref{sec:discussion} we discuss the extent to which this computation generalizes in the TQFT associated with dS$_3$ quantum gravity. The generalization is nontrivial because of the infinite-dimensionality of the torus Hilbert space and the fact that the identity character is a non-normalizable state in the Hilbert space. This leads us to a conjectural proposal for $\mathcal{Z}_{\text{TQFT}}^{\text{S}^3}$ which we identify with $\mathcal{Z}_{\text{grav}}^{\text{S}^3}$ as in the $\text{AdS}_3$ case \cite{Collier:2023fwi}.

Regardless of the specific value of $\mathcal{Z}_{\text{grav}}^{\text{S}^3}$, we notice that both the left- and the right-hand side of \eqref{eq:gs ZS3 TQFT} remain well-defined, provided we still interpret $\mathcal{Z}_{\text{TQFT}}^{\text{S}^3}$ as $\mathcal{Z}_{\text{grav}}^{\text{S}^3}$. We thus assume that it continues to hold in the irrational case which determines the value of $g_\text{s}^{-2}$. As stressed above, the normalization in the inner product \eqref{eq:3d naive inner product} involves a renormalization scheme of the $\mathfrak{bc}$-ghosts. Thus we will assume that $\mathcal{Z}_{\text{grav}}^{\text{S}^3}=g_\text{s}^{-2}$ holds only up to a $b$-independent constant which represents the freedom of choosing different schemes, which we will denote by a $\sim$ as in the introduction~\ref{sec:introduction}. To summarize, we learn that
\be 
\mathcal{Z}_{\text{grav}}^{\text{S}^3} \sim g_\text{s}^{-2}~. \label{eq:three sphere partition function string coupling}
\ee
Compared to the rational case, we now have the slightly weaker statement that the sphere partition function is determined by the normalization of the inner product up to an overall $b$-independent constant. 

\paragraph{Reality.} 
We take in the following $\mathcal{Z}^{\text{S}^3}_{\text{grav}}=\e^{S_\text{dS}}$ as an input (see however the discussion~\ref{sec:discussion} for a speculative proposal of its value). Its value is known up to one-loop order in the semiclassical expansion as in \eqref{eq:sphere partition function}, where it reads in these variables
\begin{align} 
\log \mathcal{Z}^{\text{S}^3}_{\text{grav}}
&=-2\pi i b^2-3 \log(-i b^2)+2 \log(2\pi) \pm \frac{5\pi i}{2}+ \mathcal{O}(b^{-2})~.
\end{align}
In particular, it is completely universal up to this order and does not suffer from scheme ambiguities. However, it is not real and $\arg(\mathcal{Z}^{\text{S}^3}_{\text{grav}})=\pm\frac{\pi}{2}$ as a consequence of the conformal mode problem \cite{Polchinski:1988ua}. The inner products \eqref{eq:3d naive inner product} and \eqref{eq:3d naive inner product after Map} should of course be positive definite and thus we should have $g_\text{s}^{-2} \sim |\mathcal{Z}_{\text{grav}}^{\text{S}^3}|$.

\section{Wavefunction of the universe} \label{sec:wavefunction}
We now discuss what state should play the role of the cosmological wavefunction in a $(2{+}1)$-dimensional universe. The main conclusion will be that there is a preferred state $\Psi_{g,n}^{(b)} \in \mathcal{H}_{g,n}^{(b)}$ that is prepared by the gravitational path integral on the expanding cosmology \eqref{eq:inflating universe metric} involving massive scalar particles, whose wavefunction is given by the Liouville CFT correlation function\footnote{Here we label this privileged state by the Liouville momenta $p_i$ which are proxies for the conformal dimensions and spins $(\Delta_i,s_i)$ of the external vertex operators. These two parameterizations are related via
\begin{equation}
    \Delta_i = 1 + \frac{c-13}{12} - p_i^2 - \tilde p_i^2, \qquad s_i = -p_i^2 + \tilde p_i^2\, .
\end{equation}
The Liouville CFT correlators are only defined for scalar external primaries, $s_i = 0$, so one only needs to specify $p_i$. Throughout we will reserve the Liouville momentum variables $p_i$ to refer specifically to the preferred Liouville state in the Hilbert space.}
\be 
\Psi_{g,n}^{(b)}(p_1,\ldots,p_n)=\big\langle V_{p_1} \cdots V_{p_n} \big \rangle^{(b)}_g \in \mathcal{H}_{g,n}^{(b)}(\Delta_1,0,\ldots,\Delta_n,0)~. \label{eq:wavefunction of the universe}
\ee

\subsection{Hartle-Hawking wavefunction}

We start by following the Hartle-Hawking prescription \cite{Hartle:1983ai}. It suggests that the state of the universe $\ket{\text{HH}}$ is prepared by summing over all complex 3-manifolds with given boundary topology on $\mathcal{I}^+$.
Let us consider the case without punctures for simplicity.
For the Hartle-Hawking state, we would fill in the Riemann surface $\Sigma_g$ with a Euclidean bulk that caps off. The simplest choice is a handlebody $\mathsf{S}\Sigma_g$. For $g \ge 1$ (or $g=0$ with a sufficient number of punctures), such a topology does not solve the Einstein equations, but rather is given by the analytic continuation of the corresponding $\mathrm{AdS}_3$ saddle to $(-,-,-)$ signature. In particular, this is a complex geometry that violates the Kontsevich-Segal-Witten criterion for the admissibility of complex saddles in the gravitational path integral \cite{Kontsevich:2021dmb, Witten:2021nzp}. Let us anyway proceed for the moment. By analytic continuation from $\mathrm{AdS}_3$, the path integral on the handlebody should compute the corresponding Virasoro vacuum conformal block $|\mathcal{F}_{\mathds{1}}(\mathsf{S}\Sigma_g)|^2$ on $\Sigma_g$ in the channel specified by the handlebody. Since the choice of handlebody breaks crossing symmetry, we have to sum over all possible handlebodies in order to implement the gauging of the mapping class group. Thus the naive analogue of the Hartle-Hawking wavefunction is\footnote{Here it is understood that the absolute value on the right-hand side acts on the moduli of $\mathcal{I}^+$ and $\emph{not}$ on the central charge or conformal weights that define the conformal block.}
\be 
\ket{\text{HH}}=\sum_{\text{handlebodies }\mathsf{S}\Sigma_g} |\mathcal{F}_{\mathds{1}}^{(b)}(\mathsf{S}\Sigma_g)|^2+ \dots~. \label{eq:HH state Sigmag}
\ee
The dots represent more complicated topologies than handlebodies with a single $\Sigma_g$ boundary that one might consider including in the gravitational path integral.
This discussion is entirely analogous to the computation of CFT partition functions in $\mathrm{AdS}_3$, but analytically continued to complex central charge. However we emphasize that the analytic continuation to complex central charge invalidates the systematic topological expansion in $\mathrm{e}^{-\# c}$ that is present in $\mathrm{AdS}_3$ gravity and hence there is no hierarchy of suppression of higher topologies in \eqref{eq:HH state Sigmag}. Instead, these higher topologies lead to very rapidly oscillating contributions to the wavefunction. 

The Hartle-Hawking state as defined through \eqref{eq:HH state Sigmag} is unpleasant to work with. Besides being ill-defined, every term in \eqref{eq:HH state Sigmag} is also non-normalizable,
\be 
\lVert \ket{\text{HH}} \rVert^2=\infty~,
\ee
because the vacuum block diverges at the boundaries of moduli space. In other words, it behaves like a CFT partition function with a normalizable ground state in its Hilbert space on the circle. The divergence encountered in the inner product is then simply the familiar divergence from the tachyon in bosonic string theory. 
As already mentioned above, similar comments would apply to any cosmological wavefunction defined by the partition function of a compact CFT with complex central charge and a normalizable ground state.

The Hartle-Hawking state as defined by the sum over handlebodies in pure dS$_3$ gravity was recently studied in the special case of a torus spatial slice in \cite{Godet:2024ich}. A similar observation about the non-normalizability of the Hartle-Hawking state as defined this way (in the case of a torus spatial slice) was also made in \cite{Castro:2012gc}, where it was interpreted as an instability of the de Sitter vacuum in three-dimensional Einstein gravity.

Thus the naive Hartle-Hawking wavefunction is pathological from a variety of points of view: it requires complex geometries violating the Kontsevich-Segal-Witten criterion, the contribution from a fixed topology is non-normalizable, and the sum over topologies is completely uncontrollable since there is no small parameter suppressing topology fluctuation.
We take this as a strong indication that $\ket{\text{HH}}$ is in fact \emph{not} the natural choice for the wavefunction of the three-dimensional universe. We will now propose an alternative.

\subsection{Gravitational path integral on the inflating universe}

To compute the wavefunction of the universe, we should perform the gravitational path integral over spacetimes with a given topology $\Sigma_{g,n}$ at future infinity. This will require a particular prescription of how to deal with the big bang singularity, such as the no-boundary proposal as schematically indicated on the left in figure~\ref{fig:factorizability}.

\begin{figure}[ht]
    \centering
    \begin{tikzpicture}
        \draw[very thick, red, out=-60, in=90] (0,-.5) to (1,-3);
        \draw[very thick, red, out=-120, in=90] (2,-.5) to (1,-3);        \draw[very thick, red, out=-120, in=90] (2,-.5) to (1,-3);
      \fill[color=gray!15] (0,1) to[out=180,in=180,looseness=2] (0,-1) to[out=0,in=180] (1,-0.75) to[out=0,in=180] (2,-1) to[out=0,in=0,looseness=2] (2,1) to[out=180, in =0] (1,.75) to[out=180,in=0] (0,1);
         \draw[very thick, dashed, red!60, out=-60, in=120] (0,-.5) to (.3,-1);  
         \draw[very thick, red!60, dashed, out=-120, in=60] (2,-.5) to (1.7,-1);       
         \draw[white,fill=white] (-1/3,-.08+.06) to[bend left=30] (+1/3,-.08+.06) to[bend left=20] (-1/3,-.08+.06);
        \draw[white,fill=white] (2-1/3,-.08+.06) to[bend left=30] (2+1/3,-.08+.06) to[bend left=20] (2-1/3,-.08+.06);
        
        \draw[very thick, out=180, in=180, looseness=2] (0, 1) to (0,-1);
        \draw[very thick, out=0, in = 180] (0,1) to (1,0.75) to (2, 1);
        \draw[very thick, out = 0, in=180] (0,-1) to (1,-0.75) to (2,-1);
        \draw[very thick, out=0, in=0,looseness=2] (2,1) to (2,-1);
         \draw[very thick, bend right=30] (-1/2,0+.06) to (1/2, 0+.06);
        \draw[very thick, bend right=30] (2-1/2,0+.06) to (2+1/2, 0+.06);
        \draw[very thick, bend left = 30] (-1/3,-.08+.06) to (+1/3,-.08+.06);
        \draw[very thick, bend left = 30] (2-1/3,-.08+.06) to (2+1/3,-.08+.06);
        \node[cross out, thick, draw=red, scale=.75] at (0,-.5) {}; 
        \node[cross out, thick, draw=red, scale=.75] at (2,-.5) {}; 
        \node[scale=1.2] at (1,0) {$\Sigma_{g,n}$};
        \node[scale=1.2] at (3.75,0) {$\mathcal{I}^+$};
        \node[scale=1.2,white] at (-1.75,0) {$\mathcal{I}^+$};    
        \begin{scope}[scale=.33, shift={(2,-5)}]
             \fill[color=gray!15](0,1) to[out=180,in=180,looseness=2] (0,-1) to[out=0,in=180] (1,-0.75) to[out=0,in=180] (2,-1) to[out=0,in=0,looseness=2] (2,1) to[out=180, in =0] (1,.75) to[out=180,in=0] (0,1);
            \draw[thick, densely dotted] (0,-1) to[out=180, in = 180, looseness=2] (0,1) to[out=0, in=180] (1,0.75) to[out=0, in=180] (2, 1) to[out=0, in=0, looseness = 2] (2,-1) to[out=180, in = 0] (1,-0.75) to [out=180, in=0] (0,-1);
            \draw[thick, bend right=30, densely dotted] (-1/2,0+.06) to (1/2, 0+.06);
            \draw[thick, bend right=30, densely dotted] (2-1/2,0+.06) to (2+1/2, 0+.06);
            \draw[thick, bend left = 30, densely dotted] (-1/3,-.08+.06) to (+1/3,-.08+.06);
            \draw[thick, bend left = 30, densely dotted] (2-1/3,-.08+.06) to (2+1/3,-.08+.06);
            \node[cross out, thick, draw=red, scale=.75/1.8] at (.27,-.5) {}; 
            \node[cross out, thick, draw=red, scale=.75/1.8] at (2-.27,-.5) {};
        \end{scope}
          \draw[very thick, red, out=-60, in=120] (.5,-1.33) to (.75,-1.8);  
          \draw[very thick, red, out=-120, in=60] (1.5,-1.33) to (1.25,-1.8);           
        \draw[thick] (3.17,0) to[out=-90, in = 90] (1,-3);
        \draw[thick] (-1.17,0) to[out=-90, in = 90] (1,-3);
        \draw[thick, draw=red,fill=red] (1,-3) circle (1/10);
    
\end{tikzpicture}
\caption{A cartoon of the inflating universe (\ref{eq:inflating universe metric}). There is a big bang singularity (shown in red) at $t=0$, and constant $t$ slices are given by hyperbolic surfaces $\Sigma_{g,n}$. The gravitational path integral computes the wavefunction of the universe on a late-time slice with the topology of $\Sigma_{g,n}$ to be given by the Liouville correlation function on $\Sigma_{g,n}$. The massive particles correspond to Wilson lines in the three-dimensional bulk, shown in red. }\label{fig:inflating universe} 
\end{figure}
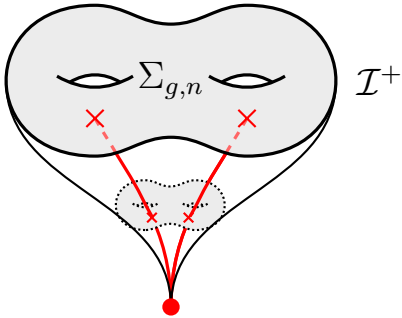

\paragraph{Inflating spacetime.} However, our set up differs in an important way from the usual discussion in the literature in the sense that our spatial slices are compact hyperbolic surfaces rather than spheres. In fact, there are no non-singular complete Euclidean on-shell topologies with such a Cauchy slice \cite{Anninos:2012ft}. Thus any on-shell manifold necessarily involves a singularity, \emph{even in Euclidean signature}!\footnote{This readily follows from known mathematical results. Manifolds with a constant positive curvature metric are called spherical manifolds in Thurston's classification. Assuming that they are geodesically complete, such manifolds necessarily must be closed as a consequence of the Bonnet-Myers theorem. They are then necessarily of the form $\mathrm{S}^3/\Gamma$ for some finite group $\Gamma$. None of these admit an embedding of a Riemann surface $\Sigma_g$ with $g \ge 2$ and constant negative curvature by Hilbert's theorem. Thus any on-shell topology cannot be geodesically complete which indicates the presence of a singularity.} We can evolve $\mathcal{I}^+$ backwards in time and reconstruct the on-shell solution:
\be 
\d s^2=-\d t^2+\sinh(t)^2 \d s_{\Sigma_{g,n}}^2~, \label{eq:inflating universe metric}
\ee
where $\d s_{\Sigma_{g,n}}^2$ is the hyperbolic metric on the Riemann surface. In the case where the constant-$t$ slices are given by all of hyperbolic 2-space $\mathbb{H}^2$ rather than a compact Riemann surface, this metric describes the hyperbolic patch of global dS$_3$ \cite{Strominger:2001pn, Anninos:2012qw, Anninos:2012ft}. One may think of this as a de Sitter version of the Maldacena-Maoz two-boundary wormhole in Euclidean AdS \cite{Maldacena:2004rf}. Indeed, the metric (\ref{eq:inflating universe metric}) is related to that of the Maldacena-Maoz wormhole in $(-,-,-)$ signature by a change of contour for $t$
\begin{equation}
    -\d t^2 + \sinh(t)^2\d s^2_{\Sigma_{g,n}} \, \stackrel{t = \rho+\tfrac{\pi i}{2}}{\longrightarrow} -\left(\d \rho^2 + \cosh(\rho)^2 \d s^2_{\Sigma_{g,n}}\right)\, .
\end{equation}
It evidently has a Milne-type big bang singularity at $t=0$ and represents an inflating universe, see figure \ref{fig:inflating universe}.

This solution is much better behaved than the handlebodies that we discussed above. We actually don't know of any other natural topology to include in the sum over topologies and thus tentatively take the gravitational path integral over the inflating universe topology \eqref{eq:inflating universe metric} to identify a good state of the universe. We will in the following discuss how the gravitational path integral can be computed by interpreting the big bang as a topological boundary. We will see in particular that the resulting wavefunction is crossing-symmetric on its own and hence in principle there is no need for a further sum over bulk spacetime topologies. 

One could alternatively attempt to extend the topology in $t$ to $t \le 0$ by deforming the $t$ contour slightly into the complex plane $t \in \RR+i \varepsilon$. This would avoid the singularity and satisfy the Kontsevich-Segal-Witten criterion. However, this creates then also a second boundary in the infinite past and we are not sure how to interpret it. Thus we will let the $t$-contour end at $t=0$ with the boundary condition that we will discuss below.

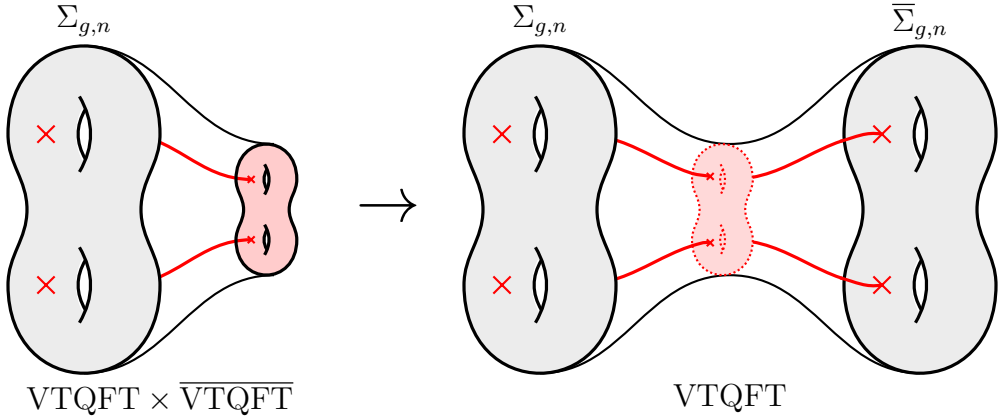
\begin{figure}[ht]
    \centering
\begin{tikzpicture}
    \draw[very thick, red] (1/2,1) to[out=0, in = 180] (2.5-3/10,.4);
    \draw[very thick, red] (1/2,-1) to[out=0, in = 180] (2.5-3/10,-.4);

    \fill[color=gray!15] (-1,1) to[out=90, in=90, looseness=2] (1,1) to[out=-90,in=90] (0.75,0) to[out=-90,in=90] (1,-1) to[out=-90,in=-90,looseness=2] (-1,-1) to[out=90, in = -90] (-0.75,0) to[out=90, in =-90] (-1,1);
    \draw[draw=white,fill=white] (-.02,1+1/3) to[bend left = 30] (-.02,1-1/3) to[bend left = 30, looseness=.8] (-.02,1+1/3);
    \draw[draw=white,fill=white] (-.02,-1+1/3) to[bend left = 30] (-.02,-1-1/3) to[bend left = 30, looseness=.8] (-.02,-1+1/3);
    
    \draw[very thick] (-1,1) to[out=90, in=90, looseness=2] (1,1) to[out=-90,in=90] (0.75,0) to[out=-90,in=90] (1,-1) to[out=-90,in=-90,looseness=2] (-1,-1) to[out=90, in = -90] (-0.75,0) to[out=90, in =-90] (-1,1);
    \begin{scope}[xscale=-1]
    \draw[very thick, bend right = 30] (+.06,1+1/2) to (+.06,1-1/2);
    \draw[very thick, bend left = 30] (-.02,1+1/3) to (-.02,1-1/3);
    \end{scope}
    \begin{scope}[xscale=-1]
    \draw[very thick, bend right = 30] (+.06,-1+1/2) to (+.06,-1-1/2);
    \draw[very thick, bend left = 30] (-.02,-1+1/3) to (-.02,-1-1/3);
    \end{scope}

    \begin{scope}[scale=.4, shift = {(6,0)}]
        \draw[draw=red, fill=red, opacity=.2] (-1,1) to[out=90, in=90, looseness=2] (1,1) to[out=-90,in=90] (0.75,0) to[out=-90,in=90] (1,-1) to[out=-90,in=-90,looseness=2] (-1,-1) to[out=90, in = -90] (-0.75,0) to[out=90, in =-90] (-1,1);
        \draw[draw=white,fill=white] (-.02,1+1/3) to[bend left = 30] (-.02,1-1/3) to[bend left = 30, looseness=.8] (-.02,1+1/3);
        \draw[draw=white,fill=white] (-.02,-1+1/3) to[bend left = 30] (-.02,-1-1/3) to[bend left = 30, looseness=.8] (-.02,-1+1/3);
        \draw[very thick] (-1,1) to[out=90, in=90, looseness=2] (1,1) to[out=-90,in=90] (0.75,0) to[out=-90,in=90] (1,-1) to[out=-90,in=-90,looseness=2] (-1,-1) to[out=90, in = -90] (-0.75,0) to[out=90, in =-90] (-1,1);
        \begin{scope}[xscale=-1]
        \draw[very thick, bend right = 30] (+.06,1+1/2) to (+.06,1-1/2);
        \draw[very thick, bend left = 30] (-.02,1+1/3) to (-.02,1-1/3);
        \end{scope}
        \begin{scope}[xscale=-1]
        \draw[very thick, bend right = 30] (+.06,-1+1/2) to (+.06,-1-1/2);
        \draw[very thick, bend left = 30] (-.02,-1+1/3) to (-.02,-1-1/3);
        \end{scope}
    \end{scope}

    \draw[thick] (0, 2.17) to[out = 0, in = 180] (5/2,.87);
    \draw[thick] (0, -2.17) to[out=0, in=180] (5/2,-.87);

    \node[scale=1] at (1,-2.5) {$\text{VTQFT}\times\overline{\text{VTQFT}}$};
    \node[scale=1] at (0,2.5) {$\Sigma_{g,n}$};
    \node[scale=2] at (4,0) {$\rightarrow$};

    \node[cross out,draw=red, thick, scale=.75](a) at (-1/2,1) {};
    \node[cross out,draw=red, thick, scale=.75](b) at (-1/2,-1) {};
    \node[cross out,draw=red, thick, scale=.3](a) at (2.5-3/10,.4) {};
    \node[cross out,draw=red, thick, scale=.3](b) at (2.5-3/10,-.4) {};

    \begin{scope}[shift={(6,0)}]

  \draw[very thick, red] (1/2,1) to[out=0, in = 180] (2.5,.4) to[out=0, in = 180] (4+1/2,1);
        \draw[very thick, red] (1/2,-1) to[out=0, in = 180] (2.5,-.4) to[out=0, in = 180] (4+1/2,-1);      

        \fill[color=gray!15] (-1,1) to[out=90, in=90, looseness=2] (1,1) to[out=-90,in=90] (0.75,0) to[out=-90,in=90] (1,-1) to[out=-90,in=-90,looseness=2] (-1,-1) to[out=90, in = -90] (-0.75,0) to[out=90, in =-90] (-1,1);
        \draw[draw=white,fill=white] (-.02,1+1/3) to[bend left = 30] (-.02,1-1/3) to[bend left = 30, looseness=.8] (-.02,1+1/3);
        \draw[draw=white,fill=white] (-.02,-1+1/3) to[bend left = 30] (-.02,-1-1/3) to[bend left = 30, looseness=.8] (-.02,-1+1/3);
        
        \draw[very thick] (-1,1) to[out=90, in=90, looseness=2] (1,1) to[out=-90,in=90] (0.75,0) to[out=-90,in=90] (1,-1) to[out=-90,in=-90,looseness=2] (-1,-1) to[out=90, in = -90] (-0.75,0) to[out=90, in =-90] (-1,1);
        \begin{scope}[xscale=-1]
        \draw[very thick, bend right = 30] (+.06,1+1/2) to (+.06,1-1/2);
        \draw[very thick, bend left = 30] (-.02,1+1/3) to (-.02,1-1/3);
        \end{scope}
        \begin{scope}[xscale=-1]
        \draw[very thick, bend right = 30] (+.06,-1+1/2) to (+.06,-1-1/2);
        \draw[very thick, bend left = 30] (-.02,-1+1/3) to (-.02,-1-1/3);
        \end{scope}
        \begin{scope}[shift={(5,0)}]
            \fill[color=gray!15] (-1,1) to[out=90, in=90, looseness=2] (1,1) to[out=-90,in=90] (0.75,0) to[out=-90,in=90] (1,-1) to[out=-90,in=-90,looseness=2] (-1,-1) to[out=90, in = -90] (-0.75,0) to[out=90, in =-90] (-1,1);
            \draw[draw=white,fill=white] (-.02,1+1/3) to[bend left = 30] (-.02,1-1/3) to[bend left = 30, looseness=.8] (-.02,1+1/3);
            \draw[draw=white,fill=white] (-.02,-1+1/3) to[bend left = 30] (-.02,-1-1/3) to[bend left = 30, looseness=.8] (-.02,-1+1/3);
            
            \draw[very thick] (-1,1) to[out=90, in=90, looseness=2] (1,1) to[out=-90,in=90] (0.75,0) to[out=-90,in=90] (1,-1) to[out=-90,in=-90,looseness=2] (-1,-1) to[out=90, in = -90] (-0.75,0) to[out=90, in =-90] (-1,1);
            \begin{scope}[xscale=-1]
            \draw[very thick, bend right = 30] (+.06,1+1/2) to (+.06,1-1/2);
            \draw[very thick, bend left = 30] (-.02,1+1/3) to (-.02,1-1/3);
            \end{scope}
            \begin{scope}[xscale=-1]
            \draw[very thick, bend right = 30] (+.06,-1+1/2) to (+.06,-1-1/2);
            \draw[very thick, bend left = 30] (-.02,-1+1/3) to (-.02,-1-1/3);
            \end{scope}
        \end{scope}

        \draw[thick] (0, 2.17) to[out = 0, in = 180] (5/2,.87) to[out=0, in = 180] (5,2.17);
        \draw[thick] (0, -2.17) to[out=0, in=180] (5/2,-.87) to[out=0, in=180] (5,-2.17);

\draw[very thick, red] (3.95,.905) to[out=15, in = 170] (4+1/2,1);

\draw[very thick, red] (3.95,-.905) to[out=-15, in = 190] (4+1/2,-1);   

        \begin{scope}[scale=.4, shift = {(6,0)}]
            \fill[color=red!15] (-1,1) to[out=90, in=90, looseness=2] (1,1) to[out=-90,in=90] (0.75,0) to[out=-90,in=90] (1,-1) to[out=-90,in=-90,looseness=2] (-1,-1) to[out=90, in = -90] (-0.75,0) to[out=90, in =-90] (-1,1);
\draw[thick, densely dotted, red] (-1,1) to[out=90, in=90, looseness=2] (1,1) to[out=-90,in=90] (0.75,0) to[out=-90,in=90] (1,-1) to[out=-90,in=-90,looseness=2] (-1,-1) to[out=90, in = -90] (-0.75,0) to[out=90, in =-90] (-1,1);   
    \node[cross out,draw=red, thick, scale=.3](b) at (-.4,1.1) {};
    \node[cross out,draw=red, thick, scale=.3](b) at (-.4,-1.1) {};

\draw[very thick, red] (-1,1.2) to[out=-15, in = 170] (-.4,1.1) ;
\draw[very thick, red] (-1,-1.2) to[out=15, in = 170] (-.4,-1.1) ;
\begin{scope}[xscale=-1]
            \draw[thick, bend right = 30, densely dotted, red] (+.06,1+1/2) to (+.06,1-1/2);
            \draw[thick, bend left = 30, densely dotted, red] (-.02,1+1/3) to (-.02,1-1/3);
            \end{scope}
\begin{scope}[xscale=-1]            
            \draw[thick, bend right = 30, densely dotted, red] (+.06,-1+1/2) to (+.06,-1-1/2);
            \draw[thick, bend left = 30, densely dotted, red] (-.02,-1+1/3) to (-.02,-1-1/3);
            \end{scope}
        \end{scope}

        \node[scale=1] at (5/2,-2.5) {$\text{VTQFT}$};
        \node[scale=1] at (0,2.5) {$\Sigma_{g,n}$};
        \node[scale=1] at (5,2.5) {$\overline{\Sigma}_{g,n}$};

        \node[cross out,draw=red, thick, scale=.75](a) at (-1/2,1) {};
        \node[cross out,draw=red, thick, scale=.75](b) at (-1/2,-1) {};
        \node[cross out,draw=red, thick, scale=.75](c) at (4+1/2,1) {};
        \node[cross out,draw=red, thick, scale=.75](d) at (4+1/2,-1) {};

    \end{scope}
    
\end{tikzpicture}
\caption{A sketch of the folding trick. On the left we have the product theory $\text{VTQFT}\times\overline{\text{VTQFT}}$ on $\Sigma_{g,n}\times I$ with dynamical boundary conditions (gray) on one end, and topological or gapped boundary conditions (red) on the other. We unfold this to a single copy of VTQFT on $\Sigma_{g,n}\times I$, but now with dynamical boundary conditions on both ends of the interval. The VTQFT partition function on the latter is computed by the correlation function of Liouville CFT on $\Sigma_{g,n}$ \cite{Collier:2023fwi}. 
}\label{fig:folding trick}
\end{figure}

\paragraph{The wavefunction from TQFT.} We can evaluate the wavefunction produced by the inflating universe with TQFT techniques. Thus, let us consider the description in terms of $\SL(2,\CC)$ Chern-Simons theory, while keeping the caveats discussed in section~\ref{subsec:relation CS} in mind. For this we have to interpret the boundary conditions created at $t=0$ in the framework of TQFT. We take the boundary condition to be of Dirichlet type where we impose that the universe has zero size. 
We take this to mean that at the big bang we should impose the following in terms of the gauge fields
\be 
\mathcal{A}^a=\bar{\mathcal{A}}^a~. \label{eq:big bang boundary conditions}
\ee
This is a valid boundary condition, since the boundary term of the variation of the action \eqref{eq:PSL(2,C) CS action}
\be 
\delta S_\partial=\frac{k}{4\pi} \int_{\partial \mathcal{M}} \tr \big(\mathcal{A} \wedge \delta \mathcal{A}-\bar{\mathcal{A}} \wedge \delta \bar{\mathcal{A}} \big)=0
\ee
vanishes. Since it doesn't require the introduction of a dynamical edge mode, such a boundary condition is called topological or gapped in the TQFT literature. We thus propose that we can translate the computation of the gravitational path integral into a TQFT computation on the topology $\Sigma_{g,n} \times I$, where we impose the gapped boundary conditions on one side representing the big bang and the dynamical (or ``gapless'') dS boundary conditions on the other side representing $\mathcal{I}^+$. See the left of figure \ref{fig:folding trick}.

Let us first discuss the corresponding computation in AdS$_3$ gravity, where we use Virasoro TQFT (VTQFT) \cite{Collier:2023fwi, Collier:2024mgv}. In the AdS framework, such a topological boundary condition corresponds to a kind of end-of-the-world brane.\footnote{It is arguably more natural to impose Neumann boundary conditions there which lead to the universe ending on an extremal surface. } The boundary conditions \eqref{eq:big bang boundary conditions} correspond to the trivial or diagonal boundary conditions in the doubled theory $\text{VTQFT} \times \overline{\text{VTQFT}}$.\footnote{The bar denotes orientation reversal.} One can `unfold' this geometry to $\Sigma_{g,n} \times I$ with dynamical boundary conditions on both sides, but computed in a single copy of VTQFT. This translation is known as folding trick in the TQFT literature. The computation of the VTQFT partition function on $\Sigma_{g,n} \times I$ was discussed in \cite{Collier:2023fwi} and is given by the corresponding Liouville partition function or correlation function with left-moving moduli associated to one boundary and right-moving moduli associated to the other boundary:
\begin{equation}
    Z_{\text{VTQFT}}(\Sigma_{g,n}\times I|\boldsymbol{m_1},\boldsymbol{m_2}) = Z_{\text{Liouville}}^{(b)}(\Sigma_{g,n}|\boldsymbol{m_1},\boldsymbol{m_2}) = \langle V_{p_1}\cdots V_{p_n}\rangle_g^{(b)}[\boldsymbol{m_1},\boldsymbol{m_2}]\, .
\end{equation}
Here $\boldsymbol{m_1},\boldsymbol{m_2}$ collectively refer to the moduli of the two boundaries; in the case of interest they are related by orientation reversal, $\boldsymbol{m_2} = \boldsymbol{\bar m_1}$, so that the TQFT partition function computes the ordinary Liouville correlator in Euclidean signature.
Translating back to the folded geometry, we conclude that the state prepared by the gravitational path integral on this slab with one topological and one dynamical boundary is the Liouville correlation function. 

The dS computation on the inflating universe can then be obtained by analytic continuation in the central charge of the AdS computation on the slab with topological and dynamical boundaries. Indeed, the canonical quantization as discussed in section~\ref{subsec:wavefunction} and appendix~\ref{app: canonical quantization} is related by analytic continuation. The obtained state should therefore still be the Liouville partition function, but now with $b^2 \in i \RR$. Possible punctures lead to vertex operator insertions and hence produce the Liouville correlation function on the dynamical boundary.\footnote{This argument is perhaps a bit fast. To make it more rigorous, we would need to develop a notion of Virasoro partial waves. See the discussion~\ref{sec:discussion} for further comments about this.}

Thus the wavefunction of the universe as produced by the gravitational path integral over this inflationary universe gives precisely the Liouville correlator and therefore physically motivates the choice for the wavefunction of the universe made in \eqref{eq:wavefunction of the universe}.

\paragraph{Spinning wavefunctions.} 
The previous discussion only determines the cosmological wavefunction in the situation that the particles at future infinity all have vanishing spin, $s_i = 0$. 
Indeed the expanding spacetime (\ref{eq:inflating universe metric}) only exists for spinless worldlines corresponding to massive scalar particles. 
Similarly the correlation functions of Liouville CFT are only defined in the case that the external operators are scalars, and do not admit an analytic continuation to the case of spinning external operators (the CFT data does not factorize into holomorphic and anti-holomorphic components, and in any case the spins should be quantized). Thus based on what we have discussed so far it is not clear what the wavefunction of the universe should be when there are spinning particles at $\mathcal{I}^+$.

In the absence of any explicit spacetime topology with spinning punctures at future infinity, in the following we will be guided by the fact that the expanding universe geometry only exists for scalar worldlines and thus propose that the spinning wavefunctions simply vanish:
\begin{equation}\label{eq:spinning wavefunctions vanish}
    \Psi_{g,n}(\Delta_1,s_1,\ldots,\Delta_n,s_n) \stackrel{!}{=} 0,\quad \text{\emph{any} } s_i \ne 0\, .
\end{equation}

\subsection{Bootstrapping the wavefunction}
There is another route towards motivating the choice \eqref{eq:wavefunction of the universe} more abstractly.
While the Liouville correlator is perhaps a natural state in the Hilbert space $\mathcal{H}^{(b)}_{g,n}(\boldsymbol{p})$, it is far from unique: in principle many more states can be constructed, for example by acting with Verlinde line operators and summing their mapping class group orbits as described in \cite{Gaiotto:2024tpl}.\footnote{The wavefunctions of \cite{Gaiotto:2024tpl} were not assumed to be crossing symmetric and hence summation of the mapping class orbit is not necessary.} Here we will outline an argument that shows that the Liouville correlator (\ref{eq:wavefunction of the universe}) is plausibly the unique solution to physically-motivated constraints and hence is the natural candidate for the state of the three-dimensional universe. 

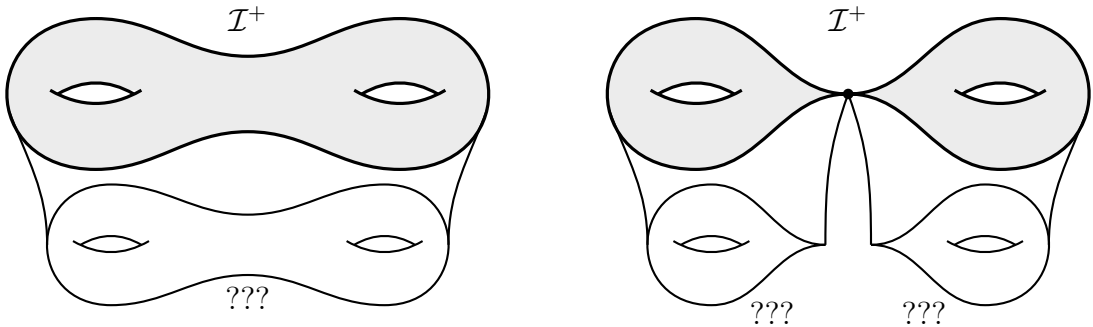
\begin{figure}
    \centering
    \begin{tikzpicture}
    \begin{scope}
        \draw[very thick, fill=gray!15!white] (-2,1) to[out=0, in=180] (0,.5) to[out=0, in=180] (2,1) to[out=0, in=0, looseness=2] (2,-1) to[out=180, in=0] (0,-.5) to[out=180,in=0] (-2,-1) to[out=180,in=180,looseness=2] (-2,1);
        \draw[very thick, bend right=30,fill=white] (-2.6,0.05) to (-1.4,0.05);
        \draw[very thick, bend left=30,fill=white] (-2.5,0) to (-1.5,0);
        \draw[very thick, bend right=30,fill=white] (1.4,0.05) to (2.6,0.05);
        \draw[very thick, bend left=30,fill=white] (1.5,0) to (2.5,0);
        \draw[thick] (-3.15,-.2) to[out=-70, in=90] (-2.64,-2);
        \draw[thick] (3.15,-.2) to[out=-110, in=90] (2.64,-2);
        \draw[thick] (-1.8,-1.2) to[out=0, in=180] (0,-1.6) to[out=0, in=180] (1.8,-1.2) to[out=0, in=0, looseness=1.8] (1.8,-2.8) to[out=180, in=0] (0,-2.4) to[out=180,in=0] (-1.8,-2.8) to[out=180,in=180,looseness=1.8] (-1.8,-1.2);
        \draw[thick, bend right=30] (-2.3,-1.95) to (-1.3,-1.95);
        \draw[thick, bend left=30] (-2.2,-2) to (-1.4,-2);
        \draw[thick, bend left=30] (2.3,-1.95) to (1.3,-1.95);
        \draw[thick, bend right=30] (2.2,-2) to (1.4,-2);
        \node at (0,1) {$\mathcal{I}^+$};
        \node at (0,-2.7) {???};
    \end{scope}
    \begin{scope}[shift={(7.9,0)}]
        \draw[very thick, fill=gray!15!white] (-2,1) to[out=0, in=180] (0,0) to[out=0, in=180] (2,1) to[out=0, in=0, looseness=2] (2,-1) to[out=180, in=0] (0,0) to[out=180,in=0] (-2,-1) to[out=180,in=180,looseness=2] (-2,1);
        \fill (0,0) circle (0.07);
        \draw[very thick, bend right=30,fill=white] (-2.6,0.05) to (-1.4,0.05);
        \draw[very thick, bend left=30,fill=white] (-2.5,0) to (-1.5,0);
        \draw[very thick, bend right=30,fill=white] (1.4,0.05) to (2.6,0.05);
        \draw[very thick, bend left=30,fill=white] (1.5,0) to (2.5,0);
        \draw[thick] (-3.15,-.2) to[out=-70, in=90] (-2.64,-2);
        \draw[thick] (3.15,-.2) to[out=-110, in=90] (2.64,-2);
        \draw[thick] (-1.8,-1.2) to[out=0, in=180] (-.3,-2) to[out=180,in=0] (-1.8,-2.8) to[out=180,in=180,looseness=1.8] (-1.8,-1.2);
        \draw[thick] (.3,-2) to[out=0, in=180] (1.8,-1.2) to[out=0, in=0, looseness=1.8] (1.8,-2.8) to[out=180, in=0] (.3,-2);
        \draw[thick, bend right=30] (-2.3,-1.95) to (-1.3,-1.95);
        \draw[thick, bend left=30] (-2.2,-2) to (-1.4,-2);
        \draw[thick, bend left=30] (2.3,-1.95) to (1.3,-1.95);
        \draw[thick, bend right=30] (2.2,-2) to (1.4,-2);
        \draw[thick] (0,0) to[out=-110, in=90] (-.3,-2);
        \draw[thick] (0,0) to[out=-70, in=90] (.3,-2);
        \node at (0,1) {$\mathcal{I}^+$};
        \node at (-1,-2.9) {???};
        \node at (1,-2.9) {???};
    \end{scope}
    \end{tikzpicture}
    \caption{The gravitational path integral prepares a state $\ket{\Psi_{g,n}}$ on $\mathcal{I}^+$. In the degeneration limit, the three-dimensional preparing geometry splits as in the right figure, which leads to the expectation of factorizability \eqref{eq:factorizability}. We are agnostic in this picture how the geometry behaves at very early times. }
    \label{fig:factorizability}
\end{figure}

The first constraint that we will apply is \emph{factorizability} of the cosmological wavefunction. Recall that in this discussion future infinity is a (possibly punctured) Riemann surface, which carries some complex structure moduli. Hence we expect that in limits where $\mathcal{I}^+$ degenerates, the cosmological wavefunction should factorize appropriately into products of lower-point wavefunctions. See for example the case of a separating degeneration of $\mathcal{I}^+$ shown in figure \ref{fig:factorizability}. In the degeneration limit, also the 3d topology factorizes and prepares the state $\ket{\Psi_{h,1+|\mathcal{J}|}(\boldsymbol{\Delta}_\mathcal{J},\boldsymbol{s}_\mathcal{J},\Delta,s)}$ on the left component and $\ket{\Psi_{g-h,1+|\mathcal{J}^c|}(\boldsymbol{\Delta}_{\mathcal{J}^c},\boldsymbol{s}_{\mathcal{J}^c},\Delta,s)}$ on the right component. Here $\mathcal{J}$ represents a subset of the $n$ momenta. Thus it is natural to expect that the wavefunction factorizes in the following way in the degeneration limit\footnote{Here the sum and integral over integral spins and dimensions runs over a complete basis of normalizable states in the Hilbert space, corresponding to dimensions in the principal series $\Delta\in 1+i\mathbb{R}$ and integer spins $s\in\mathbb{Z}$.} 
\begin{multline}
    \Psi_{g,n}(\boldsymbol{\Delta},\boldsymbol{s}) \to \\
    \sum_{s\in\mathbb{Z}} \int_{1+i\mathbb{R}}\d \Delta\, \mu(\Delta,s) \Psi_{h,1+|\mathcal{J}|}(\boldsymbol{\Delta}_\mathcal{J},\boldsymbol{s}_\mathcal{J},\Delta,s)\Psi_{g-h,1+|\mathcal{J}^c|}(\boldsymbol{\Delta}_{\mathcal{J}^c},\boldsymbol{s}_{\mathcal{J}^c},\Delta^*,-s)\, ,\label{eq:factorizability}
\end{multline}
for a suitable measure $\mu$.
In other words, the cosmological wavefunction should behave like the correlation function of a \emph{local} conformal field theory with central charge $ c= 13 + i \mathbb{R}$. This is a nontrivial constraint that ties together the wavefunctions associated with different topologies of future infinity. Combined with the gauging of the mapping class group, it implies that all cosmological wavefunctions may be computed from a set of basic structure constants associated with pairs of pants $\{\Psi_{0,3}(\Delta_1,s_1,\Delta_2,s_2,\Delta_3,s_3)\}$ that solve the bootstrap equations, precisely analogously to the correlation functions of crossing-symmetric conformal field theory.

We should mention that one might have expected to correct \eqref{eq:factorizability} by wormhole contributions, which were argued to be relevant in the de Sitter context in \cite{Chen:2020tes}. Since the inclusion of such wormhole contributions will make the story much more complicated, we will not include them.

Assuming this constraint, it is still far from obvious that this is sufficient to uniquely characterize the Liouville correlator. Indeed, the correlation functions of any local CFT with central charge $c \in 13+i \RR$ would produce a set of wavefunctions compatible with this constraint. The Liouville correlator is further distinguished by sufficiently mild behaviour near the boundaries of moduli space, which leads to \emph{normalizability} of the corresponding state: 
\begin{equation}
    \lVert\Psi_{g,n}^{(b)}(p_1,\ldots,p_n)\rVert^2 < \infty\, .
\end{equation}

The normalizability of the cosmological wavefunction defined by the Liouville correlator has its origin in the fact that Liouville CFT has an effective central charge with real part equal to one 
\begin{equation}
    \re(c_{\text{eff}}) = 1,\qquad c_{\text{eff}} \equiv c -12\Delta_{\text{min}}\, ,
\end{equation}
where $\Delta_{\text{min}}$ is the lowest-lying conformal dimension in the spectrum of the theory. In particular the identity operator does not define a normalizable state in the Hilbert space of the CFT on the circle. This leads to a much more mild growth of states at high energies and hence milder behaviour of correlation functions at the boundaries of moduli space than in ordinary compact CFTs. A CFT with effective central charge any greater than one will have correlation functions that necessarily violate the normalizability condition on the cosmological wavefunction, since the moduli space integral that defines the norm diverges; for example, while the correlation functions of any compact CFT with a normalizable vacuum would factorize as required above, they would represent non-normalizable states in the Hilbert space associated to future infinity, and it is from this point of view that the correlators of Liouville CFT are distinguished.

However the extent to which Liouville CFT represents a \emph{unique} solution to the constraints of factorizability and normalizability of the cosmological wavefunction is still not entirely clear. Our knowledge of the space of non-rational conformal field theories (even with complex central charge and relaxed unitarity constraints) is embarrassingly sparse, which limits our ability to quantify the uniqueness of the Liouville wavefunctions. For example, in the case that the effective central charge is equal to one, it suggests that the torus partition function of the theory is that of a $c=1$ free boson CFT, where we reinterpret the spectrum of operator weights $(h,\tilde h)$ in the free boson as shifted weights $(\tfrac{c-1}{24}+h,\tfrac{c-1}{24}+\tilde h)$ in the candidate wavefunction. The free boson at any finite radius is characterized by a discrete spectrum of operator weights with increasing real part, which in particular lie off the $\SL(2,\mathbb{C})$ principal series.\footnote{On the other hand, the non-compact free boson admits a continuous spectrum, which in any physical observable we may freely take to run over the principal series. Moreover the spectrum of the non-compact boson contains only scalar Virasoro primaries, and there is evidence that the solutions to the scalar-only crossing equations are governed uniquely up to operator normalization by the structure constants of Liouville CFT; see \cite{Ribault:2014hia,Collier:2017shs,Collier:2019weq}. Thus if we take our proposal (\ref{eq:spinning wavefunctions vanish}) that spinning wavefunctions vanish seriously, then the correlators of Liouville CFT represent the unique crossing-symmetric solutions to the constraints.}

We are not certain how to rule out such spectra, but let us note that there is in principle a final constraint that we have not yet leveraged, associated with \emph{unitarity} of the 3d bulk gravity theory. Although the candidate CFT correlation functions for the wavefunction of the universe need not be unitary in the usual sense of 2d CFTs (after all, they are characterized by complex central charge and complex conformal dimensions), they should encode cosmological correlators that are consistent with bulk unitarity, which is distinct from the usual CFT notion of unitarity \cite{Hogervorst:2021uvp,DiPietro:2021sjt}. However sharpening this into a precise constraint on the cosmological wavefunction requires further developing the appropriate notion of Virasoro conformal partial waves, which we discuss in more detail in the discussion section \ref{sec:discussion}. It seems plausible to us that the correlation functions of Liouville CFT represent the unique solutions to the combined constraints of factorizability, normalizability and (bulk) unitarity, but we will not attempt to prove this here.

\section{A microscopic realization of \texorpdfstring{$\text{dS}_3$}{dS3}} \label{sec:dual}
After having identified a Hilbert space together with a suitable state of the universe, we now discuss observables in this cosmology and propose a version of de Sitter holography. 
\subsection{Integrated cosmological correlators}

\paragraph{Observables.} Let us accept that the Liouville correlators are suitable wavefunctions of the universe. We are treating a model of pure de Sitter quantum gravity coupled to non-dynamical massive particles but without any matter fields. What are good observables in such a cosmology?

As in our own universe such observables are given by correlation functions in the CMB, i.e.\ cosmological correlators. Of course, there are no matter fields and gravitational waves in our model, so such correlation functions measure the correlation of moduli fluctuations in the wavefunction. In cosmology, such cosmological correlators are computed by integrating over metrics on $\mathcal{I}^+$, possibly with the insertion of some probe operators \cite{Maldacena:2002vr}, 
\be 
\left\langle \prod_{i=1}^n \mathcal{O}_i(z_i) \right\rangle =\int_{\text{metrics on $\mathcal{I}^+$}} \frac{[\mathcal{D} g]}{\text{Diff} \times \text{Weyl}}\ \big|\Psi[g]\big|^2\, \prod_{i=1}^n \mathcal{O}_i(z_i)~ , \label{eq:general cosmological correlators}
\ee
which are simply the matrix elements of the operators $\mathcal{O}_i(z_i)$.\footnote{The wavefunction $\Psi[g]$ in (\ref{eq:general cosmological correlators}) is the wavefunction before imposing any constraints as in the beginning of section \ref{subsec:wavefunction}.} In our context, there are no such probe operators. Rather, the insertions of massive particles are treated as sufficiently massive so that they backreact on the background and hence modify the cosmological wavefunction itself. 
This is in contrast to the cosmological correlators that are typically considered in the literature, which involve perturbative quantum fields on a rigid de Sitter background \cite{Hogervorst:2021uvp, DiPietro:2021sjt}.  

\paragraph{Cosmological correlators.} In quantum gravity where we integrate over the metrics on $\mathcal{I}^+$, we should also include a sum over the topology on $\mathcal{I}^+$. Thus the observables in this cosmology are\footnote{We could define an extended Hilbert space $\bigoplus_{g=0}^\infty \mathcal{H}_{g,n}^{(b)}$ and define different superselection sectors to be orthogonal. Then the right hand side of \eqref{eq:cosmological correlator as norm} can be written as a single norm. However, this should be taken with a grain of salt since as we mentioned the sum over the genus actually becomes an alternating sum.}
\be 
\sum_{g=0}^\infty g_\text{s}^{2g-2} \int_{\text{metrics on $\mathcal{I}^+$}} \frac{[\mathcal{D} g]}{\text{Diff} \times \text{Weyl}}\ \big|\Psi_{g,n}^{(b)}[g]\big|^2=\sum_{g=0}^\infty \big\lVert \Psi_{g,n}^{(b)}(p_1,\dots,p_n) \big\rVert^2~ ,\label{eq:cosmological correlator as norm}
\ee
where we take the wavefunction $\Psi_{g,n}^{(b)}$ to be the Liouville correlator \eqref{eq:wavefunction of the universe} as explained in the previous section. 
Thus, integrated cosmological correlation functions are simply the norms of the wavefunction of the universe. We should emphasize that norms appearing in the sum on the right-hand side are weighted by the factor $g_{\text{s}}^{2g-2}$ included in the definition of the inner product as in (\ref{eq:3d naive inner product}). We will see that it is natural to choose the string coupling $g_{\text{s}}$ to be purely imaginary so that the right hand side is actually an alternating sum. The phase of the string coupling was in particular not determined by the argument above which suppressed order 1 factors in \eqref{eq:ZS3 in terms of string coupling}.

\paragraph{The complex Liouville string.} Remarkably, \eqref{eq:cosmological correlator as norm} has precisely the structure of a string amplitude. The worldsheet theory is given by two coupled Liouville theories with complex central charge $c \in 13+i \RR$. Vertex operators are labelled by scalar principal series representations $\Delta_j=\frac{(b+b^{-1})^2}{2}-2p_j^2 \in 1+i \RR$. This string theory was analyzed in detail in \cite{paper1, paper2, paper3}, where it was called the complex Liouville string. See also \cite{Collier:2024kmo} for a short overview. In the conventions of \cite{paper1, paper3}, the (effective) string coupling is parametrized as 
\be 
 g_\text{s}^{-2} \sim \mathrm{e}^{2S_0} C_{\mathrm{S}^2}^{(b)}~, \label{eq:string coupling fixing}
\ee
where
\be 
C_{\mathrm{S}^2}^{(b)}=32\pi^4 \left(\frac{\sin(\pi b^2)\sin(\pi b^{-2})}{b^2-b^{-2}}\right)^2 \label{eq:CS2}
\ee
is the normalization of the string theory path integral on the sphere.  The notation of $S_0$ originates from 2d dilaton gravity \cite{Saad:2019lba} and we will also use it in this paper.
Recall that we used consistency of the three-dimensional description to fix the string coupling, which itself provides the inner product in the 3d gravity Hilbert space in terms of the gravitational path integral of pure dS$_3$ quantum gravity on the sphere \eqref{eq:three sphere partition function string coupling}
\begin{equation}\label{eq:ZS3 in terms of string coupling}
    \mathcal{Z}_{\text{grav}}^{\text{S}^3} \sim g_{\text{s}}^{-2} \sim \mathrm{e}^{2S_0} \frac{\sin(\pi b^2)^2\sin(\pi b^{-2})^2}{(b^{-2}-b^2)^2}\, ,
\end{equation}
up to an overall constant independent of $b$ (or equivalently $G_{\text{N}}$).

As was discussed in \cite{paper1}, it is also natural to include a leg factor into the definition of the string amplitude. This simply corresponds to an operator renormalization in the cosmological correlator. We choose it to have the form
\be 
\mathcal{N}_b(p)=-\frac{(b^2-b^{-2})\sin(2\pi b p)\sin(2\pi b^{-1}p)}{2\pi p \sin(\pi b^2)\sin(\pi b^{-2})}
\ee
After translating conventions, we see that the norms of the cosmological wavefunctions, which encode the contributions of fixed topologies to the integrated cosmological correlators, are precisely computed by the perturbative string amplitudes of the complex Liouville string, which were denoted by $\mathsf{A}_{g,n}^{(b)}(p_1,\dots,p_n)$ in \cite{paper1}, 
\begin{equation}\label{eq:psign norm squared}
    \prod_{i=1}^n \big(\mathrm{e}^{-S_0}\mathcal{N}_b(p_i)\big)\lVert \Psi_{g,n}^{(b)}(p_1,\ldots,p_n)\rVert^2 = \mathrm{e}^{-S_0(2g-2+n)}\mathsf{A}_{g,n}^{(b)}(p_1,\ldots,p_n)\, .
\end{equation}
Hence the main upshot is that the principal observables of the theory --- cosmological correlators of non-dynamical massive particles integrated over the metric and summed over the topology of $\mathcal{I}^+$ --- precisely correspond to the genus resummed string amplitudes of the complex Liouville string, which we denoted by $\mathsf{A}_n^{(b)}(p_1,\dots,p_n)$ in \cite{paper1,paper2,paper3}:
\begin{align}
    \Big\langle\prod_{i=1}^n\mathcal{O}_i\Big\rangle &= \prod_{i=1}^n \big(\mathrm{e}^{-S_0}\mathcal{N}_b(p_i)\big)\sum_{g=0}^\infty \lVert\Psi^{(b)}_{g,n}(p_1,\ldots,p_n)\rVert^2\nonumber\\
    &= \sum_{g=0}^\infty \mathrm{e}^{-S_0(2g-2+n)}\mathsf{A}_{g,n}^{(b)}(p_1,\ldots,p_n)\nonumber\\
    &\equiv \mathsf{A}_n^{(b)}(p_1,\ldots,p_n)\,.
\end{align}
Let us pause to emphasize an important point. The complex Liouville string and its dual matrix model both have two independent parameters: a genus-counting parameter $\mathrm{e}^{-S_0}$ or $g_{\text{s}}$ as usual in string theory, together with an independent continuous parameter $b$ that characterizes the central charge of the worldsheet theory. In dS$_3$ gravity there is only one dimensionless parameter $c=1+6(b+b^{-1})^2$ related to $\ell_\text{dS}/G_\text{N}$ semiclassically via $c \sim 6k \sim \frac{3 i \ell_\text{dS}}{2 G_\text{N}}$. Indeed we argued that in the application to 3d gravity the string coupling $g_\text{s}$ (which plays the role of normalizing the inner product as in (\ref{eq:3d naive inner product})) is determined in terms of $b$ via \eqref{eq:three sphere partition function string coupling} through consistency of the 3d description, and so in writing the genus-resummed string amplitudes above we dropped the label of the topological expansion parameter. This quantifies the sense in which higher topology contributions to the integrated cosmological correlators are suppressed in the semiclassical limit, despite the absence of a genus-counting parameter intrinsic to 3d gravity. Notice in particular that the string coupling is tiny for large universes since 
\be 
|g_\text{s}|\sim |\mathcal{Z}^{\text{S}^3}_\text{grav}|^{-\frac{1}{2}} \sim \exp\left(-\frac{\pi \ell_{\text{dS}}}{4G_\text{N}}\right)~,
\ee
where we used the leading value of $\mathcal{Z}_{\text{grav}}^{\text{S}^3}$ in terms of the area of the cosmological horizon, which we will review below, see \eqref{eq: gravity PI}.
Thus topology fluctuations are strongly suppressed for large universes as expected. The exponential behaviour in $\ell_\text{dS}/G_\text{N}$ is expected from estimating the on-shell action on a tunneling instanton from one topology to another \cite{Coleman:1980aw, Anninos:2012qw}.

\paragraph{Light states?} Punctures in $\Sigma_{g,n}$ extend to Wilson lines in the three-dimensional bulk and are interpreted as the worldlines of massive non-dynamical scalar particles of mass \eqref{eq:conformal weight mass}. One may also wonder whether one can consider light states with mass $m<\ell_{\text{dS}}^{-1}$ corresponding to the complementary series representations with $0<\Delta<1$ via \eqref{eq:conformal weight mass}. 
Even though they are unitary, such representations do not appear from the quantization of coadjoint orbits and do not naturally appear in the quantization of the phase space. Similarly, conformal blocks with internal dimensions in the complementary series are non-normalizable with respect to the inner product (\ref{eq:3d naive inner product}).

Nevertheless, one might wish to consider the analytic continuation of the string amplitudes $\mathsf{A}_{g,n}^{(b)}$ in the Liouville momenta away from the branch $p\in \mathrm{e}^{-\frac{\pi i}{4}}\mathbb{R}_{>0}$. The  values corresponding to complementary series representations lie within the domain of analyticity of the string amplitudes of the complex Liouville string, so the analytic continuation of the integrated cosmological correlators is in principal straightforward. However in this regime it is clear that the analytically continued string amplitudes can no longer be interpreted as norms, since $\Delta$ and $2-\Delta$ are no longer complex conjugates. Thus the analytic continuation of the complex Liouville string amplitudes does not seem to capture the cosmological correlators of particles with masses in the complementary series. 
\paragraph{Analytic structure of the integrated cosmological correlators.}

In the bootstrap approach to cosmological correlators, the analytic structure of the cosmological wavefunction and of the correlators themselves provides an important physical input. As emphasized in \cite{paper1}, the string amplitudes of the complex Liouville string $\mathsf{A}_{g,n}^{(b)}(\boldsymbol{p})$ exhibit a rich analytic structure, characterized by an infinite set of poles and discontinuities when viewed as complex functions of the Liouville momenta $p_i$ of the external vertex operators. This analytic structure leads to very stringent constraints on the string amplitudes, which together with symmetry considerations provides a means to bootstrap the string amplitudes (which in simple cases yields a solution that is unique up to an assumption about the asymptotic growth). The integrated cosmological correlators of massive particles in dS$_3$ thus inherit this analytic structure (viewed as functions of the particle masses via (\ref{eq:conformal weight mass})) by the de Sitter/matrix model duality discussed here.

For example, the Liouville correlator and hence the cosmological wavefunction $\Psi^{(b)}_{g,n}(\boldsymbol{p})$ itself exhibits an infinite set of poles in the Liouville momenta of the external vertex operators when
\begin{equation}\label{eq:Liouville correlator poles}
    \pm p_1 \pm p_2 \cdots \pm p_n = \frac{2g-2+n}{2}(b+b^{-1}) + rb+sb^{-1},\quad r,s\in\mathbb{Z}_{\geq 0}
\end{equation}
for any choice of $\pm$ signs.
They arise when the Liouville background charge is saturated and the correlator that defines the cosmological wavefunction reduces to a Coulomb gas/linear dilaton correlator together with a divergent zero mode integral.
These poles of the cosmological wavefunction, together with another infinite set of complex-conjugated poles from the bra, thus descend to poles of the full integrated cosmological correlator. 

At first glance the poles (\ref{eq:Liouville correlator poles}) of the cosmological wavefunction are difficult to interpret in the context of three-dimensional gravity. Recall that the Liouville momenta are related to the particle masses via
\begin{equation}
    \frac{b^2+b^{-2}}{2}-2p_i^2 = \pm \sqrt{1-m_i^2\ell_{\text{dS}}^2}\, .
\end{equation}
In the semiclassical $b\to 0$ limit (in which the imaginary part of the central charge is taken to infinity), we can think of the massive particles as sourcing conical defects of deficit angle $4\pi b p_i$.\footnote{Recall that before analytic continuation the combination $bp_i$ is taken to be real.} In this limit, the first ($r=s=0$) pole occurs precisely when the Cauchy surface $\Sigma_{g,n}$ no longer admits a hyperbolic metric, in other words, when the expanding universe (\ref{eq:inflating universe metric}) goes off-shell. From this perspective it is not surprising that the cosmological wavefunction exhibits a singularity. These poles are reminiscent of the ``total energy'' singularities of the cosmological wavefunction as discussed for example in \cite{Baumann:2021fxj}. That the singularities occur at configurations that are additive in the Liouville momenta rather than in the particle masses is loosely evocative of the structure of multi-particle bound states in AdS$_3$ quantum gravity \cite{Collier:2018exn}, where the Liouville momentum defines a sort of quantum deficit angle that is the additive parameter in the spectrum of multi-particle bound states.

The integrated cosmological correlators also exhibit an infinite set of discontinuities that are intrinsically related to the integration over metrics at $\mathcal{I}^+$. In dialing the momenta $p_i$ away from the regime $p_i \in \mathrm{e}^{\mp\frac{\pi i}{4}}\mathbb{R}$ associated with principal series conformal dimensions, it may be that the integral over metrics that defines the cosmological correlators no longer converges. In this case the integrated correlators must be defined by analytic continuation from the region where the integral converges, leading to branch cuts. The discontinuities are associated with boundary divisors of moduli space, where the late-time surface $\mathcal{I}^+$ either divides into two surfaces connected at a nodal point or a surface of lower genus with two joining nodal points. The formula for the discontinuity is given by \cite{paper1}
\begin{multline} 
\Disc_{p_*=0} \mathsf{A}_{g,n}^{(b)}(\boldsymbol{p})=2\pi i \Bigg[\Res_{p=p_*} \sum_{\begin{subarray}{c} 
h=0,\dots,g \\ 
I \subset \{1,2,\dots,n\} \\ \text{ stable}
\end{subarray}} 2p\, \mathsf{A}^{(b)}_{h,|I|+1}(\boldsymbol{p}_I,p)\mathsf{A}^{(b)}_{h,|I^c|+1}(\boldsymbol{p}_{I^c},p)\\
+\Res_{p=\frac{1}{2}p_*}2p\,  \mathsf{A}^{(b)}_{g-1,n+2}(\boldsymbol{p},p,p)\Bigg]~.\label{eq:discontinuity cutting}
\end{multline}
Here $p_*$ corresponds to a pole of the simpler string amplitudes that appear on the right-hand side. These discontinuities are elegantly captured by cutting rules precisely analogous to those of perturbative quantum field theory \cite{paper2}. They are computed by cutting stable graphs --- the Feynman diagrams of the closed string theory corresponding to degenerations of the worldsheet surface --- along internal lines (corresponding to nodal points of the degenerated worldsheet) such that one of the resulting sub-diagrams develops a singularity, and computing the residue at the pole.

We notice that the cutting rules (\ref{eq:discontinuity cutting}) that determine the discontinuities of the integrated cosmological correlators are structurally very similar to cutting rules for cosmological correlators that have been discussed in the literature on the cosmological bootstrap \cite{Baumann:2021fxj,Melville:2021lst,Goodhew:2021oqg}. These cutting rules follow from the ``cosmological optical theorem,'' an infinite set of relations among the coefficients of the cosmological wavefunction that follow from perturbative unitarity of time evolution in de Sitter space \cite{Goodhew:2020hob}. We have not attempted to make a direct connection between the above cutting rules and unitarity of the three-dimensional bulk, but it would be interesting to better understand this point. They seem more directly related to perturbative unitarity of the two-dimensional target space of the complex Liouville string. 

\subsection{de Sitter holography}
We have found a new interpretation of the quantities $\mathsf{A}_{g,n}^{(b)}(\boldsymbol{p})$ as the cosmological correlators of massive non-dynamical particles in 3d de Sitter quantum gravity. Remarkably, the main point of our paper \cite{paper2} was to demonstrate that $\mathsf{A}_{g,n}^{(b)}(\boldsymbol{p})$ are related to the resolvents of a double scaled matrix model in a precise way. We are thus led to a relation between the cosmological correlators and a matrix model. 
The matrix model captures \emph{all} observables in this cosmology and thus constitutes a complete dual. We included a short review on some aspects of the matrix model in appendix~\ref{app:matrix model dual}.

This constitutes a precise version of de Sitter holography, even though the dual theory is not a Euclidean CFT, but a matrix model. Relatedly, the dual theory does not compute the wavefunction of the universe but rather the cosmological correlators.  The wavefunction itself does not appear as a gauge invariant quantity in the correspondence since one is naturally led to integrate over the metric on $\mathcal{I}^+$ in de Sitter space. At least in this context, it is thus misguided to think of the dual theory as `living on the boundary of spacetime'. In fact, it does not make reference to spacetime at all, but only computes the observables $\mathsf{A}_{g,n}^{(b)}$. Questions about the bulk reconstruction and emergence of time from the cosmological correlators thus seem much more daunting than in the AdS setting.

\paragraph{The phase of the string coupling.} We mentioned above that we chose the string coupling in \eqref{eq:cosmological correlator as norm} to be purely imaginary. We are not aware of a precise bulk argument dictating this choice.\footnote{We would expect it to be fixed in principle from the bulk, but it would require a somewhat subtle off-shell calculation.} However, as discussed in \cite{paper2}, this choice is necessary in the matrix model and arises from the requirement that the density of eigenvalues is positive, which is why we will assume it in the following.

\paragraph{(Doubly) non-perturbative effects.} As we demonstrated in \cite{paper3}, the sum over topologies that defines the genus-resummed cosmological correlators is an asymptotic series; the individual cosmological correlators grow factorially with the genus of $\mathcal{I}^+$, roughly
\begin{equation}
    \lVert\Psi_{g,n}^{(b)}(\boldsymbol{p})\rVert^2 \sim (2g)!~. \label{eq:large g growth}
\end{equation} 
For a more precise formula see \cite{paper3}.
In this sense, the cosmological correlators exhibit an instability towards higher topology of future infinity. A conceptually similar observation was previously made in the context of four-dimensional higher-spin dS/CFT in \cite{Banerjee:2013mca}, although the precise details of the high-genus growth differ. 

This is also similar to what was observed in \cite{Maloney:2015ina, Eberhardt:2022wlc} for chiral 3d gravity, where one computes the dimension of the analogous Hilbert space $\mathcal{H}_{g,n}^{(b)}$ obtained by quantizing $\mathcal{M}_{g,n}$ (as opposed to $T^*\mathcal{M}_{g,n}$ as we did in this paper). The dimension of the Hilbert space is given semiclassically by the volume of phase space, which in turn is given by the Weil-Petersson volumes. This has the same large $g$ growth as \eqref{eq:large g growth}. For a sufficiently random state with coordinate entries of order 1, one thus expects the norm to grow like $(2g)!$.

This instability is in tension with the supposed Hilbert space of finite dimension $\mathrm{e}^{S_\text{dS}}$ that describes de Sitter space. Strictly speaking, to reproduce the genus expansion and the behavior \eqref{eq:large g growth}, we need to consider a double-scaled matrix model with an infinite $N$. However, via resurgence, \eqref{eq:large g growth} also tells us that the string amplitudes receive non-perturbative corrections from ZZ-instantons in order to get a well-defined non-perturbative completion. Such corrections are of order
\be 
\exp(\# i |g_\text{s}|^{-1})\sim\exp\left(\# i \exp\left(\frac{\ell_\text{dS}}{4 \ell_\text{Planck}}\right)\right) \label{eq:non perturbative corrections} 
\ee
with $\#$ representing an order 1 real number. Because of the $i$, they are actually of order 1, but extremely rapidly oscillating in the value of $\ell_\text{dS}/\ell_\text{Planck}$. This rapid oscillation signals the discreteness of the Hilbert space and can be detected precisely from the plateau in the spectral form factor \cite{Cotler:2016fpe}. The $i$ in \eqref{eq:non perturbative corrections} stems from the fact that the tensions of the ZZ-instantons are all purely imaginary in this model \cite{paper3}.

While we do not know how to compute corrections of the form \eqref{eq:non perturbative corrections} directly within 3d gravity, we used string theory technology in \cite{paper3} to compute the leading non perturbative corrections and matched them to the matrix model. 

Non-perturbative corrections are ambiguous. There are different inequivalent non-perturbative completions of the perturbative data of the double scaled matrix model. These completions amount to picking Stokes constants in the matrix model, or a steepest descent contour for the integral over the eigenvalues. These ambiguities exist both on the string theory side and the matrix model side and thus presumably also in 3d gravity. We will find some evidence below that there is a preferred non-perturbative completion that predicts the correct de Sitter entropy.

\subsection{Review of Gibbons-Hawking de Sitter entropy proposal}\label{subsec:subsectionGH}

We now turn to the second main claim of the paper: that the dual matrix model possesses the right number of states as predicted by the de Sitter entropy. Let us hence begin by reviewing the Gibbons-Hawking entropy proposal. The reader familiar with this may safely jump to section~\ref{subsec:microstates}.

\paragraph{Entropy from the gravitational path integral.}
The question of whether the de Sitter event horizon carries microscopic degrees of freedom akin to a de Sitter entropy has been a long standing question. According to a conjecture of Gibbons and Hawking \cite{Gibbons:1976ue, Gibbons:1977mu}, the de Sitter entropy can, at least macroscopically be extracted from the gravitational path integral
\begin{equation}\label{eq: SdS}
    S_{\mathrm{dS}} = \log \mathcal{Z}_{\mathrm{grav}}~,\quad \mathcal{Z}_{\mathrm{grav}} = \sum_{\mathcal{M} ~\mathrm{compact}}\int [\mathcal{D}g]\,\mathrm{e}^{-S_{\mathrm{EH}}[g,\Lambda]}~,\quad \Lambda >0~.
\end{equation}
In the above equation $S_{\mathrm{EH}}[g,\Lambda]$ denotes the Euclidean Einstein Hilbert action with positive cosmological constant. The leading contribution to the path integral (\ref{eq: SdS}) in $d$ dimensions stems from the $d$-dimensional sphere. The $d$-sphere is a saddle of the Euclidean Einstein Hilbert action and reproduces the Gibbons-Hawking area law
\begin{equation}
  \mathcal{Z}_{\mathrm{grav}} \approx \mathrm{e}^{\frac{A_{\text{h}}}{4G_{\text{N}}}}~,  
\end{equation}
where $A_{\text{h}}$ denotes the area of the cosmological horizon in dS$_d$.

\paragraph{The three-sphere partition function.}
Restricting now to three dimensions, we can study the three-sphere partition function
\begin{equation}
 \mathcal{Z}_{\mathrm{grav}}^{\text{S}^3} = \int \frac{[\mathcal{D}g]}{\mathrm{vol}(\text{Diff}(\mathrm{S}^3))}\, \mathrm{e}^{- S_{\mathrm{EH}}[g,\Lambda]}~,\quad -S_{\mathrm{EH}}[g,\Lambda] =\frac{1}{16\pi G_{\text{N}}}\int_{\text{S}^3} \d^3x\sqrt{g}\left(\mathcal{R}- 2\Lambda\right)~,
\end{equation}
where $\Lambda>0$ and $\mathrm{vol}(\text{Diff}(\mathrm{S}^3))$ denotes the volume of the three-dimensional diffeomorphism group. The equations of motion admit the round sphere saddle with
\begin{equation}
 \mathcal{R} = 6\Lambda~,
\end{equation}
 and the on-shell action leads to the saddle point approximation
 \be 
\mathcal{Z}_{\mathrm{grav}}^{\text{S}^3} \approx \mathrm{e}^{\frac{\pi \ell_{\mathrm{dS}}}{2G_{\text{N}}}}~. \label{eq: gravity PI}
\ee
Using that the area of the dS$_3$ horizon is $A_{\text{h}} = 2\pi \ell_{\mathrm{dS}}$, to leading order the gravitational path integral on a three sphere agrees with the Gibbons-Hawking area law \cite{Gibbons:1976ue, Gibbons:1977mu}.

One should also study fluctuations around the three-sphere saddle $g_{\mu\nu} = g_{\mu\nu}^{\text{S}^3} + h_{\mu\nu}$. We decompose the fluctuation metric $h_{\mu\nu}$ as
\begin{equation}\label{eq: hmunu decomposition}
    h_{\mu\nu} = h_{\mu\nu,\mathrm{TT}} + \frac{1}{\sqrt{2}}(\nabla_\mu\xi_\nu + \nabla_\nu\xi_\mu) +\frac{1}{\sqrt{3}} g_{\mu\nu} \tilde{h}~,
\end{equation}
where TT denotes the transverse traceless components, and  $\tilde{h}= h_\lambda^{~\lambda}$ is the trace. The vector fields $\xi_\mu$ denote the pure gauge components. For (\ref{eq: hmunu decomposition}) to be unique we further require that 
\begin{equation}\label{eq: orthogonality}
    \xi_\mu \perp \xi_\mu^{\mathrm{KV}}~\quad \quad \mathrm{and}\quad\quad   \tilde{h} \perp \nabla^\mu\xi_\mu^{\mathrm{CKV}}~,
\end{equation}
where KV and CKV denote the Killing and conformal Killing vectors respectively. 
For the transverse traceless and the trace component we find that the quadratic fluctuations takes the form
\begin{subequations}
\begin{align}\label{eq: hTT}
        S_{\mathrm{TT}}[h]&= -\frac{1}{96\pi G_{\text{N}}}\int_{\text{S}^3} \d^3 x\sqrt{{g}^{\text{S}^3}} \,h^{\mu\nu,\mathrm{TT}}\!\left(-{\nabla}_{(2)}^2 + 2\right)\!h_{\mu\nu,\mathrm{TT}}\\ \label{eq: conformal factor}
        S_{\mathrm{trace}}[h]&= -\frac{1}{96\pi G_{\text{N}}}\int_{\text{S}^3} \d^3 x\sqrt{{g}^{\text{S}^3}} \,\tilde{h}\!\left(-{\nabla}_{(0)}^2 - 3\right)\!\tilde{h}~,
\end{align}
\end{subequations}
where $\nabla_{(s)}^{2}$ denotes the spherical Laplacian for spin $s$ fields. 
In particular we see that the trace component leads to an unbounded Gaussian action in the gravitational path integral $\mathcal{Z}_\mathrm{grav}^{\text{S}^3}$ and, taken at face value $\mathcal{Z}_\mathrm{grav}^{\text{S}^3}$ would be infinite. 

To define the path integral $\mathcal{Z}_{\mathrm{grav}}^{\text{S}^3}$ one defines an alternative contour for the conformal mode \cite{Gibbons:1978ac, Polchinski:1988ua}, which renders the quadratic action (\ref{eq: conformal factor}) bounded. To illustrate this strategy we consider the expansion of $\tilde{h}$ into eigenfunctions of the three-sphere Laplacian 
\begin{equation}\label{eq: mode expansion}
    \tilde{h}(\Omega) = \sum_{l,m,n}\mathsf{h}_{l,m,n} Y_{l,m,n}(\Omega)~
\end{equation}
where $\Omega$ is a point on $\text{S}^3$ and
\begin{equation}
    -{\nabla}_{(0)}^2 Y_{l,m, n}(\Omega) = l(l+2)Y_{l, m, n}(\Omega)~,\quad m=0,\ldots , l~,\quad n= -l,\ldots ,l~.
\end{equation}
Each $l$ is $(l+1)^2$-fold degenerate; It is now clear that only the $l=0$ mode is Gaussian suppressed, whereas eigenfunctions with $l\geq 2$ lead to a Gaussian unsuppressed contribution to $\mathcal{Z}_{\mathrm{grav}}^{\text{S}^3}$. The four-fold degenerate $l=1$ modes are zero modes of (\ref{eq: conformal factor}). They correspond to the four conformal Killing vectors of $\text{S}^3$ and hence do not satisfy \eqref{eq: orthogonality}, meaning that they should be excluded from the path integral. To deal with the Gaussian unsuppressed modes with $l \ge 2$, we Wick rotate $\tilde{h}\rightarrow \pm i\tilde{h}$, leading to\footnote{In this equation, we have not kept track of the overall normalization of the Gaussian integral. This will not influence the discussion of the phase that follows.} 
\begin{equation}\label{eq:infinite product}
    \int [\mathcal{D}g]\, \mathrm{e}^{-S[g]}\bigg|_{\text{one-loop, unsuppressed}} =  \prod_{l \ge 2}\left(\frac{\mathrm{e}^{\pm {\pi i}}}{l(l+2)-3}\right)^{\frac{(l+1)^2}{2}}~,
\end{equation}
where the exponential is the Jacobian of the Wick rotation $\tilde{h}\rightarrow \pm i \tilde{h}$. We included both directions for the Wick rotations, although a proposal was made in \cite{Maldacena:2024spf} for a definite choice. 

The infinite product in (\ref{eq:infinite product}) in particular contributes a phase to the gravitational path integral \cite{Polchinski:1988ua} which we can obtain from a zeta-function regularization. For this we introduce a zeta function for the unsuppressed modes,
\begin{equation}\label{eq:zetas}
    \zeta_{\text{S}^3}(s) = \mathrm{e}^{\pm \pi i s}\sum_{l\geq 2}\frac{(l+1)^2}{(l(l+2)-3)^s}~,\quad s>0~.
\end{equation}
As usual in zeta-function regularization $\mathrm{e}^{\frac{1}{2}\zeta_{\text{S}^3}'(s)}|_{s=0}$ gives the infinite product in (\ref{eq:infinite product}):
\begin{equation}
  \prod_{l\ge 2}\left(\frac{\mathrm{e}^{\pm {\pi i}}}{l(l+2)-3}\right)^{\frac{(l+1)^2}{2}} = \mathrm{e}^{\frac{1}{2}\zeta_{\text{S}^3}'(s)}|_{s=0} = \mathrm{e}^{\pm \frac{\pi i}{2}\zeta_{\text{S}^3}(0)} \times \mathrm{real ~contributions}~. 
\end{equation}
Thus the phase of the answer can be computed from $\zeta_{\text{S}^3}(0)$, which in turn can be related to the standard Riemann zeta-function by expanding in large $\ell$ as follows,
\begin{multline}
     \frac{(l+1)^2}{(l(l+2)-3)^s} \\
     = \frac{1}{l^{2s}}\left(l^2+2(1-s)l+(1+s+2s^2)-\frac{2s(1+9s+2s^2)}{3l} + s \, \mathcal{O}(l^{-2})\right)~.
\end{multline}
The error term is absolutely convergent for $s=0$ and its prefactor vanishes. Thus it does not contribute to $\zeta_{\text{S}^3}(0)$. In the remaining terms we can put $s=0$, except in the harmonically divergent term where we need to keep terms linear in $s$, since those will combine with the pole of $\zeta(s)$ at $s=1$. Thus
\begin{align}
    \zeta_{\text{S}^3}(0)&= \lim_{s\rightarrow 0}\sum_{l \geq 2} \frac{1}{l^{2s}}\left(\!(1+l)^2 - \frac{2s}{3l}\!\right) \nonumber\\
    &=-4+ \lim_{s \to 0} \big(\zeta(2s-2)+2 \zeta(2s-1)+\zeta(2s)-\frac{2}{3} s \zeta(2s+1)\big)
    = -5~.
\end{align}
Consequently the trace fluctuations (\ref{eq: conformal factor}) contribute a phase $\mathrm{e}^{\pm \frac{5\pi i}{2}}$ to the gravitational path integral $\mathcal{Z}_{\mathrm{grav}}$.\footnote{This phase can also be obtained from a more slick argument. If we would have rotated all modes in the trace fluctuations including the $l=1$ and $l=0$ modes, we would have produced a formal factor $i^\infty$, which should be real by ultralocality of the path integral measure and the fact that there are no appropriate counter terms in odd that could spoil that reality. The $i^{\pm 5}$ then precisely originates from the non Wick-rotated modes \cite{Polchinski:1988ua}. However, it is gratifying to see that this somewhat abstract argument is confirmed in the explicit zeta-function calculation.} 
Due to the $+2$  the transverse traceless action (\ref{eq: hTT}) is Gaussian suppressed and putting everything together we finally obtain
\begin{equation}\label{eq: Zgrav S3}
    \mathcal{Z}_{\mathrm{grav}}^{\text{S}^3} \approx \mathrm{e}^{\pm \frac{5\pi i}{2}}\,\mathrm{e}^{\frac{\pi \ell_{\mathrm{dS}}}{2G_{\text{N}}}}\left(\frac{2 G_{\text{N}}}{\pi \ell_{\mathrm{dS}}}\right)^{3}\frac{1}{\mathrm{vol}(\mathrm{SO}(4))}\frac{\sideset{}{'}\det (-\nabla_{(1)}^2-2)^{1/2}}{\det(-\nabla_{(2)}^2+2)^{1/2}}~.
\end{equation}
The volume of SO$(4)$ and the term $\left(\frac{2 G_{\text{N}}}{\pi \ell_{\mathrm{dS}}}\right)^{3}$ come from a careful consideration of the volume of the three-dimensional diffeomorphism group $\text{Diff}(\mathrm{S}^3)$ \cite{Anninos:2020hfj}. In particular, the exponent 3 originates from the dimension of the isometry group, $3= \frac{1}{2}\mathrm{dim}(\SO(4))$. The functional determinant in the numerator comes from the Fadeev-Popov fields for the gauge component of the metric, and the prime indicates that we omit the $l=1$ eigenvalue of the spherical Laplacian $\nabla^2_{(1)}$ of spin one fields. The $l=1$ modes are six fold degenerate, corresponding exactly to the Killing vectors in (\ref{eq: orthogonality}) which we need to omit in the decomposition of $h_{\mu\nu}$ (\ref{eq: hmunu decomposition}).  We refer to \cite{Law:2020cpj} for more details.

\subsection{Counting microstates} \label{subsec:microstates}
We now count the number of microstates in the matrix model.
Before double scaling, we consider a two-matrix model of the form
\begin{equation}
    \int [\d M_1][\d M_2]\, \mathrm{e}^{-N \tr(V_1(M_1) + V_2(M_2) - M_1 M_2)}~, \label{eq:two matrix integral main text}
\end{equation}
where $M_1$ and $M_2$ are two $N\times N$ hermitian matrices.
\paragraph{What is the entropy?} A priori it is not clear what this number should be since the matrix model is not a conventional quantum mechanical system. 
To simplify thinking about the system, it is useful to integrate out $M_2$ in \eqref{eq:two matrix integral main text} so that we only need to think about a single matrix model (albeit with a non-analytic potential). This is always possible since we do not consider observables with respect to the second matrix.

If we think about the matrix model in the spirit of Saad, Shenker and Stanford \cite{Saad:2019lba}, we would interpret the $N \times N$ matrix $M_1$ as a Hamiltonian for a quantum mechanical system with Hilbert space of dimension $N$. This would suggest that we should define the entropy of the system as $\log N$.

However, we will see that this does not lead to the correct answer in this context. Considering the matrix $M_1\in \RR^{N\times N}$ instead as a classical random variable would lead one to associate the entropy $\log (N^2)=2 \log(N)$. We will take this as a working assumption for now and discuss possible interpretations in the discussion section~\ref{sec:discussion}.

\paragraph{The effective number of eigenvalues.} With these preparations, we can now count the number of microstates predicted by the matrix model. As already briefly mentioned in the introduction section~\ref{sec:introduction}, this number is naively infinite since we are dealing with a double-scaled matrix integral and the integral over the density of eigenvalue given by \eqref{eq: eigenvalue distribution} is not normalized. However, the density of eigenvalues is oscillating and a straightforward way to define the total number of eigenvalues is to only include those in the first positive region of the eigenvalue density. See figure~\ref{fig:eigenvalue density} for a schematic picture of the eigenvalue density. This resonates also with the discussion of the non-perturbative completion of the matrix model, since the contour of the integral over the eigenvalues has to be deformed away from the real axis for sufficiently large energies in order to ensure convergence of the matrix integral \cite{paper3}. The simplest choice is to deform the contour at the first extremum of the effective potential, which corresponds to the first ZZ-instanton. Remarkably, the location of the first such extremum precisely coincides with the first zero of the density of eigenvalues! This allows us to precisely consider only eigenvalues in the first interval of positivity as actual `states' of the matrix model. In our conventions of the density of eigenvalues \eqref{eq: eigenvalue distribution}, this interval is $2\le E \le E_0=2 \cos(\pi b^{-2})$.
 This hard cut off might be a bit of a brutal truncation, but we will see that it leads to sensitive results (even non-perturbatively in $b$!).

In any case, we take the effective number of eigenvalues in the matrix model to be given by
\be \label{eq:Neff def}
N_\text{eff}= \int_2^{E_0}\d E \ \mathrm{e}^{S_0} \rho_0(E)~,
\ee
where $E_0=2 \cos(\pi b^{-2})$ is the first zero of the density of eigenvalues. 

Thus the microscopic de Sitter entropy should be, 
\begin{align} 
S_{\text{dS}}^\text{micro}&=2\log \int_2^{E_0} \d E\ \mathrm{e}^{S_0}\, \rho_0(E) \nonumber\\
&=2S_0+ 2\log\left( \frac{-4i \sin(\pi b^2) \sin(\pi b^{-2})}{\pi(b^{-2}-b^2)}\right) \nonumber\\ 
&=2\log \left(\frac{1}{2\pi i} T_{1,1}^{(b)}\right)~.\label{eq:SdS matrix}
\end{align}
In the last equality we made use of the fact that the integral is precisely the integral around a closed cycle on the spectral curve that defines the tension $T_{1,1}^{(b)}$ of the first ZZ-instanton, as discussed in appendix~\ref{app:matrix model dual} and \cite{paper3}. This tension is explicitly given by
\begin{equation}\label{eq:tension explicit form}
    T_{1,1}^{(b)} = \mathrm{e}^{S_0}\frac{8b^2 \sin(\pi b^2)\sin(\pi b^{-2})}{1-b^4} \equiv \mathrm{e}^{S_0} \widehat{T}_{1,1}^{(b)}\, .
\end{equation}

\paragraph{Microscopic de Sitter entropy.} To claim victory, we have to verify that this agrees with the entropy as computed from the sphere partition function in the bulk, which in turn is computed by the three-sphere partition function, i.e.
\be \label{eq:SdS tension logZS3}
S_{\text{dS}}^\text{micro}=2\log \left(\frac{1}{2\pi i} T_{1,1}^{(b)}\right) \overset{?}{=} \log |\mathcal{Z}_{\text{grav}}^{\text{S}^3}|=S_{\text{dS}}~.
\ee
Of course, the left hand side still contains $S_0$, so we have to use (\ref{eq:ZS3 in terms of string coupling}) and \eqref{eq:string coupling fixing}, which writes the sphere partition function in terms of the string coupling together with the explicit form of the string coupling in the matrix model, to eliminate it from the equation.

When we insert \eqref{eq:string coupling fixing} and use the explicit expression in \eqref{eq:SdS matrix}, we are led to the condition
\be 
(\widehat{T}_{1,1}^{(b)})^2 \overset{?}{\sim} C_{\mathrm{S}^2}^{(b)}~, \label{eq:tension normalization test}
\ee
where recall that the tilde $\sim$ means equality up to order 1 factors and that $\widehat{T}_{1,1}^{(b)}$ was defined in \eqref{eq:tension explicit form}.
The explicit form of $C_{\mathrm{S}^2}^{(b)}$ given in \eqref{eq:CS2} and of the tension in (\ref{eq:tension explicit form}) shows that this equality indeed holds! We want to emphasize that this is a genuinely nontrivial check that the count of degrees of freedom in the matrix model (as defined in (\ref{eq:Neff def})) reproduce the de Sitter entropy as defined by the logarithm of the sphere partition function, regardless of the latter's specific value. The condition (\ref{eq:tension normalization test}) \emph{does not} hold in the Virasoro minimal string \cite{Collier:2023cyw}, for example. The reason we say that it holds regardless of the specific value of the sphere partition function is because the factor $\mathrm{e}^{S_0}$ that appears in (\ref{eq:SdS tension logZS3}) also appears in (\ref{eq:string coupling fixing}) and hence drops out so that one needs only to confirm (\ref{eq:tension normalization test}). Thus, the matching with the de Sitter entropy is a structural feature of the duality that doesn't depend on the specific value of $\mathcal{Z}_{\text{grav}}^{\text{S}^3}$. This is both fortunate and unfortunate. It is fortunate since we are led to a sharp prediction in the matrix model which is independent of the value of $\mathcal{Z}_{\text{grav}}^{\text{S}^3}$ and which we will test momentarily. It is however also unfortunate because the duality works irrespective of the value of $\mathcal{Z}_{\text{grav}}^{\text{S}^3}$. We might have hoped that the dual description has the de Sitter entropy baked in as an essential ingredient.

This matching is exact in $G_{\text{N}}$ and in particular includes all perturbative corrections around $\mathrm{S}^3$. Extending the matching also to the (doubly) non-perturbative sector is more ambiguous since it depends on how we precisely define the cutoff on the eigenvalues. In particular, it reproduces to leading order the Gibbons-Hawking entropy for de Sitter, see eq.~\eqref{eq:sphere partition function}. We view this as major evidence that the matrix model indeed constitutes a complete dual to $\mathrm{dS}_3$ gravity.

We can use the matching also to fix the order 1 constants in \eqref{eq:string coupling fixing} and \eqref{eq:ZS3 in terms of string coupling} and put
\be 
g_s^{-1}=T_{1,1}^{(b)}=2\pi i |\mathcal{Z}_{\text{grav}}^{\text{S}^3}|^{\frac{1}{2}}~. \label{eq:T11 ZS3 normalization}
\ee
This is the natural definition for the string coupling since it controls the large genus asymptotics of the string amplitudes via resurgence \cite{paper3} and thus provides an intrinsic definition of the string coupling. We called \eqref{eq:T11 ZS3 normalization} the effective string coupling in \cite{paper3}. In particular, the string coupling is purely imaginary (the sign is ambiguous).

\section{Discussion} \label{sec:discussion}
We will now discuss a few open questions and future directions.

\paragraph{The nature of holographic duality for dS$_3$.}
The main result of this work is a novel form of  holographic duality between a two-matrix integral and late-time cosmological correlators of massive particles in dS$_3$ quantum gravity, which automatically incorporates the integration over metrics of the late-time surface. This is a somewhat unfamiliar paradigm and it is reasonable to wonder whether this the only available holographic interpretation of dS$_3$ quantum gravity. 

On the gravity side, many technical elements of the story (particularly the discussion of the conformal block Hilbert space defined by quantizing the gravity phase space on an initial value surface) closely mirrored recent developments in AdS$_3$ quantum gravity \cite{Collier:2023fwi,Collier:2024mgv}. 
These recent developments systematically facilitate the exact computation of the Euclidean gravitational path integral on any fixed topology that solves Einstein's equations, which is delicately reproduced by statistical moments of boundary CFT quantities.
The full gravitational path integral then involves a sum over topologies consistent with the boundary conditions.  

In the present dS$_3$ discussion, our considerations are inherently Lorentzian, and the duality with the dual matrix integral provides access to the cosmological correlators of massive particles with a fixed topology of the late-time slice. In the more conventional picture of dS/CFT holography \cite{Strominger:2001pn}, the wavefunction of the universe is computed by the partition function of a dual CFT; here, the dual matrix model computes the norm of the cosmological wavefunction, integrated over metrics on $\mathcal{I}^+$.
It would be very interesting to better understand how to compute the contribution of a fixed spacetime topology to the de Sitter gravitational path integral and to understand its contribution to the holographic dual. Notice that in our computation of the gravitational path integral on the inflating de Sitter universe \eqref{eq:inflating universe metric}, the resulting wavefunction was crossing-symmetric on its own, and so does not require a further sum over bulk topologies. A better understanding of the sum over topologies in dS$_3$ gravity will require further developing complex Virasoro TQFT in its own right, which we have only begun to undertake in this paper.

\paragraph{Including an observer.}
The importance of the presence of an observer in defining meaningful observables in the static patch of de Sitter quantum gravity has recently been emphasized in \cite{Anninos:2011af, Chandrasekaran:2022cip, Loganayagam:2023pfb}. In this paradigm observables are gravitationally dressed to the observer's worldline. Since what the matrix model naturally computes are late-time cosmological correlators, our discussion of the duality between the dS$_3$ cosmological correlators and the dual two-matrix model seemingly makes no reference to this idea. It would be very interesting to understand how to augment this duality to account for the presence of an observer in the static patch of dS$_3$ quantum gravity. 

\paragraph{Comparison with $\mathrm{AdS}_3$.} Let us contrast our findings with the holographic picture that crystallized in the last few years for pure Einstein gravity with negative cosmological constant. The main finding was that the path integral of pure 3d gravity computes \emph{universal} contributions to statistical moments of CFT data which can be computed from vacuum dominance and crossing symmetry, see \cite{Belin:2020hea, Chandra:2022bqq} and \cite{Collier:2024mgv,deBoer:2024mqg} for further developments. Thus the holographic description is not a single CFT, and in fact a  matrix-tensor model has been proposed for the non-perturbative description of such a putative ensemble of CFTs \cite{Belin:2023efa,Jafferis:2024jkb}.
Even though the observables in both theories are quite different, this is somewhat similar to the matrix model that we identified in this paper in that Einstein gravity seems only to have access to certain moments and not the matrix itself. The general lesson seems to be that we can only hope for a truly microscopic description in a top-down construction coming from string theory. 

A perhaps closer parallel is with chiral AdS$_3$ quantum gravity \cite{Eberhardt:2022wlc}. Partition functions on the off-shell topologies $\Sigma_{g,n}\times \text{S}^1$ in chiral 3d gravity are computed by taking the trace of the Hilbert space defined by quantization of the moduli space of $\Sigma_{g,n}$ \cite{Maloney:2015ina}, and were shown using intersection theory techniques to be captured by topological recursion of a double-scaled matrix integral \cite{Eberhardt:2022wlc}.
These partition functions may also be computed by gauging the mapping class group after quantization, upon which it becomes clear that they are structurally identical to a string worldsheet path integral. Indeed they are precisely computed by the string amplitudes (``quantum volumes'') of the Virasoro minimal string \cite{Collier:2023cyw} and its matrix integral dual, thereby establishing a duality between resolvents of the VMS matrix model and the off-shell partition functions of chiral AdS$_3$ gravity. The absence of an asymptotic boundary of these off-shell topologies renders this duality somewhat non-standard compared to the usual AdS/CFT paradigm, but it seems more closely related to the dS$_3$/matrix model duality of the present paper.

\paragraph{Complex Virasoro TQFT and Virasoro conformal partial waves.} 
In this paper we constructed the Hilbert space of dS$_3$ gravity on an initial value surface $\Sigma_{g,n}$ defined by canonical quantization of the gravity phase space. The Hilbert space is spanned by products of left- times right-moving Virasoro conformal blocks with central charge $c = 13 + i\mathbb{R}$ and internal conformal dimensions in the principal series $\Delta\in 1+ i\mathbb{R}$ of $\SL(2,\mathbb{C})$. In gravity, large diffeomorphisms are gauged, and for the application to cosmological correlators we were mostly interested in the Hilbert space defined by gauging the mapping class group before quantization. States in this Hilbert space correspond to crossing-symmetric combinations of conformal blocks, and the inner product is defined as in (\ref{eq:3d naive inner product after Map}) by integrating the corresponding wavefunctions (CFT correlation functions) over the moduli space of the late-time Cauchy surface. Indeed we argued that the correlation functions of Liouville CFT are physically well-motivated cosmological wavefunctions and showed that their norms in this Hilbert space, which we may interpret as cosmological correlators of massive particles in dS$_3$, are precisely computed by the perturbative string amplitudes of the complex Liouville string.

In the TQFT approach to AdS$_3$ quantum gravity \cite{Collier:2023fwi, Collier:2024mgv}, it was computationally very useful to work with the Hilbert space defined by gauging the mapping class group \emph{after} quantization. In this case wavefunctions are simply linear combinations of products of conformal blocks of the appropriate central charge, which need not be crossing symmetric. An essential computational tool is the inner product between individual conformal blocks, which is defined by integrating the conformal blocks over the Teichm\"uller space of the Cauchy surface as in (\ref{eq:3d naive inner product}). Equipped with this inner product we would be able to compute the gravitational path integral on fixed spacetime topologies by applying standard TQFT surgery techniques.

A preliminary analysis shows that unlike in the Virasoro TQFT approach to AdS$_3$ quantum gravity, individual conformal blocks \emph{do not} form an orthogonal basis for the Hilbert space equipped with the inner product (\ref{eq:3d naive inner product}). This perhaps could have been anticipated. In higher dimensions, cosmological correlators of quantum fields on rigid de Sitter space admit a spectral decomposition into a complete basis of conformal partial waves associated with the unitary representations of the Euclidean conformal group $\SO(1,d+1)$. For example, in the case of the four-point function in three bulk dimensions, the spectral decomposition is given by\footnote{Below the arguments of the conformal partial waves refer to the internal conformal dimensions and spins, and the dependence on the external dimensions and spins is kept implicit.}
\begin{equation}\label{eq:spec decomp of cosmo correlator}
    \langle\mathcal{O}_1\mathcal{O}_2\mathcal{O}_3\mathcal{O}_4\rangle = \sum_{s = 0}^\infty \int_{1+i\mathbb{R}}\frac{\d\Delta}{2\pi i}\rho^{(1234)}_{0,4}(\Delta,s)\psi_{0,4}(\Delta,s)
\end{equation}
Here all of the dynamical information about the cosmological correlator is contained in the spectral function $\rho^{(1234)}_{0,4}(\Delta,s)$. The integral on the right-hand side runs over principal series representations of $\SO(1,3)$\footnote{In principle the right-hand side may receive contributions from other unitary representations of the Euclidean conformal group such as the complementary series, but for simplicity of notation here we omit them.} and the conformal partial wave $\psi_{0,4}(\Delta,s)$ is a particular linear combination of the ordinary global conformal block $F_{0,4}(\Delta,s)$ and the corresponding shadow block $F_{0,4}(2-\Delta,s)$ that is tuned to be single-valued in Euclidean kinematics. The conformal partial waves form a complete basis of eigenfunctions of the conformal Casimir, and are orthogonal with respect to an inner product defined by integrating the partial waves over the conformal cross-ratio of the four insertion points on the late-time two-sphere \cite{Caron-Huot:2017vep}. The conformal Casimir is self-adjoint with respect to this inner product, so we may extract the spectral density by taking the inner product between the correlator and the appropriate conformal partial wave, which is known as the ``Euclidean inversion formula'' \cite{Caron-Huot:2017vep,Karateev:2018oml}. 

Let us note however that the decomposition of the Liouville wavefunction of the universe (\ref{eq:wavefunction of the universe}) into conformal blocks is tantalizingly structurally similar to the spectral decomposition of the cosmological correlator (\ref{eq:spec decomp of cosmo correlator}). Indeed, the conformal block decomposition of any Liouville correlation function with $c\in13 + i\mathbb{R}$ can be taken to run over scalar principal series $\Delta \in 1 + i\mathbb{R}$ Virasoro conformal blocks:
\begin{equation}
    \Psi_{g,n}^{(b)}(p_1,p_2,p_3,p_4) = \int_{1+i\mathbb{R}}\frac{\d \Delta}{2\pi i}\, \rho^{(b)}_{0,4}(\Delta,s=0)\mathcal{F}^{(b)}_{0,4}(\Delta,s=0)\, ,
\end{equation}
where $\mathcal{F}^{(b)}_{0,4}(\Delta,s)$ corresponds to the appropriate product of left- times right-moving Virasoro conformal blocks and $\rho_{0,4}^{(b)}(\Delta,s=0)$ is the OPE density corresponding to the DOZZ \cite{Dorn:1994xn,Zamolodchikov:1995aa} solution for the structure constants of Liouville CFT. Notably, when written this way the spectral integral actually only runs over part of the scalar principal series due to the $\frac{c-13}{12}$ shift in the dimensions of Liouville primaries.

To proceed in formulating complex Virasoro TQFT we seek a complete basis for the 3d gravity Hilbert space that orthogonalizes the inner product (\ref{eq:3d naive inner product}), akin to the conformal partial waves in the higher-dimensional/global case. It seems that single-valuedness should not be taken as a guiding principle for the Virasoro conformal partial waves since Virasoro conformal blocks transform in a much more complicated way under monodromy than ordinary global conformal blocks. Instead we simply seek a combination of Virasoro conformal blocks that form an orthogonal basis for the Hilbert space. This basis should in particular be compatible with the Hilbert space carrying a unitary representation of the mapping class group of the Cauchy surface $\mathop{\text{Map}}(\Sigma_{g,n})$. By assuming an asymptotic behaviour of the putative Virasoro partial waves near the boundaries of moduli space of the same form as conformal blocks, it seems to us that one wants to demand that the inner product of two Virasoro partial waves contains the delta function $\delta(s-s') \delta(i(\Delta-\Delta'))$ putting the internal spins and scaling dimensions equal, likely together with a second reflected ``shadow'' term as for global conformal blocks.

\paragraph{The three-sphere partition function.}
A better understanding of the complete basis of wavefunctions for the 3d gravity Hilbert space (before gauging of the mapping class group) would allow us to compute the gravitational path integral on fixed topologies using standard TQFT surgery techniques, as in AdS$_3$ \cite{Collier:2023fwi,Collier:2024mgv}. One spacetime topology of particular interest is the three-sphere, due to the fact that it appears to encode the entropy of the cosmological horizon associated with the static patch of dS$_3$ \cite{Gibbons:1976ue,Gibbons:1977mu}. In confirming that the matrix model enumerates the microscopic degrees of freedom that account for the de Sitter entropy in this paper, the three-sphere partition function dropped out of the computation and its precise value was not needed. It would be more satisfying to compute the three-sphere partition function to all orders in the gravitational coupling directly in complex Virasoro TQFT. 

For instance, we could compute the three-sphere partition function by splitting it along two interlinked solid tori as discussed in section~\ref{subsec:3d consistency} and depicted in figure \ref{fig:S3 splittings}. Then the TQFT partition function on the three-sphere would be given by the following matrix element\footnote{The gravity partition function is related to the TQFT partition function by gauging of the bulk mapping class group, which is trivial for the case of the three-sphere.} 
\begin{equation}\label{eq:ZS3 TQFT}
    \mathcal{Z}^{\text{S}^3}_{\text{TQFT}} = \braket{\id_{1,0}^{(b)}|\mathbb{S}|\id_{1,0}^{(b)}}\, .
\end{equation}
Here $\ket{\id_{1,0}^{(b)}}$
corresponds to the state defined by the TQFT path integral on the solid torus --- corresponding to the identity Virasoro character --- and $\mathbb{S}$ is the representation of the modular S-transformation on the torus Hilbert space. Recall that the identity block defines a non-normalizable state in the Hilbert space $\mathcal{H}_{1,0}^{(b)}$.

In order to compute (\ref{eq:ZS3 TQFT}), we need to better understand how the mapping class group of the Cauchy surface acts on the 3d gravity Hilbert space. In the absence of this, we may attempt to compute it by analytic continuation of the computation in AdS$_3$ gravity, which is related to two copies of Virasoro TQFT. In VTQFT we formally have 
\begin{equation}
    \mathcal{Z}_{\text{VTQFT}}^{\text{S}^3} = \mathbb{S}_{\id\id}\, ,
\end{equation}
where $\mathbb{S}$ is the usual modular $S$-matrix associated with holomorphic torus Virasoro characters. In particular,
\begin{equation}
    \mathbb{S}_{\id h} = \frac{2\sqrt{2}\sinh\left(2\pi b \sqrt{h-\frac{c-1}{24}}\right)\sinh\left(2\pi b^{-1}\sqrt{h-\frac{c-1}{24}}\right)}{\sqrt{h-\frac{c-1}{24}}}\, .
\end{equation}
This expresses the identity character in a complete basis of non-degenerate characters with $h \geq \tfrac{c-1}{24}$. Notice that here we have written the modular $S$-matrix in the $h$ basis rather than the Liouville momentum $p$ basis (the difference is a Jacobian factor). This will be important in the computation that follows, and we do not have a better justification for it other than that in complex Virasoro TQFT we expect that the torus partial waves are orthonormal in the conformal dimension rather than Liouville momentum variables. Similarly, since we only fixed the torus inner product up to a numerical ($b$-independent) factor, we can hope at best that this gives the correct answer up to a numerical factor. We will fix the numerical factor by comparing to the one-loop calculation.

In order to compute $\mathbb{S}_{\id\id}$ we need to define $\mathbb{S}_{\id,0}-\mathbb{S}_{\id,1}$ by analytic continuation, since the values $h=0,1$ do not lie on the contour that defines the modular $S$-matrix. This gives 
\begin{equation}
    \mathbb{S}_{\id\id} = \begin{cases}
    \frac{4ib\sin(\pi b^2)\sin(\pi b^{-2})}{1-b^4}, & 0<|b|<1 \\ 
    \frac{4i b^3\sin(\pi b^2)\sin(\pi b^{-2})}{b^4-1}, & |b|>1
    \end{cases}\, ,
\end{equation}
where we fixed the numerical prefactor such that it will agree with the one-loop result below. We notice however that the phase and the powers of $\pi$ come naturally out of this computation.
If we accept that the sphere partition function may be computed from this by analytic continuation in $b$, then the gravity partition function takes the following form (say for $0<|b|<1$)\footnote{The identity-identity element of the modular S matrix is squared on the right-hand side because AdS$_3$ gravity is related to two copies of Virasoro TQFT. In the full gravitational theory there is a further gauging of the bulk mapping class group which is trivial in the case of the three-sphere. We are not completely sure whether this formula is missing a factor of $\frac{1}{2}$ or $2$ because of the global structure of the gauge group. The gravity computation discussed in section~\ref{subsec:subsectionGH} suggests that the relevant global gauge group at least to one-loop level is $\SO(4) \cong (\SU(2) \times \SU(2))/\ZZ_2$. See however section~\ref{subsec:relation CS} for caveats regarding this gauge group. This factor is not important in our computation since we did not keep track of the order 1 normalization of the inner product.} 
\begin{equation}
    \mathcal{Z}_{\text{grav}}^{\text{S}^3} = \mathbb{S}_{\id\id}^2 = -\frac{16 b^2 \sin(\pi b^2)^2\sin(\pi b^{-2})^2}{(1-b^4)^2}\, . \label{eq:conjecture ZS3}
\end{equation}
Despite the unjustified leaps involved in this computation, the logarithm of the sphere partition function defined this way reproduces the semiclassical expansion of the de Sitter entropy (\ref{eq:sphere partition function}) up to $O(S_\text{GH}^0)$ 
\begin{equation}
    \log\mathcal{Z}_{\text{grav}}^{\text{S}^3} = S_{\text{GH}} - 3\log S_{\text{GH}} +5\log(2\pi) - \frac{\pi i}{2} + O(S_{\text{GH}}^{-1})\, .
\end{equation}

The expression \eqref{eq:conjecture ZS3} is a conjectural expression for the non-perturbative three-sphere partition function. Let us mention a few caveats. It depends on the specific renormalization scheme that we used, in particular how we relate $\ell_\text{dS}/G_\text{N}$ to $b$. A similar, but different proposal was made in \cite{Anninos:2020hfj} by analytic continuation of the $\SU(2)_k$ Chern-Simons three-sphere partition function. 
In any case, the result has some similarities when identifying $b^2=\frac{1}{k+2}$ as suggested in \eqref{eq:b kappa map}. However, we should emphasize that our expression differs in particular non-perturbatively from what was suggested in \cite{Anninos:2020hfj} due to the presence of the second sine factor in (\ref{eq:conjecture ZS3}). 

We should also note that if we accept \eqref{eq:conjecture ZS3}, then \eqref{eq:T11 ZS3 normalization} implies the particularly simple relation
\be 
\mathrm{e}^{S_0} =\frac{\pi}{|b|}~ ,
\ee
which is perhaps an aesthetic reason supporting \eqref{eq:conjecture ZS3}.

\paragraph{Lens spaces.}

Beyond the leading sphere contribution, the Gibbons-Hawking proposal 
\begin{equation}\label{eq: SdS_discussion}
    S_{\mathrm{dS}} = \log \mathcal{Z}_{\mathrm{grav}}~,\quad \mathcal{Z}_{\mathrm{grav}} = \sum_{\mathcal{M} ~\mathrm{compact}}\int [\mathcal{D}g]\,\mathrm{e}^{-S_{\mathrm{EH}}[g,\Lambda]}~,\quad \Lambda >0~.
\end{equation}
does not specify which contributions $\mathcal{M}$ we should sum over in $\mathcal{Z}_{\mathrm{grav}}$ \cite{Anninos:2022ujl}. In two dimensions one might wonder whether all compact manifolds, labelled by their Euler characteristic should be added, motivating the work in \cite{Anninos:2021eit, Anninos:2022ujl}. In four dimensions the Nariai geometry, $\text{S}^2 \times \text{S}^2$, \cite{Volkov:2000ih} is a subleading solution to the gravitational path integral (\ref{eq: SdS}) and likely has to be included. This raises the question: what topologies are we supposed to sum over in three-dimensions? Beyond the $\text{S}^3$ saddle it has been conjectured that lens spaces $L(p,q)$ contribute to $\mathcal{Z}_{\mathrm{grav}}$ \cite{Castro:2011xb}. Lens spaces are quotients of $\text{S}^3/\Gamma$ where $\Gamma$ is a discrete subgroup of $\SO(4)$. For simplicity we focus only on the subgroup $\mathbb{Z}_p$ which yields the lens spaces $L(p,q) = \text{S}^3/\mathbb{Z}_p$. Here $q<p$ are coprime integers with $q$ labelling the different ways of embedding the cyclic group into $\SO(4)$, the isometry group of $\text{S}^3$; for example $L(1,0)$ is $\text{S}^3$ whereas $L(2,1) = \mathbb{RP}^3 = \text{S}^3/\mathrm{antipodal~points}$. Lens spaces are defined through the following Euclidean metric together with the identifications
\begin{equation}\label{eq: Hopf metric}
    \frac{ds_{\mathrm{lens}}^2}{\ell_{\mathrm{dS}}^2} = \cos^2\rho\, \d t^2 + \d\rho^2 + \sin^2\rho\,  \d\varphi^2~,\quad (t,\varphi) \sim (t,\varphi) + 2\pi\left(\frac{m}{p},\, m\frac{q}{p}+n\right)~,
\end{equation}
for $m,n\in \mathbb{Z}$. For $(p,q)=(1,0)$, (\ref{eq: Hopf metric}) is the metric of $\text{S}^3$ in Hopf coordinates. It is the Wick rotation of the static patch de Sitter geometry. For $(p,q)\neq (1,0)$ lens spaces loosely correspond to a static patch decorated by mass and angular momentum along a worldline or conical defect, resonating to some extend with the situation in AdS$_3$ and the rotating BTZ black hole. This suggests that they shouldn't be included in the computation of the deSitter entropy, since they necessitate the inclusion of an observer in the static patch.\footnote{Recall also that our specific choice for the state of the universe does not allow for spinning massive worldlines.} Their on-shell volume is $\frac{1}{p}$ times the volume of the three-sphere. Thus at tree-level, their partition functions scale like
\be 
\mathcal{Z}_\text{grav}^{L(p,q)} \approx \left(\mathcal{Z}_\text{grav}^{\text{S}^3}\right)^{\frac{1}{p}} \sim g_\text{s}^{-\frac{2}{p}}~ .
\ee
We conclude that lens spaces formally contribute fractional terms in the genus expansion and clearly don't appear in the dual matrix model. Moreover the sum over such contributions badly diverges \cite{Castro:2011xb}.

\section*{Acknowledgement}
We would like to thank Dionysios Anninos, Yiming Chen, Alessandro Fumagalli, Victor Godet, Victor Gorbenko, Kristan Jensen, Shota Komatsu, Jonah Kudler-Flam, Adam Levine, Keivan Namjou, Juan Maldacena, Alex Maloney,  Shu-Heng Shao, Jon Sorce, Zimo Sun, Kamran Salehi Vaziri, Erik Verlinde, Herman Verlinde and Edward Witten.
We especially thank Victor Rodriguez for initial collaboration and discussions about related topics and Victor Godet for helpful discussions leading to the relation of the string amplitudes and 3d de Sitter. We would like to particularly thank Dionysios Anninos for many illuminating and helpful discussions and for useful comments on the draft. We also thank Herman Verlinde for his interest and comments about this work.
We thank l’Institut Pascal
at Universit\'e Paris-Saclay, with the
support of the program ``Investissements d’avenir'' ANR-11-IDEX-0003-01, 
and SC thanks the Kavli Institute for Theoretical Physics (KITP), which is supported in part by grant NSF PHY-2309135, for hospitality during the course of this work.  
SC is supported by the U.S. Department of Energy, Office of Science, Office of High Energy Physics of U.S. Department of Energy under grant Contract Number DE-SC0012567 (High Energy Theory research), DOE Early Career Award  DE-SC0021886 and the Packard Foundation Award in Quantum Black Holes and Quantum Computation. LE is supported by the European Research Council (ERC) under the European Union’s Horizon 2020 research
and innovation programme (grant agreement No 101115511).
BM gratefully acknowledges funding provided by the Sivian Fund at the Institute for Advanced Study and the National Science Foundation with grant number PHY-2207584.

\appendix

\section{de Sitter geometry}\label{app:dS}

In this appendix we give a very short overview of some basic properties of the de Sitter geometry, for more details we refer e.g.\ to \cite{Anninos:2012qw}. 3d global de Sitter can be embedded in four-dimensional Minkowski space $\d s^2 = -\d X_0^2 + \sum_{i=1}^3 \d X_i^2$ as a hyperboloid
\begin{equation}
    -X_0^2 +X_1^2 +X_2^2 + X_3^2 = \ell_{\mathrm{dS}}^2~.
\end{equation}
Various coordinate systems of de Sitter cover some or even all of this hyperboloid. The global Lorenzian geometry of de Sitter can be parametrized by the metric 
\begin{equation}
    \frac{ds_{\mathrm{gl}}^2}{\ell_{\mathrm{dS}}^2} = -\d\tau^2 +  \cosh^2 \tau\, (\d\psi^2 + \sin^2 \psi \, \d\theta^2)~,
\end{equation}
where $\tau \in \mathbb{R}$. Constant $\tau$-slices are two-spheres, which shrink from $\mathcal{I}^-$ at $\tau\rightarrow -\infty$ to a minimum at $\tau=0$, and re-expand from $\tau=0$ to $\tau\rightarrow +\infty$ ($\mathcal{I}^+$). Global de Sitter corresponds to the full Penrose diagram depicted in Figure~\ref{fig:penrose}.

The global patch is not the patch of our universe visible to an observer. The exponential acceleration of spacetime creates a cosmological event horizon. The visible universe is the static patch, depicted as the blue triangle in the Penrose diagram. Lorentzian static patch coordinates for dS$_3$ are given by
\begin{equation}
    \frac{ds_{\mathrm{st}}^2}{\ell_{\mathrm{dS}}^2} = -\cos^2\rho\, \d t^2 + \d\rho^2 + \sin^2\rho \, \d\varphi^2~,
\end{equation}
where $\varphi \sim \varphi +2\pi$, $t\in \mathbb{R}$ and $0\leq \rho\leq \pi/2$. The cosmological event horizon is located at $\rho =\pi/2$, whereas the observers worldline is at $\rho=0$. The area of the cosmological event horizon is 
\begin{equation}\label{eq: horizon area}
    A_{\text{h}} = 2\pi \ell_{\mathrm{dS}}~.
\end{equation}
Although the static and the global patch of de Sitter describe distinct portions of the de Sitter geometry, Wick rotating them to Euclidean signature leads in either case to a full three sphere metric. For the global patch, we take $\tau \rightarrow i \tau_{\text{E}}$, for the static $t\rightarrow i t_{\text{E}}$. In the latter case, smoothness imposes the identification $t_{\text{E}} \sim t_{\text{E}} + 2\pi$: 
\begin{subequations}
\begin{align}
 \frac{\d s_{\mathrm{gl, Eucl}}^2}{\ell_{\mathrm{dS}}^2} &= \d\tau_{\text{E}}^2 +  \cos^2 \tau_{\text{E}}\, (\d\psi^2 + \sin^2 \psi \, \d\theta^2)\\
    \frac{\d s_{\mathrm{st, Eucl}}^2}{\ell_{\mathrm{dS}}^2} &= \cos^2\rho\, \d t_{\text{E}}^2 + \d\rho^2 + \sin^2\rho \, \d\varphi^2~,\quad t_{\text{E}} \sim t_{\text{E}}+ 2\pi~.
\end{align}
\end{subequations}

\section{Phase space and constraints} \label{app: canonical quantization}
 \subsection{Canonical quantization of \texorpdfstring{dS$_3$}{dS3} gravity}
We will now give some of the more technical details of the quantization of 3d Einstein gravity with $\Lambda>0$. Since there are a substantial amount of wrong or imprecise statements in the literature, our discussion we will be rather careful. The main result of this discussion is what we motivated heuristically in the main text: the Hilbert space of dS$_3$ quantum gravity on an initial value surface $\Sigma_{g,n}$ is spanned by wavefunctions transforming like CFT correlation functions of central charge $\in 13+i \RR$ and vertex operators with conformal weight $\in \frac{1}{2}+i \RR$ with inner product given by \eqref{eq:3d naive inner product after Map}. We will consider 3d quantum gravity on orientable manifolds; there should also exist a version in which orientation reversal is gauged. 
We will translate some technology from the gauge theory setting to the gravitational setting by using the dictionary \eqref{eq:SL(2,C) gauge fields}, though we will be able to sidestep the subtlety about invertibility that we mentioned in subsection~\ref{subsec:relation CS}.

\paragraph{Poisson brackets and conventions.} Consider an (orientable) initial value surface $\Sigma_g$ of genus $g$. We will first assume that there are no punctures.
The unconstrained phase space of Chern-Simons theory consists of all gauge fields on an initial value surface $\Sigma_g$ of genus $g$. This is enough data to uniquely solve the equations of motion
\be 
\mathcal{F}^{(3)}(\mathcal{A})=\d \mathcal{A}+\mathcal{A} \wedge \mathcal{A}=0 \label{eq:flatness 3d}
\ee
everywhere. The superscript $(3)$ emphasizes that this is the three-dimensional curvature. 
We can read off the symplectic form from the action \eqref{eq:PSL(2,C) CS action}. Choose complex coordinates on $\Sigma_g$ and decompose
\be 
\mathcal{A}=\mathcal{A}_z \d z + \mathcal{A}_{\bar{z}}
 \d \bar{z}+\mathcal{A}_t \d t~ .
 \ee
We can partially fix the gauge by setting $\mathcal{A}_t=0$. Plugging this into the action \eqref{eq:PSL(2,C) CS action} gives
\be 
S=\frac{k}{2\pi} \int \d z \wedge \d \bar{z} \wedge \d t\, \tr \mathcal{A}_{\bar{z}} \partial_t \mathcal{A}_z+(\mathcal{A} \leftrightarrow \bar{\mathcal{A}}, k \leftrightarrow \bar{k})~.
\ee
This has the standard form $\int \d t\, \sum_i p_i \partial_t q_i$ for the Hamiltonian of canonically conjugate variables. The standard Poisson brackets $\{q_i,p_j\}=\delta_{ij}$ read in this case
\begin{subequations}
\begin{align} 
\{\mathcal{A}_z^a(z), \mathcal{A}_{\bar{z}}^b(w)\}&=\frac{4\pi}{k} K^{ab}\, \delta^{(2)}(z-w)~, \\
\{\bar{\mathcal{A}}_z^a(z), \bar{\mathcal{A}}_{\bar{z}}^b(w)\}&=\frac{4\pi}{\bar{k}} K^{ab}\, \delta^{(2)}(z-w)=-\frac{4\pi}{k} K^{ab}\, \delta^{(2)}(z-w)~,
\end{align} \label{eq:SL(2,C) Poisson brackets}%
\end{subequations}
where $K^{ab}=2\tr(t^a t^b)$ is the Killing form in the fundamental representation. We normalize generators such that
\be 
[t^3,t^\pm]=\pm t^\pm~, \quad [t^+,t^-]=t^3~, \qquad K^{33}=K^{+-}=K^{-+}=1~. \label{eq:sl2C structure constants and Killing form}
\ee
The delta function that appears is the natural delta function in complex coordinates, it satisfies
\be 
\int \d w \wedge \d \bar{w}\, f(w) \, \delta^{(2)}(z-w)=f(z)~ .
\ee

\paragraph{Constraints.} Flatness of the three-dimensional gauge field \eqref{eq:flatness 3d} requires flatness of the two-dimensional one and thus the initial gauge field has to satisfy the Gauss law $\mathcal{F}(\mathcal{A})=0$, where $\mathcal{F}$ denotes the curvature of the gauge field on $\Sigma_g$. Relatedly, gauge-equivalent gauge fields on $\Sigma_g$ and thus the physically relevant phase space consists of all \emph{flat} gauge fields modulo gauge equivalence on $\Sigma_g$. This is also known as the constrained phase space. Its symplectic structure is induced from \eqref{eq:SL(2,C) Poisson brackets} via symplectic reduction. The relevant group is the gauge group $\mathcal{G}$ consisting of maps $\Sigma \to \PSL(2,\CC)$ and the moment map is given by the curvature $\mathcal{F}(\mathcal{A}) \in \text{Lie}(\mathcal{G})^*$ \cite{AtiyahBott}. This is naturally in the dual of the Lie algebra of the gauge group since it can be paired with Lie-algebra valued functions and integrated over the surface.

Let us explain the structure of the constrained phase space in the Chern-Simons language in more detail. Fix a point $x_0 \in \Sigma_g$.
Every flat $\PSL(2,\CC)$ connection gives rise to a holonomy representation $\rho(\gamma) \in \PSL(2,\CC)$ which associates the holonomy of the gauge field along the closed loop $\gamma$ anchored at $x_0$. Flatness of the gauge field ensures that $\rho(\gamma)$ is independent of small deformations of the path. Thus $\rho$ is a homomorphism $\rho: \pi_1(\Sigma_g,x_0) \longrightarrow \PSL(2,\CC)$. Vice versa, such a holonomy representation determines the gauge field uniquely. Choosing another base point $x_0$ conjugates the representation and thus the moduli space of flat $\PSL(2,\CC)$ connections can be identified with the character variety consisting of all such homomorphisms,
\be 
\mathcal{M}(\Sigma_g)=\big\{ \rho: \pi_1(\Sigma_g) \longrightarrow \PSL(2,\CC) \text{ homomorphism} \big\} / \PSL(2,\CC)~. \label{eq:character variety}
\ee
This space is famously hyperk\"ahler \cite{Hitchin:1986vp}.
However, as mentioned in section~\ref{subsec:relation CS} this phase space is potentially slightly too big for gravity as it contains pathological gauge field configurations such as $\mathcal{A}=0$. We will see how this issue is naturally adressed in the quantization below.

\subsection{Quantization}

One can try to quantize \eqref{eq:character variety} directly, meaning that one imposes the Gauss law on the classical level and then quantizes. This is quite difficult because \eqref{eq:character variety} is a very non-linear space. In Chern-Simons theory, it is usually simpler to first quantize and then impose the Gauss law as an operator constraint on the wavefunction. Thus we will now go back to the phase space before the Gauss law constraint, which consists of \emph{all} $\PSL(2,\CC)$ connections (without identifying gauge-equivalent configurations). This space is also hyperk\"ahler which will be helpful in the following.\footnote{This is easy to understand, since the components of the gauge field $\mathcal{A}_z^a$, $\mathcal{A}_{\bar{z}}^a$, $\bar{\mathcal{A}}_z^a$ and $\bar{\mathcal{A}}_{\bar{z}}^a$ admit two different complex structures, either by complex conjugating the gauge field or the component.} 

Let us compare with the metric formalism that we discussed in section~\ref{subsec:wavefunction}. Since one is working in a second order formalism, the three Hamiltonian constraints $\mathcal{H}=\mathcal{H}_i=0$ expressing invariance under spatial diffeomorphisms and time evolution are second order. This makes them hard to solve explicitly, see e.g.\ \cite{Chakraborty:2023yed} for a recent discussion. Instead, in the Chern-Simons formulation, there are six first order constraints.

\paragraph{A holographic polarization.} We follow the quantization procedure sketched in \cite{Verlinde:2024zrh} that is most suited for a holographic discussion (with a small twist). It is essentially analogous to the situation with $\Lambda<0$, where one encounters two copies of $\PSL(2,\RR)$ as the gauge group. To quantize one has to pick a polarization which specifies the coordinates on which the wavefunction depends. 
These coordinates need to form a Lagrangian submanifold of the phase space, i.e.\ Poisson commute. 
In the holographic setup, the correct choice is determined by looking at the boundary condition of the gauge field $\mathcal{A}^a$. Near future infinity $\mathcal{I}^+$ of dS$_3$, one off-diagonal component of the gauge field decays much fast than the other off-diagonal component, see e.g.\ \cite{Cotler:2024xzz}. The wave-function thus depends only on say the $-$-components, but not the $+$-components. For the $3$-components, the gauge field becomes chiral near the boundaries and depends only on $\mathcal{A}_z^3$ and $\bar{\mathcal{A}}_{\bar{z}}^3$. Thus we take the wave-function to depend on
\be 
\Psi[\mathcal{A}_z^-,\mathcal{A}_{\bar{z}}^-,\bar{\mathcal{A}}_z^-, \bar{\mathcal{A}}_{\bar{z}}^-,\mathcal{A}_z^3, \bar{\mathcal{A}}_{\bar{z}}^3]~.
\ee
These coordinates indeed Poisson-commute and is the analogous choice to \cite{Verlinde:1989ua}. Notice also that this choice is invariant under simultaneous conjugation of the $z$-coordinate ($z \to \bar{z}$) and the complex structure of $\mathcal{A}$ ($\mathcal{A} \to \bar{\mathcal{A}}$). This invariance is shared by the symplectic structure \eqref{eq:SL(2,C) Poisson brackets} that picks up an additional minus sign when exchanging $z$ and $\bar{z}$. Thus we only have to discuss $\mathcal{A}^a$ going forward since everything regarding $\bar{\mathcal{A}}$ can be obtained by complex conjugating both $z$ and $\mathcal{A}$. In particular, we can momentarily assume that the wavefunction factorizes, $\Psi[\mathcal{A}_z^-,\mathcal{A}_{\bar{z}}^-,\mathcal{A}_z^3] \otimes \widetilde{\Psi}[\bar{\mathcal{A}}_z^-, \bar{\mathcal{A}}_{\bar{z}}^-, \bar{\mathcal{A}}_{\bar{z}}^3]$ and discuss the constraints on the `chiral' wavefunction $\Psi$. 
This also means that we can regard this polarization either as real or as complex, depending on which complex structure we consider. This is the bonus we get from a hyperk\"ahler phase space and we will exploit it for the inner product on the Hilbert space below.

This choice is different from the usual Chern-Simons boundary conditions that were for example used in \cite{Witten:1989ip} in which the wavefunction depends on $\mathcal{A}_z^a$ and $\bar{\mathcal{A}}_{\bar{z}}^a$. We explain the relation to that quantization scheme below. Canonical quantization replaces the Poisson brackets by commutators as usual, $\{ \bullet, \bullet \} \longrightarrow -i\,  [ \bullet, \bullet]$. We thus realize the remaining coordinates as operators at the quantum level as follows,  
\be 
\mathcal{A}^+_z(z)=-\frac{4\pi i}{k}\frac{\delta}{\delta \mathcal{A}^-_{\bar{z}}(z)}~,\quad 
\mathcal{A}^+_{\bar{z}}(z)=\frac{4\pi i}{k}\frac{\delta}{\delta \mathcal{A}^-_{z}(z)}~,\quad
\mathcal{A}^3_{\bar{z}}(z)=\frac{4\pi i}{k}\frac{\delta}{\delta \mathcal{A}^3_{z}(z)} ~. \label{eq:A functional derivative}
\ee
Let us remark that we can hence think of $k^{-1}$ as playing the role of $\hbar$ in this quantization. Thus corrections in $\frac{1}{k}$ below can be viewed as quantum corrections.
On the chiral wave-function one has to impose the flatness constraints
\be 
\mathcal{F}(\mathcal{A})^{\pm,3}\Psi[\mathcal{A}_z^-,\mathcal{A}_{\bar{z}}^-,\mathcal{A}_z^3]=0~. \label{eq:chiral constraints}
\ee
Explicitly, the three constraints read
\begin{subequations}
\begin{align}
    \mathcal{F}(\mathcal{A})^-&=\partial \mathcal{A}_{\bar{z}}^--\bar{\partial} \mathcal{A}_z^-+\mathcal{A}_z^3 \mathcal{A}_{\bar{z}}^--\frac{4\pi i}{k} \mathcal{A}_z^- \frac{\delta}{\delta \mathcal{A}_z^3}~ , \label{eq:F- constraint}\\
    \mathcal{F}(\mathcal{A})^3&=\frac{4\pi i}{k} \partial \frac{\delta}{\delta \mathcal{A}_z^3}-\bar{\partial} \mathcal{A}_z^3+\frac{4\pi i}{k} \mathcal{A}_z^- \frac{\delta}{\delta \mathcal{A}_z^-}+\frac{4\pi i}{k} \mathcal{A}_{\bar{z}}^- \frac{\delta}{\delta \mathcal{A}_{\bar{z}}^-}~, \label{eq:F3 constraint}\\
    \mathcal{F}(\mathcal{A})^+&=\frac{4\pi i}{k}\partial \frac{\delta}{\delta \mathcal{A}_z^-}+\frac{4\pi i}{k} \bar{\partial} \frac{\delta}{\delta \mathcal{A}_{\bar{z}}^-}-\frac{4\pi i}{k} \mathcal{A}_z^3\frac{\delta}{\delta \mathcal{A}_z^-}+\left(\frac{4\pi}{k}\right)^2 \frac{\delta}{\delta \mathcal{A}_z^3} \frac{\delta}{\delta \mathcal{A}_{\bar{z}}^-}~, \label{eq:F+ constraint}
\end{align}
\end{subequations}
Notice that there is a normal ordering ambiguity in $\mathcal{F}(\mathcal{A})^3$. For the moment, we will use the ordering as indicated in $\mathcal{F}(\mathcal{A})^3$, but this is not entirely natural and we will correct it below.
Everything from this point onward is very similar to the quantization of Teichm\"uller space arising as a component of the phase space of $\PSL(2,\RR)$ gauge theory described in \cite{Verlinde:1989ua}, except that the level is purely imaginary. 

\paragraph{Restricting the phase space.} \eqref{eq:chiral constraints} gives three constraints. One can write them out explicitly in terms of the gauge field by using $\mathcal{F}(\mathcal{A})=\d \mathcal{A}+\mathcal{A} \wedge \mathcal{A}$ and replacing the gauge fields that do not appear in the wavefunction by the functional derivatives \eqref{eq:A functional derivative}. They are operator valued constraints when using \eqref{eq:A functional derivative}. The constraints $\mathcal{F}(\mathcal{A})^-$ and $\mathcal{F}(\mathcal{A})^3$ contain only first order derivatives, while $\mathcal{F}(\mathcal{A})^+$ contains second order derivatives since the quadratic term $\mathcal{A}_z^+ \mathcal{A}_{\bar{z}}^3$ appears in $\mathcal{F}(\mathcal{A})^+$. As explained in \cite{Verlinde:1989ua}, the first order constraints can be explicitly solved. For this, it is convenient to use the parametrization
\be 
\mathcal{A}^-=\mathrm{e}^{\varphi}(\d z+\mu \d \bar{z}) ~, \quad \mathcal{A}_z^3=\omega
\ee
with $\mu$ a Beltrami differential and $\omega$ a holomorphic differential \cite{Moncrief:1989dx, Krasnov:2005dm}.

Importantly, not every gauge field can be written in this form, since $\mathcal{A}^-$ cannot vanish thanks to the $\d z$ component being non-vanishing. Going back to the map \eqref{eq:SL(2,C) gauge fields}, this parametrization alters the phase space and cures the problem of invertibility discussed in section~\ref{subsec:relation CS}. Thus, going forward, this quantization procedure \emph{does not} quantize $\SL(2,\CC)$ Chern-Simons theory. To quantize all of the $\SL(2,\CC)$ phase space, we would need to allow for logarithmic singularities in $\varphi$.

\paragraph{Solving the first constraint.} Even though we don't write subscripts, everything is understood to be written in components to avoid confusions of anti-commutativity of forms.
Let us now solve the $-$ constraint \eqref{eq:F- constraint}. It reads in this parametrization
\be 
\left(\mu \partial \varphi +\partial \mu - \bar{\partial} \varphi+\mu \omega -\frac{4\pi i}{k} \frac{\delta}{\delta \omega}\right)\Psi[\mu,\varphi,\omega]=0
\ee
and is solved by
\be 
\Psi[\mu,\varphi,\omega]=\mathrm{e}^{S[\mu,\varphi,\omega]} \Psi[\mu,\varphi]
\ee
with
\be 
S[\mu,\varphi,\omega]=-\frac{ik}{4\pi}\int \d z \wedge \d \bar{z}\ \left(\frac{1}{2} \mu \omega^2-\omega (\bar{\partial} \varphi-\mu \partial \varphi-\partial \mu)\right) ~.
\ee
Notice the measure in the action comes from the wedge product. This is the natural measure since we are working in holomorphic coordinates and is the measure in which the right hand side of \eqref{eq:SL(2,C) Poisson brackets} act as delta-functions.

\paragraph{Solving the second constraint.} We can then express the other two constraints as constraints on the reduced wavefunction $\Psi[\mu,\varphi]$. This is simple for $\mathcal{F}(\mathcal{A})^3$, since the last two functional derivatives in \eqref{eq:F3 constraint} conspire to give $\frac{4\pi i}{k} \frac{\delta}{\delta \varphi}$. We hence obtain 
\begin{align} \nonumber
0&=-\bigg(-\frac{4\pi i}{k} \partial \frac{\delta}{\delta \omega}+\bar{\partial} \omega-\frac{4\pi i}{k} \frac{\delta}{\delta \varphi}\bigg) \big(\mathrm{e}^{S[\mu,\varphi,\omega]} \Psi[\mu,\varphi]\big)  \nonumber \\
&= -\mathrm{e}^{S[\mu,\varphi,\omega]} \bigg( -\frac{4\pi i}{k} \partial \frac{\delta S}{\delta \omega}+\bar{\partial} \omega-\frac{4\pi i}{k} \frac{\delta S}{\delta \varphi}-\frac{4\pi i}{k} \frac{\delta}{\delta \varphi} \bigg) \Psi[\mu,\varphi]  \nonumber \\
&=-\mathrm{e}^{S[\mu,\varphi,\omega]} \bigg(\partial \big( \bar{\partial} \varphi-\mu \partial \varphi-\partial \mu-\mu \omega \big)+\bar{\partial} \omega
-\bar{\partial} \omega+\partial(\mu \omega)-\frac{4\pi i}{k} \frac{\delta}{\delta \varphi} \bigg) \Psi[\mu,\varphi] \nonumber \\
&=-\mathrm{e}^{S[\mu,\varphi,\omega]} \bigg(\partial \big( \bar{\partial} \varphi-\mu \partial \varphi-\partial \mu\big)-\frac{4\pi i}{k} \frac{\delta}{\delta \varphi} \bigg) \Psi[\mu,\varphi] ~. \label{eq:F3 constraint after first reduction}
\end{align}
A similar, but more tedious computation gives 
\begin{align} 
    0&=\big(\mathcal{F}(\mathcal{A})^++\omega \mathrm{e}^{-\varphi} \mathcal{F}(\mathcal{A})^3\big) \big(\mathrm{e}^{S[\mu,\varphi,\omega]} \Psi[\mu,\varphi]\big)  \nonumber \\
    &=\frac{4\pi i}{k} \mathrm{e}^{-\varphi} \bigg((\partial-\partial \varphi) \frac{\delta}{\delta \varphi}+\bigg(\bar{\partial}-\bar{\partial} \varphi+\mu \omega-\frac{4\pi i}{k} \frac{\delta}{\delta \omega}\bigg) \frac{\delta}{\delta \mu} +\omega \partial \frac{\delta}{\delta \omega}-\frac{k}{4\pi i} \omega \bar{\partial} \omega \bigg) \nonumber\\
    &\qquad\times \big(\mathrm{e}^{S[\mu,\varphi,\omega]} \Psi[\mu,\varphi]\big)  \nonumber \\
    &=\frac{4\pi i}{k} \mathrm{e}^{-\varphi} \mathrm{e}^{S[\mu,\varphi,\omega]}\bigg((\partial-\partial \varphi) \bigg(\frac{\delta S}{\delta \varphi}+\frac{\delta}{\delta \varphi}\bigg)\nonumber\\
    &\qquad+\bigg(\bar{\partial}-\bar{\partial} \varphi+\mu \omega-\partial \mu+\mu \partial \varphi-\mu \partial-\frac{4\pi i}{k} \frac{\delta S}{\delta \omega}\bigg) \bigg(\frac{\delta S}{\delta \mu}+\frac{\delta}{\delta \mu}\bigg) \nonumber\\
    &\qquad+\omega \partial \frac{\delta S}{\delta \omega}-\frac{k}{4\pi i} \omega \bar{\partial} \omega \bigg) \Psi[\mu,\varphi]  \nonumber \\
    &=\frac{4\pi i}{k} \mathrm{e}^{-\varphi} \mathrm{e}^{S[\mu,\varphi,\omega]}\bigg((\partial-\partial \varphi) \bigg(-\frac{i k}{4\pi} \big(\bar{\partial} \omega-\partial(\mu \omega)\big)+\frac{\delta}{\delta \varphi}\bigg)\nonumber\\
    &\qquad+\big(\bar{\partial}-2 \partial \mu-\mu \partial \big) \bigg(-\frac{i k}{4\pi} \bigg(\frac{1}{2} \omega^2+\omega \partial \varphi-\partial \omega \bigg)+\frac{\delta}{\delta \mu}\bigg) \nonumber\\
    &\qquad-\frac{i k \omega}{4\pi} \partial \big( \mu \omega-\bar{\partial}\varphi+\mu \partial \varphi+\partial \mu \big) -\frac{k}{4\pi i} \omega \bar{\partial} \omega \bigg) \Psi[\mu,\varphi]  \nonumber \\
    &=\frac{4\pi i}{k} \mathrm{e}^{-\varphi} \mathrm{e}^{S[\mu,\varphi,\omega]}\bigg( (\partial -\partial \varphi) \frac{\delta}{\delta \varphi}+(\bar{\partial}-2 \partial \mu-\mu \partial) \frac{\delta}{\delta \mu} \bigg) \Psi[\mu,\varphi]~. \label{eq:F+ constraint after first reduction}
\end{align}
We can solve \eqref{eq:F3 constraint after first reduction} for $\varphi$ explicitly,
\be 
\Psi[\mu,\varphi]=\mathrm{e}^{\tilde{S}[\mu,\varphi]} \Psi[\mu]~,
\ee
with 
\be 
\tilde{S}[\mu,\varphi]=\frac{i k}{4\pi} \int \d z \wedge \d \bar{z} \left(\frac{1}{2} \partial \varphi\bar{\partial} \varphi-\mu \left(\frac{1}{2}(\partial \varphi)^2-\partial^2 \varphi \right) \right)~. \label{eq:tilde S anomaly}
\ee
We can finally translate \eqref{eq:F+ constraint after first reduction} to a constraint on the fully reduced wavefunction $\Psi[\mu]$. The computation is again similar as above,
\begin{align} \nonumber
    0&=\bigg( (\partial -\partial \varphi) \frac{\delta}{\delta \varphi}+(\bar{\partial}-2 \partial \mu-\mu \partial) \frac{\delta}{\delta \mu} \bigg)\big(\mathrm{e}^{\tilde{S}[\mu,\varphi]} \Psi[\mu]\big) \\
    &=\mathrm{e}^{\tilde{S}[\mu, \varphi]}\bigg( (\partial -\partial \varphi) \frac{\delta \tilde{S}}{\delta \varphi}+(\bar{\partial}-2 \partial \mu-\mu \partial) \bigg(\frac{\delta \tilde{S}}{\delta \mu}+\frac{\delta}{\delta \mu}\bigg) \bigg) \Psi[\mu]  \nonumber \\
    &=\mathrm{e}^{\tilde{S}[\mu, \varphi]}\bigg(- \frac{ik}{4\pi}(\partial -\partial \varphi) \partial(\bar{\partial} \varphi-\mu \partial \varphi-\partial \mu)\nonumber\\
    &\qquad+(\bar{\partial}-2 \partial \mu-\mu \partial) \bigg(-\frac{i k}{4\pi} \bigg(\frac{1}{2}(\partial \varphi)^2-\partial^2 \varphi\bigg)+\frac{\delta}{\delta \mu}\bigg) \bigg) \Psi[\mu]  \nonumber \\  
    &=\mathrm{e}^{\tilde{S}[\mu, \varphi]}\bigg((\bar{\partial}-2 \partial \mu-\mu \partial) \frac{\delta}{\delta \mu}+\frac{i k}{4\pi} \partial^3 \mu\bigg) \Psi[\mu]~.
\end{align}
Thus the remaining constraint is
\be 
\bigg((\bar{\partial}-2 \partial \mu-\mu \partial) \frac{\delta}{\delta \mu}+\frac{i c}{24\pi} \partial^3 \mu\bigg) \Psi[\mu]=0~, \label{eq:Virasoro Ward identity}
\ee
with $c=6k$.

We can take the chiral wavefunction to be only $\Psi[\mu]$, which we refer to as the reduced wavefunction. The exponential factors $S$ and $\tilde{S}$ merely tell us how $\Psi$ changes if we change the explicit metric or spin connection. In particular \eqref{eq:tilde S anomaly} is the standard holomorphic Weyl anomaly of a CFT with central charge $c=6k$. 

\paragraph{Virasoro Ward identities.} As observed in \cite{Verlinde:1989ua}, the remaining constraint \eqref{eq:Virasoro Ward identity} can be identified with the Virasoro Ward identities of a 2d CFT of central charge $c$.

To see why this is true, consider a conformal block and deform the complex structure via the Beltrami differential. This can be achieved by formally inserting $\exp\left(\frac{1}{2\pi i} \int \d z \wedge \d \bar{z} \, \mu \,T\right)$ into a correlation function. This is just the usual coupling of the stress tensor to the complex structure and can be viewed as coupling to a background gauge field. Let us work out the constraints implied by Virasoro symmetry. We compute
\be 
    \partial_{\bar{z}} \frac{\delta}{\delta \mu(z)} \left \langle \exp\left(\frac{1}{2\pi i} \int \d z \wedge \d \bar{z}\, \mu\,T\right) \right \rangle  =\frac{1}{2\pi i} \, \partial_{\bar{z}}\left\langle  T(z)\, \exp\left(\frac{1}{2\pi i} \int \d z \wedge \d \bar{z}\, \mu\, T\right) \right \rangle~.
\ee
Naively this vanishes, except for contact terms. They arise when $T(z)$ collides with $T(w)$ present in the exponent. We can thus formally expand the exponential to apply the $TT$-OPE to get
\begin{align}
    &T(z)\, \exp\left(\frac{1}{2\pi i} \int \d w \wedge \d \bar{w}\, \mu\, T\right)\nonumber\\
    &\quad=T(z)\, \sum_{n=0}^\infty \frac{1}{n!}\left(\frac{1}{2\pi i} \int \d w \wedge \d \bar{w}\, \mu\, T\right)^n \nonumber\\
    &\quad\sim \frac{1}{2\pi i} \sum_{n=0}^\infty \frac{1}{(n-1)!}\left(\frac{1}{2\pi i} \int \d w \wedge \d \bar{w}\, \mu\, T\right)^{n-1}\nonumber\\
    &\qquad\times\int \d w \wedge \d \bar{w}\, \mu(w) \left(\frac{\frac{c}{2}}{(z-w)^4}+\frac{2 T(w)}{(z-w)^2}+\frac{\partial T(w)}{z-w} \right)  \nonumber\\
    &\quad=\frac{1}{2\pi i}\int \d w \wedge \d \bar{w}\, \mu(w) \left(\frac{\frac{c}{2}}{(z-w)^4}+\frac{2 T(w)}{(z-w)^2}+\frac{\partial T(w)}{z-w} \right) \nonumber\\
    &\qquad\times \exp\left(\frac{1}{2\pi i} \int \d w \wedge \d \bar{w}\, \mu\, T\right)~,
\end{align}
where we omitted regular terms and hence holomorphic terms by applying the OPE. We thus have with the help of the standard distributional identity $\partial_{\bar{z}} \frac{1}{z}=-2\pi i \delta^{(2)}(z)$,
\begin{align}
    &\partial_{\bar{z}} \frac{\delta}{\delta \mu(z)} \left \langle \exp\left( \int \d w \wedge \d \bar{w}\, \mu\, T\right) \right \rangle\nonumber\\
    &\quad=\frac{1}{(2\pi i)^2}\,  \partial_{\bar{z}} \int \d w \wedge \d \bar{w}\, \mu(w) \bigg \langle\!\!\left(\frac{\frac{c}{2}}{(z-w)^4}+\frac{2 T(w)}{(z-w)^2}+\frac{\partial T(w)}{z-w} \right) \nonumber\\
    &\qquad\qquad\times \exp\left(\frac{1}{2\pi i} \int \d w \wedge \d \bar{w}\, \mu\, T\right)\!\!\bigg \rangle \nonumber\\
    &\quad=\frac{-2\pi i}{(2\pi i)^2}  \int \d w \wedge \d \bar{w}\,\mu(w) \bigg \langle\!\Big(\frac{c}{12} \partial_w^3\delta^{(2)}(z-w)+2 T(w)\partial_w \delta^{(2)}(z-w)\nonumber\\
    &\qquad\qquad+\partial_w T(w)\delta^{(2)}(z-w) \Big) \exp\left(\frac{1}{2\pi i} \int \d w \wedge \d \bar{w}\, \mu\, T\right)\!\!\bigg \rangle \nonumber\\
    &\quad=\frac{1}{2\pi i}\bigg\langle \!\Big(\frac{c}{12} \partial^3 \mu(z)+2 \partial \mu(z) T(z)+\mu(z) \partial T(z)\Big)  \exp\left(\frac{1}{2\pi i} \int \d w \wedge \d \bar{w}\, \mu\, T\right)\!\!\bigg \rangle \nonumber \\
    &\quad=\left(\frac{c\, \partial^3 \mu(z)}{24 \pi i} +2 \partial \mu(z) \frac{\delta}{\delta \mu(z)}+\mu(z) \partial_z \frac{\delta}{\delta \mu(z)}\right)\left\langle \exp\left(\frac{1}{2\pi i} \int \d w \wedge \d \bar{w}\, \mu\, T\right)\right \rangle ~.
\end{align}
Subtracting the right hand side from the left hand side of this equation hence leads to the constraint 
\be 
\bigg((\bar{\partial}-2 \partial \mu-\mu \partial) \frac{\delta}{\delta \mu}+\frac{i c}{24\pi} \partial^3 \mu\bigg) \exp\left(\frac{1}{2\pi i} \int \d w \wedge \d \bar{w}\, \mu\, T\right)=0~.
\ee
We condensed the notation and see that this coincides with \eqref{eq:Virasoro Ward identity}, which are hence indeed the Virasoro Ward identities. 
It then makes also sense that we precisely got the holomorphic conformal anomaly in \eqref{eq:tilde S anomaly} since $\Psi[\mu]$ carries this anomaly.

Let us now go back to the non-chiral wavefunction, which we can similarly reduce to $\Psi[\mu,\bar{\mu}]$. $\bar{\mu}$ arises in the decomposition of $\bar{\mathcal{A}}^-$. The reality conditions of $\PSL(2,\CC)$ now imply that we should view $\bar{\mu}$ as the complex conjugate of $\mu$ and not as an independent Beltrami differential.
$\Psi[\mu,\bar{\mu}]$ satisfies an identical constraints as \eqref{eq:Virasoro Ward identity} with $\mu \to \bar{\mu}$. In particular, the right-moving central charge $c$ is identical to the left-moving central charge. Given that $k$ is purely imaginary, the central charge $c$ is also purely imaginary.

\paragraph{One-loop correction.} This result receives a quantum correction. The central charge should actually take values in $c \in 13+i \RR$. The $+13$ is a one-loop correction to the central charge. Indeed, as mentioned above, $k$ plays the role of $\hbar^{-1}$ and the leading central charge that found dominates over an order 1 correction. It is relatively well-known in the context of $\mathrm{AdS}_3$ gravity see \cite{Giombi:2008vd}. It was discussed from a path integral perspective in \cite{Cotler:2019nbi} and is independent of the sign of the cosmological constant.

Let us explain how it arises from the canonical quantization perspective analogous to what was explained in \cite{Eberhardt:2022wlc} for $\mathrm{AdS}_3$ gravity. Following the geometric quantization procedure in the form that we explained for say the harmonic oscillator would not correctly reproduce the ground state energy $\frac{1}{2}  \hbar \omega$ of the harmonic oscillator. Indeed, we normal ordered the constraint in \eqref{eq:F3 constraint} such that there is no ground state energy. We could attempt to also put in half of the ordering $\frac{\delta}{\delta \mathcal{A}_z^+} \mathcal{A}_z^+$ and $\frac{\delta}{\delta \mathcal{A}_{\bar{z}}^+} \mathcal{A}_{\bar{z}}^+$ and regularize, but this is a very divergent procedure.
Instead it is better to discuss this on the level of the constrained phase space, which is finite dimensional. The wavefunction $\Psi[\mu,\bar{\mu}]$ modulo the constraint \eqref{eq:Virasoro Ward identity} can be viewed as a `function' on the moduli space of complex structures on $\Sigma_g$. To be precise, we actually get the universal covering space of moduli space, which is Teichm\"uller space $\mathcal{T}_g$. We discussed this already in section~\ref{subsec:relation CS}. This is a global issue which doesn't play a role for the local constraint such as the Virasoro Ward identity \eqref{eq:Virasoro Ward identity}. The wavefunction is also not quite a function because of the conformal anomaly which tells us that we have to choose an explicit trivialization of a line bundle over $\mathcal{T}_g$. In other words, it is a section of a hermitian line bundle $|\mathscr{L}|^{2c}$ over Teichmuller space. The line bundle is hermitian because for purely imaginary central charge, the mod squared of the wavefunction $|\Psi|^2$ does not carry a conformal anomaly and is thus a function on $\mathcal{T}_g$. Thus the norm on the Hilbert space for such a real polarization should be the standard $\mathrm{L}^2$-inner product on $\mathcal{T}_g$,
\be 
\lVert \Psi \rVert\overset{?}{=} \int_{\mathcal{T}_g} |\Psi|^2~. \label{eq:norm before metaplectic correction}
\ee
However, we don't have a natural integration measure on $\mathcal{T}_{g}$ appropriate for this quantization. Thus we ultimately want to view the wavefunction as a density on $\mathcal{T}_g$, i.e.~the `improved' wave-function is $\Psi\sqrt{\Omega}$, where $\Omega$ transforms as a top form on $\mathcal{T}_g$. Then the integral \eqref{eq:norm before metaplectic correction} makes sense. Since $\mathcal{T}_g$ is a complex manifold, the line bundle of top forms can be written as $\mathscr{K} \otimes \mathscr{K}^*$ with $\mathscr{K}$ the holomorphic canonical line bundle, whose secions are holomorphic top forms.
This modification solves two issues. On the one hand, it reinstates the natural Casimir energy ($\frac{1}{2}\hbar \omega$ for the harmonic oscillator). Furthermore, it makes it possible to define a natural norm on the Hilbert space that we will discuss below. In the context of geometric quantization, this improvement is known as the metaplectic correction or half-form quantization.
In any case, the chiral wavefunction is locally modified by a square root of the canonical line bundle (the line bundle of \emph{holomorphic} top forms) on moduli space.

Physicists understand the line bundle $\mathscr{K}$ very well from string theory \cite{Friedan:1986ua}. 
In string theory, we also consider integrals $\int_{\mathcal{T}_{g,n}} \Omega$ with $\Omega \in \mathscr{K} \otimes \mathscr{K}^*$ (up to the replacement $\mathcal{T}_{g,n} \to \mathcal{M}_{g,n})$.
Cancellation of the conformal anomaly requires that the integrand carries left- and right-moving central charge $26$. Thus a holomorphic section of the canonical line bundle over Teichm\"uller space transforms like CFT partition function of holomorphic central charge $c=26$. Upon taking the square root, we learn that the modification $\sqrt{\Omega}$ transforms like a CFT partition function of central charge $13$. In the mathematical literature the isomorphism $\mathscr{K} \cong \mathscr{L}^{26}$ is known as Mumford's isomorphism.\footnote{It is usually formulated in terms of the determinant line bundle of the Hodge bundle $\mathcal{L}^2 \cong \det \EE$ which also appears in the work of Friedan and Shenker \cite{Friedan:1986ua}.} Taken together with $\Psi[\mu]$, the improved chiral wavefunction hence satisfies the Virasoro Ward identities with left moving central charge $c=13+6k$. The right-moving central charge is likewise modified. Overall, the wavefunction hence behaves like a CFT partition function of central charge $13+6k$ for both left- and right-movers, except that we haven't imposed crossing symmetry at this point since we haven't gauged the mapping class group.

\paragraph{Chern-Simons polarization.} Let us explain a different route to arrive at this result. We can quantize via the Chern-Simons polarization. In this case, the constraints on the wavefunction translate to the Ward identities for $\mathfrak{sl}(2,\CC)$ current algebra blocks of \emph{imaginary} level $k$. The current algebra takes the form in the normalization of \eqref{eq:sl2C structure constants and Killing form}
\begin{subequations}
\begin{align} 
J^3(z) J^\pm(w) &\sim \frac{\pm J^\pm(w)}{z-w}~, \\
J^3(z) J^3(w) &\sim -\frac{k}{2(z-w)^2}~, \\
J^+(z) J^-(w) &\sim -\frac{k}{2(z-w)^2}+\frac{J^3(w)}{(z-w)^2}~, 
\end{align}
\end{subequations}
together with a complex conjugated set of currents. 

Contrary to the familiar story of the quantization of $\mathrm{SU}(2)$ Chern-Simons theory \cite{Witten:1988hf, Axelrod:1989xt}, this polarization can also be viewed as real because the phase space is hyperk\"ahler. Thus we can define an inner product by integrating over a real slice of the moduli space, which we can take to be the moduli space of flat $\SU(2)$ connections as in \cite{Witten:1989ip}. We would like to define the norm as
\be 
\lVert \Psi \rVert^2=\int_{\mathcal{M}_g(\SU(2))} |\Psi|^2 \label{eq:SL(2,C) current norm before metaplectic correction}
\ee
with $\mathcal{M}_g(\SU(2))$ the moduli space of flat $\SU(2)$ connections. This again requires the inclusion of a one-loop correction to the level of the current algebra.

Thus we have to figure out what level the holomorphic canonical bundle $\mathscr{K}$ on $\mathcal{M}_g(\SU(2))$ carries.  Integration over $\mathcal{M}_g(\SU(2))$ requires 3 $\mathfrak{b}\mathfrak{c}$ ghosts of conformal weight $h(\mathfrak{b})=1$ and $h(\mathfrak{c})=0$.\footnote{The BRST current for this gauging has the standard form $j=\mathfrak{c}_a j^a-\frac{1}{2} f^{abc} \mathfrak{b}_a \mathfrak{c}_b \mathfrak{c}_c$, together with the corresponding right-moving part.} Their bilinears generate an $\mathfrak{su}(2)$ current algebra of level $-4$.\footnote{This comes from the well-known statement that three fermions $\psi^3$, $\psi^\pm$ generate an $\mathfrak{su}(2)_2$ current algebra. The $\mathfrak{b}\mathfrak{c}$ ghosts above are twice as many fields and hence generate a $\mathfrak{su}(2)_4$ current algebra. Finally, the level for an $\mathfrak{sl}(2,\RR)$ current algebra is conventionally oppositely defined.} Similar to the cancellation of the conformal anomaly, the current algebra anomaly has to cancel when integrating over the constrained phase space and the total level of an integrand has to be $+4$ for both the left- and the right-movers. Thus the section $\sqrt{\Omega}$ transforms like a current block of level $2$. We hence learn in complete analogy that the metaplectic correction shifts the level of the current algebra to $\kappa=2+k$.

\paragraph{Hamiltonian reduction.} One can then change the polarization at the quantum level by gauging the current $J^-$ of the current algebra to a constant. This is realized via the BRST charge 
\be 
 \mathcal{Q}= \oint (\mathfrak{c} (J^--1))+(\bar{\mathfrak{c}} (\bar{J}^--1))~, \label{eq:BRST operator quantum Hamiltonian reduction}
\ee
where $\mathfrak{c}$ is a ghost of conformal weight 0 (which has nothing to do with the string-theoretic ghosts). Thanks to the vanishing of the $J^- J^-$ OPE, this BRST differential squares to zero. Imposing the constraints lets us pass to the cohomology of the BRST charge. This procedure is known as Drinfel'd-Sokolov or quantum Hamiltonian reduction \cite{Drinfeld:1984qv}. The chiral algebra acting on the cohomology is generated by the Virasoro tensor 
\be 
T=\frac{1}{\kappa-2}\Big(-(J^3J^3)+\frac{1}{2}\big((J^+ J^-)+(J^- J^+)\big)\Big)- \partial J^3- (\mathfrak{b} \partial \mathfrak{c})~. \label{eq:Hamiltonian reduction Virasoro tensor}
\ee
The first term is the standard Sugawara stress tensor, the second an improvement term which makes the conformal weight of $J^-$ vanish so that it has a chance to commute with the BRST operator \eqref{eq:BRST operator quantum Hamiltonian reduction} while the final piece is the contribution from the ghosts. The total central charge coming from the three pieces is
\be 
c=\frac{3\kappa}{\kappa-2}+6\kappa-2=1+6(b+b^{-1})^2\text{ with }b^2=\frac{1}{\kappa-2}=\frac{1}{k} \in i \RR~. \label{eq:b kappa map}
\ee
Thus the remaining Ward identities acting on the cohomology are precisely the Virasoro Ward identities of the Virasoro tensor \eqref{eq:Hamiltonian reduction Virasoro tensor}. Written in terms of $k$, we have
\be 
c=6k+13+\frac{6}{k}~.
\ee
The first two terms reproduce precisely what we found above. The third term is a two-loop correction, because $\frac{1}{k} \sim \hbar$. Since the imaginary part of $k$ is not quantized, it is not protected against renormalization and in this scheme gets a two-loop correction. We could simply redefine $k+\frac{1}{k} \to k$ to precisely match with the result above. However, it importantly preserves $c \in 13+i \RR$.

\subsection{Punctures}

 We can enrich the discussion by considering a number of punctures in the surface $\Sigma_{g,n}$. 
\paragraph{Monodromies.} The $\PSL(2,\CC)$ gauge field then has some prescribed monodromy around these punctures, i.e.\ every puncture is classically labelled by a conjugacy class in $\PSL(2,\CC)$ and the phase space depends on the choice of $n$ such conjugacy classes.
Most of what we discussed about the quantization remains unchanged, but let us discuss what happens to the gauge field close to the puncture. The presence of the monodromy means that the gauge fields must have singularities near the puncture. Let us first review what happens in the usual Chern-Simons polarization \cite{Elitzur:1989nr}. 
 Focus on a disk centered at the puncture and consider the boundary values of the gauge field. This determines the gauge field anywhere inside the disk by flatness. The loop group of $\mathrm{G}=\PSL(2,\CC)$ acts on the boundary values of the gauge field by $\mathcal{A} \mapsto \mathcal{A}^g=g \mathcal{A} g^{-1}+k \d g g^{-1}$, where $g: \S^1 \longrightarrow \mathrm{G}$. This is the coadjoint action of the loop group $\widehat{\mathscr{L}\mathrm{G}}_k$. The conjugacy class of the monodromy $\text{Pexp}\left(\oint_{\S^1} \mathcal{A} \right)$ is invariant under such gauge transformations. Thus conjugacy classes of $\mathrm{G}$ are in one-to-one correspondence with coadjoint orbits of the loop group. Quantization of these coadjoint orbits in turn gives rise to representations of the corresponding current algebra $\mathfrak{g}_k$. 
 Hence in the quantum theory, punctures are labelled by representations of $\mathfrak{g}_k$.

\paragraph{Virasoro coadjoint orbits.} When employing the polarization leading to Virasoro conformal blocks, we are tempted to think that we obtain instead coadjoint orbits of the complex Virasoro group $\mathrm{Vir}_\CC$, which upon quantization lead to Virasoro representations. This is however actually not true as there is no such thing as a complex Virasoro group \cite{Lempert}. Thus we will in the following discuss only the situation for the global $\PSL(2,\CC)$ in the language of geometric quantization. The coadjoint orbits $\PSL(2,\CC)/\CC^\times$ of $\PSL(2,\CC)$ are also hyperk\"ahler manifolds \cite{Hitchin_1979,Kronheimer1990AHS} and in fact nothing else than the Eguchi-Hanson space $T^*\CP^1$. Thus geometric quantization is simple, since we can choose a polarization in which the wavefunctions only depend on the $\CP^1$-coordinates. They are sections of a non-holomorphic line bundle $\mathscr{L}^h \otimes \bar{\mathscr{L}}^{\tilde{h}}$ over $\CP^1$, where $(h,\tilde{h})$ depends on the chosen coadjoint orbit. The $\PSL(2,\CC)$ action on a non-holomorphic section of this line bundle is
\be 
f(z) \longmapsto (c z+d)^{-2h}\overline{(c z+d)}^{-2\tilde{h}} f \left(\frac{a z+b}{c z+d} \right)~, \quad \begin{pmatrix}
    a & b \\ c & d
\end{pmatrix} \in \PSL(2,\CC)~. \label{eq:PSL(2,C) representation principal series}
\ee
$h-\tilde{h}$ describes the rotational part of the monodromy of the coadjoint orbit. Its values are quantized to $h-\tilde{h} \in \ZZ$ which is required for the line bundle to be well-defined. $h+\tilde{h}$ describes the scaling part of the monodromy. It is naively required to be purely imaginary given that the rotational part is purely real. 
We again need to perform the one-loop correction by tensoring the wavefunction by the square root of the volume element $\sqrt{\d^2z}$, which transforms like $h=\tilde{h}=\frac{1}{2}$. Thus allowed values after this correction are $s=h-\tilde{h} \in \ZZ$ and $\Delta=h+\tilde{h} \in 1+i \RR$ corresponding to the principal series representation of $\PSL(2,\CC)$.

\eqref{eq:PSL(2,C) representation principal series} is precisely the action of the global conformal group on a primary field of conformal dimension $(h,\tilde{h})$. Thus as in the case of Chern-Simons theory with a compact gauge group \cite{Witten:1988hf}, the inclusion of punctures naturally incorporates punctures in the conformal blocks. This argument only probes the allowed representations for the global part of the group. As is well-known in the case of $\mathrm{SU}(2)_k$, not all $\mathrm{SU}(2)$ representations uplift to the loop group, but the algebro-geometric argument for this is somewhat non-trivial even in the case of $\mathrm{SU}(2)$ \cite{Gawedzki:1990qt}. 

We believe that there is morally no restriction on the representations which uplift to $\text{Vir}_\CC$ (even though this group doesn't exist). However, we have not been able to make this completely precise.

\paragraph{Inner product in the presence of punctures.} Much of the discussion of the metaplectic correction goes through in the presence of punctures. The inner product now naturally takes the form
\be 
\lVert \Psi \rVert^2=\int_{\mathcal{T}_{g,n}} |\Psi|^2~.\label{eq:3d naive inner product appendix}
\ee
We already analyzed the conditions for this integral to be well-defined in \eqref{eq:h htilde reality conditions}. The fact that $\Re(\Delta) = 1$ is the analogue of the metaplectic correction in this context. 

\paragraph{Equivalent polarizations.} The inner product \eqref{eq:3d naive inner product appendix} on $\hat{\mathcal{H}}_{g,n}^{(b)}(\boldsymbol{\Delta},\boldsymbol{s})$ can also be viewed as a non-degenerate bilinear form
\be 
\hat{\mathcal{H}}_{g,n}^{(b)}(\boldsymbol{\Delta},\boldsymbol{s})\otimes \hat{\mathcal{H}}_{g,n}^{(-ib)}(2-\boldsymbol{\Delta},-\boldsymbol{s}) \longrightarrow \CC~.
\ee
Dualizing the bilinear form leads to a natural isometry between $\hat{\mathcal{H}}_{g,n}^{(b)}(\boldsymbol{\Delta},\boldsymbol{s})$ and $\hat{\mathcal{H}}_{g,n}^{(-ib)}(2-\boldsymbol{\Delta},-\boldsymbol{s})$. This map is furthermore equivariant with respect to the action of the mapping class group $\text{Map}(\Sigma_{g,n})$ on the Hilbert space. This shows that these two quantizations are equivalent. This is to be expected since nothing in 3d gravity told us which sign to use in \eqref{eq:level PSL(2,C) CS gravity}.

\subsection{Normalizability}
We now want to analyze the conditions on the wavefunction $\Psi$ in order to be normalizable. In particular, we are interested in the behaviour of $\Psi$ near the boundary of $\mathcal{T}_{g,n}$ where one of the curves defining the pair of pants decomposition pinches.

\paragraph{Behaviour of the integrand under degeneration.} Consider such a degenerating region in $\mathcal{T}_{g,n}$. We consider a local plumbing coordinate $q$, so that pinching of the surface corresponds to $q \to 0$ and Dehn twists act as $q \to \mathrm{e}^{2\pi i} q$. 
 Consider now the integrand $(\Psi_{g,n}')^*\Psi_{g,n}$ of the integrand in \eqref{eq:3d naive inner product appendix}. As explained e.g.\ in \cite{Witten:2013tpa},\footnote{This fact is well-known in string field theory where $\int \d q \, q^{L_0-2}=\frac{1}{L_0-1}$ is the propagator.} such integrals degenerate in a separating limit to
\be 
(\Psi_{g,n}')^*\Psi_{g,n} \sim (\Psi_{g_1,n_1+1}')^*\Psi_{g_1,n_1+1}\,  \d^2 q\,  q^{L_0-2} \, \bar{q}^{\tilde{L}_0-2}\, (\Psi_{g_2,n_2+1}')^*\Psi_{g_2,n_2+1}~,
\ee
where $L_0$ and $\tilde{L}_0$ run over the spectrum of the worldsheet CFT. Normalizability tells us that $L_0-\tilde{L}_0 \in \RR$ and $\Re(L_0+\tilde{L}_0) \ge 2$. Translated back to the scaling dimensions of $\Psi_{g,n}$, this means that the internal scaling dimensions are bounded from below by $\Re \Delta \ge 1$ and $s \in \RR$ for normalizability of the inner product \eqref{eq:3d naive inner product}.
This leaves the hard question of finding a complete and (delta function) normalizable basis of the inner product. We anticipate that the answer is given by `Virasoro partial waves', see the discussion \ref{sec:discussion}. We in particular believe that a normalizable basis with only $\Re \Delta=1$ can be chosen.

\section{The matrix integral dual of the complex Liouville string}\label{app:matrix model dual}

In this appendix we recall some details of the dual matrix integral of the complex Liouville string \cite{paper2} and therefore of cosmological correlators in $\mathrm{dS}_3$. 

Before double scaling, we consider a two-matrix model of the form
\begin{equation}
    \int [\d M_1][\d M_2]\, \mathrm{e}^{-N \tr(V_1(M_1) + V_2(M_2) - M_1 M_2)}~, \label{eq:two matrix integral}
\end{equation}
where $M_1$ and $M_2$ are two $N\times N$ hermitian matrices. Double scaling amounts to taking $N\rightarrow \infty$ and to zoom into a particular region of the eigenvalue distribution. We also consider \eqref{eq:two matrix integral} with the insertion of resolvents
\begin{equation}\label{eq: resolvent}
    R(x) = \tr \frac{1}{x- M_1}~.
\end{equation}
Such (connected) correlators of resolvents admit a genus expansion in $1/N$.
A resolvent has a pole whenever $x$ equals one of the eigenvalues of $M_1$, which get smeared out to a branch cut once we insert them in the matrix integral. Thus, expectation values of resolvents are multivalued functions in the complex plane with branch cuts along the real axis where the eigenvalues of the matrices are located. 
This defines a multi-sheeted covering of the complex $x$-plane known as the spectral curve of the model. The spectral curve of the matrix model is a Riemann surface and encodes the whole data of the potential in \eqref{eq:two matrix integral} (together with the filling fractions of different cuts). In particular, it is possible to extract the whole perturbative genus expansion of the correlators of the resolvents from the geometry of the spectral curve via topological recursion.
Thus instead of working with the potential in \eqref{eq:two matrix integral}, it is much more convenient to specify the spectral curve of the model.
In \cite{paper2}, we showed that the amplitudes $\mathsf{A}_{g,n}^{(b)}$ of the complex Liouville string admit a dual description in terms of the connected correlators of resolvents in a matrix model with spectral curve\footnote{The form of the spectral curve makes it necessary to consider two-matrix models. Spectral curves of single matrix models are hyperelliptic, meaning that they are of the form $y^2=P(x)$ for some entire function $P(x)$. However, since we are not considering observables containing the second matrix $M_2$, we could simply integrate it out in \eqref{eq:two matrix integral}, thus effectively leading to a single matrix model, but with a non-analytic potential. \label{footnote:two matrices}} 
\begin{equation}\label{eq: MM spectral curve}
    \xx(z) = -2 \cos(\pi b \sqrt{z})~,\quad \yy(z) = 2\cos(\pi b^{-1}\sqrt{z})~.
\end{equation}
Here $z$ is a uniformizing parameter on the spectral curve $(x,y)=(\xx(z),\yy(z))$.
From the spectral curve, one can in particular extract the density of eigenvalues of $M_1$ by taking the discontinuity of the first resolvent $R_{0,1}(x)=V_1'(x)-\yy(\xx^{-1}(x))$. This leads to
\begin{equation}\label{eq: eigenvalue distribution}
    \rho_0(E) = \frac{2}{\pi}\sinh(-i \pi b^2)\sin \left(- ib^2 \mathrm{arccosh}\left(\frac{E}{2}\right)\right)~,
\end{equation}
for the first matrix. A similar expression can be obtained for the second matrix.

The Riemann surface (\ref{eq: MM spectral curve}) has infinitely many branch points where $\d \xx(z_*)=0$ and infinitely many self-intersections. The latter are parametrized by two integers ($r,s$) and are given by
\begin{equation}
    \xx(z_{(r,s)}^\pm) = \yy(z_{(r,s)}^\pm) \quad \Leftrightarrow \quad z_{(r,s)}^\pm =(rb \pm s b^{-1})^2~.
\end{equation}
Nodal singularities may be viewed as collapsed cycles of a higher genus surface.
Their main relevance for us is that they control (doubly) non-perturbative effects. Each of the nodal singularities leads to a non-perturbative correction of the form \eqref{eq:non perturbative corrections} with a different tension in the exponential. The tensions are given by the integrals of the form $-\int_{z_{r,s}^-}^{z_{r,s}^+} \yy(z) \d \xx(z)$, which are closed cycles on the spectral curve.
The locations of ZZ-instantons also correspond to saddle points in the effective potential of the eigenvalues (i.e.\ the potential generated from the explicit terms in \eqref{eq:two matrix integral}, together with the eigenvalue repulsion coming from the Vandermonde determinant).

The eigenvalue distribution (\ref{eq: eigenvalue distribution}) is initially positive for $E \ge 2$ above the lower threshold, where it has the characteristic square root behaviour of matrix models. However, it develops a maximum and goes to zero at $E_{0}= 2 \cos(\pi b^{-2})$. See figure \ref{fig:eigenvalue density} for a picture of this behaviour. For larger energies, the eigenvalue density oscillates. 

The negativities of the eigenvalue density are clearly non-sensical.
As was noticed in \cite{paper3} the first zero $E_0$ coincides with the location of the first ZZ-instanton $z_{(1,1)}$ since $E_0=\xx(z_{(1,1)})$. Since the locations of ZZ-instantons correspond to extrema of the effective potential, one can choose an arbitrary Lefschetz thimble as a steepest descent contour above $E_0$ to give a non-perturbative definition of the matrix integral. In particular, following a steepest descent contour means that the contour turns into the complex plane above $E_0$, thus evading the negativity in the eigenvalue density.
In the next section, we take this as an indication that we should only count all the eigenvalues up to $E_0$ as physical microstates and effectively cut the eigenvalue distribution off at $E_0$.

\bibliographystyle{JHEP}
\bibliography{bib}

\providecommand{\href}[2]{#2}\begingroup\raggedright\begin{thebibliography}{100}

\bibitem{Collier:2024kmo}
S.~Collier, L.~Eberhardt, B.~M\"uhlmann and V.~A. Rodriguez, \emph{{The complex
  Liouville string}},  \href{https://arxiv.org/abs/2409.17246}{{\ttfamily
  2409.17246}}.

\bibitem{Anninos:2017hhn}
D.~Anninos and D.~M. Hofman, \emph{{Infrared Realization of dS$_2$ in
  AdS$_2$}}, \href{https://doi.org/10.1088/1361-6382/aab143}{\emph{Class.
  Quant. Grav.} {\bfseries 35} (2018) 085003}
  [\href{https://arxiv.org/abs/1703.04622}{{\ttfamily 1703.04622}}].

\bibitem{Maldacena:2019cbz}
J.~Maldacena, G.~J. Turiaci and Z.~Yang, \emph{{Two dimensional Nearly de
  Sitter gravity}}, \href{https://doi.org/10.1007/JHEP01(2021)139}{\emph{JHEP}
  {\bfseries 01} (2021) 139}
  [\href{https://arxiv.org/abs/1904.01911}{{\ttfamily 1904.01911}}].

\bibitem{Cotler:2019nbi}
J.~Cotler, K.~Jensen and A.~Maloney, \emph{{Low-dimensional de Sitter quantum
  gravity}}, \href{https://doi.org/10.1007/JHEP06(2020)048}{\emph{JHEP}
  {\bfseries 06} (2020) 048}
  [\href{https://arxiv.org/abs/1905.03780}{{\ttfamily 1905.03780}}].

\bibitem{Anninos:2021ene}
D.~Anninos, T.~Bautista and B.~M\"uhlmann, \emph{{The two-sphere partition
  function in two-dimensional quantum gravity}},
  \href{https://doi.org/10.1007/JHEP09(2021)116}{\emph{JHEP} {\bfseries 09}
  (2021) 116} [\href{https://arxiv.org/abs/2106.01665}{{\ttfamily
  2106.01665}}].

\bibitem{Anninos:2021eit}
D.~Anninos and B.~M\"uhlmann, \emph{{The semiclassical gravitational path
  integral and random matrices (toward a microscopic picture of a dS$_{2}$
  universe)}}, \href{https://doi.org/10.1007/JHEP12(2021)206}{\emph{JHEP}
  {\bfseries 12} (2021) 206}
  [\href{https://arxiv.org/abs/2111.05344}{{\ttfamily 2111.05344}}].

\bibitem{Anninos:2023exn}
D.~Anninos, P.~Benetti~Genolini and B.~M\"uhlmann, \emph{{dS$_{2}$
  supergravity}}, \href{https://doi.org/10.1007/JHEP11(2023)145}{\emph{JHEP}
  {\bfseries 11} (2023) 145}
  [\href{https://arxiv.org/abs/2309.02480}{{\ttfamily 2309.02480}}].

\bibitem{Anninos:2024iwf}
D.~Anninos, C.~Baracco and B.~M\"uhlmann, \emph{{Remarks on 2D quantum
  cosmology}}, \href{https://doi.org/10.1088/1475-7516/2024/10/031}{\emph{JCAP}
  {\bfseries 10} (2024) 031}
  [\href{https://arxiv.org/abs/2406.15271}{{\ttfamily 2406.15271}}].

\bibitem{Verlinde:2024zrh}
H.~Verlinde and M.~Zhang, \emph{{SYK Correlators from 2D Liouville-de Sitter
  Gravity}},  \href{https://arxiv.org/abs/2402.02584}{{\ttfamily 2402.02584}}.

\bibitem{Verlinde:2024znh}
H.~Verlinde, \emph{{Double-scaled SYK, Chords and de Sitter Gravity}},
  \href{https://arxiv.org/abs/2402.00635}{{\ttfamily 2402.00635}}.

\bibitem{Coleman:2021nor}
E.~Coleman, E.~A. Mazenc, V.~Shyam, E.~Silverstein, R.~M. Soni, G.~Torroba
  et~al., \emph{{De Sitter microstates from $T\overline{T}+\Lambda_{2}$ and the
  Hawking-Page transition}},
  \href{https://doi.org/10.1007/JHEP07(2022)140}{\emph{JHEP} {\bfseries 07}
  (2022) 140} [\href{https://arxiv.org/abs/2110.14670}{{\ttfamily
  2110.14670}}].

\bibitem{Batra:2024kjl}
G.~Batra, G.~B. De~Luca, E.~Silverstein, G.~Torroba and S.~Yang,
  \emph{{Bulk-local dS$_{3}$ holography: the matter with $ T\overline{T}
  +\Lambda_{2}$}}, \href{https://doi.org/10.1007/JHEP10(2024)072}{\emph{JHEP}
  {\bfseries 10} (2024) 072}
  [\href{https://arxiv.org/abs/2403.01040}{{\ttfamily 2403.01040}}].

\bibitem{tHooft:1993dmi}
G.~'t~Hooft, \emph{{Dimensional reduction in quantum gravity}}, {\emph{Conf.
  Proc. C} {\bfseries 930308} (1993) 284}
  [\href{https://arxiv.org/abs/gr-qc/9310026}{{\ttfamily gr-qc/9310026}}].

\bibitem{Susskind:1994vu}
L.~Susskind, \emph{{The World as a hologram}},
  \href{https://doi.org/10.1063/1.531249}{\emph{J. Math. Phys.} {\bfseries 36}
  (1995) 6377} [\href{https://arxiv.org/abs/hep-th/9409089}{{\ttfamily
  hep-th/9409089}}].

\bibitem{Maldacena:1997re}
J.~M. Maldacena, \emph{{The Large N limit of superconformal field theories and
  supergravity}}, \href{https://doi.org/10.4310/ATMP.1998.v2.n2.a1}{\emph{Adv.
  Theor. Math. Phys.} {\bfseries 2} (1998) 231}
  [\href{https://arxiv.org/abs/hep-th/9711200}{{\ttfamily hep-th/9711200}}].

\bibitem{Strominger:2001pn}
A.~Strominger, \emph{{The dS / CFT correspondence}},
  \href{https://doi.org/10.1088/1126-6708/2001/10/034}{\emph{JHEP} {\bfseries
  10} (2001) 034} [\href{https://arxiv.org/abs/hep-th/0106113}{{\ttfamily
  hep-th/0106113}}].

\bibitem{Maldacena:2002vr}
J.~M. Maldacena, \emph{{Non-Gaussian features of primordial fluctuations in
  single field inflationary models}},
  \href{https://doi.org/10.1088/1126-6708/2003/05/013}{\emph{JHEP} {\bfseries
  05} (2003) 013} [\href{https://arxiv.org/abs/astro-ph/0210603}{{\ttfamily
  astro-ph/0210603}}].

\bibitem{Witten:2001kn}
E.~Witten, \emph{{Quantum gravity in de Sitter space}},  in \emph{{Strings
  2001: International Conference}}, 6, 2001,
  \href{https://arxiv.org/abs/hep-th/0106109}{{\ttfamily hep-th/0106109}}.

\bibitem{Anninos:2011ui}
D.~Anninos, T.~Hartman and A.~Strominger, \emph{{Higher Spin Realization of the
  dS/CFT Correspondence}},
  \href{https://doi.org/10.1088/1361-6382/34/1/015009}{\emph{Class. Quant.
  Grav.} {\bfseries 34} (2017) 015009}
  [\href{https://arxiv.org/abs/1108.5735}{{\ttfamily 1108.5735}}].

\bibitem{Anninos:2017eib}
D.~Anninos, F.~Denef, R.~Monten and Z.~Sun, \emph{{Higher Spin de Sitter
  Hilbert Space}}, \href{https://doi.org/10.1007/JHEP10(2019)071}{\emph{JHEP}
  {\bfseries 10} (2019) 071}
  [\href{https://arxiv.org/abs/1711.10037}{{\ttfamily 1711.10037}}].

\bibitem{Dong:2018cuv}
X.~Dong, E.~Silverstein and G.~Torroba, \emph{{De Sitter Holography and
  Entanglement Entropy}},
  \href{https://doi.org/10.1007/JHEP07(2018)050}{\emph{JHEP} {\bfseries 07}
  (2018) 050} [\href{https://arxiv.org/abs/1804.08623}{{\ttfamily
  1804.08623}}].

\bibitem{Shyam:2021ciy}
V.~Shyam, \emph{{$ \mathrm{T}\overline{\mathrm{T}} $ +
  \ensuremath{\Lambda}$_{2}$ deformed CFT on the stretched dS$_{3}$ horizon}},
  \href{https://doi.org/10.1007/JHEP04(2022)052}{\emph{JHEP} {\bfseries 04}
  (2022) 052} [\href{https://arxiv.org/abs/2106.10227}{{\ttfamily
  2106.10227}}].

\bibitem{Anninos:2014lwa}
D.~Anninos, T.~Anous, D.~Z. Freedman and G.~Konstantinidis, \emph{{Late-time
  Structure of the Bunch-Davies De Sitter Wavefunction}},
  \href{https://doi.org/10.1088/1475-7516/2015/11/048}{\emph{JCAP} {\bfseries
  11} (2015) 048} [\href{https://arxiv.org/abs/1406.5490}{{\ttfamily
  1406.5490}}].

\bibitem{Hogervorst:2021uvp}
M.~Hogervorst, J.~a. Penedones and K.~S. Vaziri, \emph{{Towards the
  non-perturbative cosmological bootstrap}},
  \href{https://doi.org/10.1007/JHEP02(2023)162}{\emph{JHEP} {\bfseries 02}
  (2023) 162} [\href{https://arxiv.org/abs/2107.13871}{{\ttfamily
  2107.13871}}].

\bibitem{DiPietro:2021sjt}
L.~Di~Pietro, V.~Gorbenko and S.~Komatsu, \emph{{Analyticity and unitarity for
  cosmological correlators}},
  \href{https://doi.org/10.1007/JHEP03(2022)023}{\emph{JHEP} {\bfseries 03}
  (2022) 023} [\href{https://arxiv.org/abs/2108.01695}{{\ttfamily
  2108.01695}}].

\bibitem{Baumann:2022jpr}
D.~Baumann, D.~Green, A.~Joyce, E.~Pajer, G.~L. Pimentel, C.~Sleight et~al.,
  \emph{{Snowmass White Paper: The Cosmological Bootstrap}},  in
  \emph{{Snowmass 2021}}, 3, 2022,
  \href{https://arxiv.org/abs/2203.08121}{{\ttfamily 2203.08121}}.

\bibitem{Chandrasekaran:2022cip}
V.~Chandrasekaran, R.~Longo, G.~Penington and E.~Witten, \emph{{An algebra of
  observables for de Sitter space}},
  \href{https://doi.org/10.1007/JHEP02(2023)082}{\emph{JHEP} {\bfseries 02}
  (2023) 082} [\href{https://arxiv.org/abs/2206.10780}{{\ttfamily
  2206.10780}}].

\bibitem{Gibbons:1976ue}
G.~W. Gibbons and S.~W. Hawking, \emph{{Action Integrals and Partition
  Functions in Quantum Gravity}},
  \href{https://doi.org/10.1103/PhysRevD.15.2752}{\emph{Phys. Rev. D}
  {\bfseries 15} (1977) 2752}.

\bibitem{Gibbons:1977mu}
G.~W. Gibbons and S.~W. Hawking, \emph{{Cosmological Event Horizons,
  Thermodynamics, and Particle Creation}},
  \href{https://doi.org/10.1103/PhysRevD.15.2738}{\emph{Phys. Rev. D}
  {\bfseries 15} (1977) 2738}.

\bibitem{paper2}
S.~Collier, L.~Eberhardt, B.~M\"uhlmann and V.~A. Rodriguez, \emph{{The complex
  Liouville string: the matrix integral}},
  \href{https://arxiv.org/abs/2410.07345}{{\ttfamily 2410.07345}}.

\bibitem{Brown:1986nw}
J.~D. Brown and M.~Henneaux, \emph{{Central Charges in the Canonical
  Realization of Asymptotic Symmetries: An Example from Three-Dimensional
  Gravity}}, \href{https://doi.org/10.1007/BF01211590}{\emph{Commun. Math.
  Phys.} {\bfseries 104} (1986) 207}.

\bibitem{Castro:2012gc}
A.~Castro and A.~Maloney, \emph{{The Wave Function of Quantum de Sitter}},
  \href{https://doi.org/10.1007/JHEP11(2012)096}{\emph{JHEP} {\bfseries 11}
  (2012) 096} [\href{https://arxiv.org/abs/1209.5757}{{\ttfamily 1209.5757}}].

\bibitem{Gaiotto:2024osr}
D.~Gaiotto and J.~Teschner, \emph{{Schur Quantization and Complex Chern-Simons
  theory}},  \href{https://arxiv.org/abs/2406.09171}{{\ttfamily 2406.09171}}.

\bibitem{Godet:2024ich}
V.~Godet, \emph{{Quantum cosmology as automorphic dynamics}},
  \href{https://arxiv.org/abs/2405.09833}{{\ttfamily 2405.09833}}.

\bibitem{Chakraborty:2023los}
T.~Chakraborty, J.~Chakravarty, V.~Godet, P.~Paul and S.~Raju,
  \emph{{Holography of information in de Sitter space}},
  \href{https://doi.org/10.1007/JHEP12(2023)120}{\emph{JHEP} {\bfseries 12}
  (2023) 120} [\href{https://arxiv.org/abs/2303.16316}{{\ttfamily
  2303.16316}}].

\bibitem{Hartle:1983ai}
J.~B. Hartle and S.~W. Hawking, \emph{{Wave Function of the Universe}},
  \href{https://doi.org/10.1103/PhysRevD.28.2960}{\emph{Phys. Rev. D}
  {\bfseries 28} (1983) 2960}.

\bibitem{Witten:1998qj}
E.~Witten, \emph{{Anti-de Sitter space and holography}},
  \href{https://doi.org/10.4310/ATMP.1998.v2.n2.a2}{\emph{Adv. Theor. Math.
  Phys.} {\bfseries 2} (1998) 253}
  [\href{https://arxiv.org/abs/hep-th/9802150}{{\ttfamily hep-th/9802150}}].

\bibitem{Kontsevich:2021dmb}
M.~Kontsevich and G.~Segal, \emph{{Wick Rotation and the Positivity of Energy
  in Quantum Field Theory}},
  \href{https://doi.org/10.1093/qmath/haab027}{\emph{Quart. J. Math. Oxford
  Ser.} {\bfseries 72} (2021) 673}
  [\href{https://arxiv.org/abs/2105.10161}{{\ttfamily 2105.10161}}].

\bibitem{Witten:2021nzp}
E.~Witten, \emph{{A Note On Complex Spacetime Metrics}},
  \href{https://arxiv.org/abs/2111.06514}{{\ttfamily 2111.06514}}.

\bibitem{Collier:2023fwi}
S.~Collier, L.~Eberhardt and M.~Zhang, \emph{{Solving 3d gravity with Virasoro
  TQFT}}, \href{https://doi.org/10.21468/SciPostPhys.15.4.151}{\emph{SciPost
  Phys.} {\bfseries 15} (2023) 151}
  [\href{https://arxiv.org/abs/2304.13650}{{\ttfamily 2304.13650}}].

\bibitem{paper1}
S.~Collier, L.~Eberhardt, B.~M\"uhlmann and V.~A. Rodriguez, \emph{{The complex
  Liouville string: the worldsheet}},
  \href{https://arxiv.org/abs/2409.18759}{{\ttfamily 2409.18759}}.

\bibitem{paper3}
S.~Collier, L.~Eberhardt, B.~M\"uhlmann and V.~A. Rodriguez, \emph{{The complex
  Liouville string: worldsheet boundaries and non-perturbative effects}},
  \href{https://arxiv.org/abs/2410.09179}{{\ttfamily 2410.09179}}.

\bibitem{paper4}
S.~Collier, L.~Eberhardt and B.~M\"uhlmann, \emph{{The complex Liouville
  string: the gravitational path integral}},  (to appear).

\bibitem{Gross:1989vs}
D.~J. Gross and A.~A. Migdal, \emph{{Nonperturbative Two-Dimensional Quantum
  Gravity}}, \href{https://doi.org/10.1103/PhysRevLett.64.127}{\emph{Phys. Rev.
  Lett.} {\bfseries 64} (1990) 127}.

\bibitem{Brezin:1990rb}
E.~Brezin and V.~A. Kazakov, \emph{{Exactly Solvable Field Theories of Closed
  Strings}}, \href{https://doi.org/10.1016/0370-2693(90)90818-Q}{\emph{Phys.
  Lett. B} {\bfseries 236} (1990) 144}.

\bibitem{Douglas:1989ve}
M.~R. Douglas and S.~H. Shenker, \emph{{Strings in Less Than One-Dimension}},
  \href{https://doi.org/10.1016/0550-3213(90)90522-F}{\emph{Nucl. Phys. B}
  {\bfseries 335} (1990) 635}.

\bibitem{Douglas1991}
M.~R. Douglas, \emph{The Two-Matrix Model}, pp.~77--83.
\newblock Springer US, Boston, MA, 1991.
\newblock 10.1007/978-1-4615-3772-4.

\bibitem{Collier:2023cyw}
S.~Collier, L.~Eberhardt, B.~M\"uhlmann and V.~A. Rodriguez, \emph{{The
  Virasoro Minimal String}},
  \href{https://doi.org/10.21468/SciPostPhys.16.2.057}{\emph{SciPost Phys.}
  {\bfseries 16} (2024) 057}
  [\href{https://arxiv.org/abs/2309.10846}{{\ttfamily 2309.10846}}].

\bibitem{Gibbons:1978ac}
G.~W. Gibbons, S.~W. Hawking and M.~J. Perry, \emph{{Path Integrals and the
  Indefiniteness of the Gravitational Action}},
  \href{https://doi.org/10.1016/0550-3213(78)90161-X}{\emph{Nucl. Phys. B}
  {\bfseries 138} (1978) 141}.

\bibitem{Anninos:2020hfj}
D.~Anninos, F.~Denef, Y.~T.~A. Law and Z.~Sun, \emph{{Quantum de Sitter horizon
  entropy from quasicanonical bulk, edge, sphere and topological string
  partition functions}},
  \href{https://doi.org/10.1007/JHEP01(2022)088}{\emph{JHEP} {\bfseries 01}
  (2022) 088} [\href{https://arxiv.org/abs/2009.12464}{{\ttfamily
  2009.12464}}].

\bibitem{Carlip:1992wg}
S.~Carlip, \emph{{The Sum over topologies in three-dimensional Euclidean
  quantum gravity}},
  \href{https://doi.org/10.1088/0264-9381/10/2/004}{\emph{Class. Quant. Grav.}
  {\bfseries 10} (1993) 207}
  [\href{https://arxiv.org/abs/hep-th/9206103}{{\ttfamily hep-th/9206103}}].

\bibitem{Guadagnini:1994ahx}
E.~Guadagnini and P.~Tomassini, \emph{{Sum over the geometries of three
  manifolds}}, \href{https://doi.org/10.1016/0370-2693(94)90541-X}{\emph{Phys.
  Lett. B} {\bfseries 336} (1994) 330}.

\bibitem{Anninos:2022hqo}
D.~Anninos and E.~Harris, \emph{{Interpolating geometries and the stretched
  dS$_{2}$ horizon}},
  \href{https://doi.org/10.1007/JHEP11(2022)166}{\emph{JHEP} {\bfseries 11}
  (2022) 166} [\href{https://arxiv.org/abs/2209.06144}{{\ttfamily
  2209.06144}}].

\bibitem{Anninos:2021ihe}
D.~Anninos and E.~Harris, \emph{{Three-dimensional de Sitter horizon
  thermodynamics}}, \href{https://doi.org/10.1007/JHEP10(2021)091}{\emph{JHEP}
  {\bfseries 10} (2021) 091}
  [\href{https://arxiv.org/abs/2106.13832}{{\ttfamily 2106.13832}}].

\bibitem{Polchinski:1988ua}
J.~Polchinski, \emph{{The phase of the sum over spheres}},
  \href{https://doi.org/10.1016/0370-2693(89)90387-0}{\emph{Phys. Lett. B}
  {\bfseries 219} (1989) 251}.

\bibitem{Maldacena:2024spf}
J.~Maldacena, \emph{{Real observers solving imaginary problems}},
  \href{https://arxiv.org/abs/2412.14014}{{\ttfamily 2412.14014}}.

\bibitem{Witten:1989ip}
E.~Witten, \emph{{Quantization of {Chern-Simons} Gauge Theory With Complex
  Gauge Group}}, \href{https://doi.org/10.1007/BF02099116}{\emph{Commun. Math.
  Phys.} {\bfseries 137} (1991) 29}.

\bibitem{Witten:1988hc}
E.~Witten, \emph{{(2+1)-Dimensional Gravity as an Exactly Soluble System}},
  \href{https://doi.org/10.1016/0550-3213(88)90143-5}{\emph{Nucl. Phys. B}
  {\bfseries 311} (1988) 46}.

\bibitem{Krasnov:2005dm}
K.~Krasnov and J.-M. Schlenker, \emph{{Minimal surfaces and particles in
  3-manifolds}}, \href{https://doi.org/10.1007/s10711-007-9132-1}{\emph{Geom.
  Dedicata} {\bfseries 126} (2007) 187}
  [\href{https://arxiv.org/abs/math/0511441}{{\ttfamily math/0511441}}].

\bibitem{Eberhardt:2022wlc}
L.~Eberhardt, \emph{{Off-shell Partition Functions in 3d Gravity}},
  \href{https://doi.org/10.1007/s00220-024-04963-2}{\emph{Commun. Math. Phys.}
  {\bfseries 405} (2024) 76}
  [\href{https://arxiv.org/abs/2204.09789}{{\ttfamily 2204.09789}}].

\bibitem{Witten:2010cx}
E.~Witten, \emph{{Analytic Continuation Of Chern-Simons Theory}}, {\emph{AMS/IP
  Stud. Adv. Math.} {\bfseries 50} (2011) 347}
  [\href{https://arxiv.org/abs/1001.2933}{{\ttfamily 1001.2933}}].

\bibitem{Castro:2023dxp}
A.~Castro, I.~Coman, J.~R. Fliss and C.~Zukowski, \emph{{Keeping matter in the
  loop in dS$_{3}$ quantum gravity}},
  \href{https://doi.org/10.1007/JHEP07(2023)120}{\emph{JHEP} {\bfseries 07}
  (2023) 120} [\href{https://arxiv.org/abs/2302.12281}{{\ttfamily
  2302.12281}}].

\bibitem{Hikida:2021ese}
Y.~Hikida, T.~Nishioka, T.~Takayanagi and Y.~Taki, \emph{{Holography in de
  Sitter Space via Chern-Simons Gauge Theory}},
  \href{https://doi.org/10.1103/PhysRevLett.129.041601}{\emph{Phys. Rev. Lett.}
  {\bfseries 129} (2022) 041601}
  [\href{https://arxiv.org/abs/2110.03197}{{\ttfamily 2110.03197}}].

\bibitem{Maloney:2015ina}
A.~Maloney, \emph{{Geometric Microstates for the Three Dimensional Black
  Hole?}},  \href{https://arxiv.org/abs/1508.04079}{{\ttfamily 1508.04079}}.

\bibitem{Moncrief:1989dx}
V.~Moncrief, \emph{{Reduction of the Einstein equations in (2+1)-dimensions to
  a Hamiltonian system over Teichmuller space}},
  \href{https://doi.org/10.1063/1.528475}{\emph{J. Math. Phys.} {\bfseries 30}
  (1989) 2907}.

\bibitem{Mess:2007}
G.~Mess, \emph{Lorentz spacetimes of constant curvature},
  \href{https://arxiv.org/abs/0706.1570}{{\ttfamily 0706.1570}}.

\bibitem{McGough:2016lol}
L.~McGough, M.~Mezei and H.~Verlinde, \emph{{Moving the CFT into the bulk with
  $ T\overline{T} $}},
  \href{https://doi.org/10.1007/JHEP04(2018)010}{\emph{JHEP} {\bfseries 04}
  (2018) 010} [\href{https://arxiv.org/abs/1611.03470}{{\ttfamily
  1611.03470}}].

\bibitem{Araujo-Regado:2022gvw}
G.~Araujo-Regado, R.~Khan and A.~C. Wall, \emph{{Cauchy slice holography: a new
  AdS/CFT dictionary}},
  \href{https://doi.org/10.1007/JHEP03(2023)026}{\emph{JHEP} {\bfseries 03}
  (2023) 026} [\href{https://arxiv.org/abs/2204.00591}{{\ttfamily
  2204.00591}}].

\bibitem{Giombi:2008vd}
S.~Giombi, A.~Maloney and X.~Yin, \emph{{One-loop Partition Functions of 3D
  Gravity}}, \href{https://doi.org/10.1088/1126-6708/2008/08/007}{\emph{JHEP}
  {\bfseries 08} (2008) 007} [\href{https://arxiv.org/abs/0804.1773}{{\ttfamily
  0804.1773}}].

\bibitem{Cotler:2018zff}
J.~Cotler and K.~Jensen, \emph{{A theory of reparameterizations for AdS$_3$
  gravity}}, \href{https://doi.org/10.1007/JHEP02(2019)079}{\emph{JHEP}
  {\bfseries 02} (2019) 079}
  [\href{https://arxiv.org/abs/1808.03263}{{\ttfamily 1808.03263}}].

\bibitem{Verlinde:1989ua}
H.~L. Verlinde, \emph{{Conformal Field Theory, 2-$D$ Quantum Gravity and
  Quantization of Teichmuller Space}},
  \href{https://doi.org/10.1016/0550-3213(90)90510-K}{\emph{Nucl. Phys. B}
  {\bfseries 337} (1990) 652}.

\bibitem{Marolf:2020xie}
D.~Marolf and H.~Maxfield, \emph{{Transcending the ensemble: baby universes,
  spacetime wormholes, and the order and disorder of black hole information}},
  \href{https://doi.org/10.1007/JHEP08(2020)044}{\emph{JHEP} {\bfseries 08}
  (2020) 044} [\href{https://arxiv.org/abs/2002.08950}{{\ttfamily
  2002.08950}}].

\bibitem{Iliesiu:2024cnh}
L.~V. Iliesiu, A.~Levine, H.~W. Lin, H.~Maxfield and M.~Mezei, \emph{{On the
  non-perturbative bulk Hilbert space of JT gravity}},
  \href{https://doi.org/10.1007/JHEP10(2024)220}{\emph{JHEP} {\bfseries 10}
  (2024) 220} [\href{https://arxiv.org/abs/2403.08696}{{\ttfamily
  2403.08696}}].

\bibitem{Anninos:2012ft}
D.~Anninos, F.~Denef and D.~Harlow, \emph{{Wave function of
  Vasiliev\textquoteright{}s universe: A few slices thereof}},
  \href{https://doi.org/10.1103/PhysRevD.88.084049}{\emph{Phys. Rev. D}
  {\bfseries 88} (2013) 084049}
  [\href{https://arxiv.org/abs/1207.5517}{{\ttfamily 1207.5517}}].

\bibitem{Anninos:2012qw}
D.~Anninos, \emph{{De Sitter Musings}},
  \href{https://doi.org/10.1142/S0217751X1230013X}{\emph{Int. J. Mod. Phys. A}
  {\bfseries 27} (2012) 1230013}
  [\href{https://arxiv.org/abs/1205.3855}{{\ttfamily 1205.3855}}].

\bibitem{Maldacena:2004rf}
J.~M. Maldacena and L.~Maoz, \emph{{Wormholes in AdS}},
  \href{https://doi.org/10.1088/1126-6708/2004/02/053}{\emph{JHEP} {\bfseries
  02} (2004) 053} [\href{https://arxiv.org/abs/hep-th/0401024}{{\ttfamily
  hep-th/0401024}}].

\bibitem{Collier:2024mgv}
S.~Collier, L.~Eberhardt and M.~Zhang, \emph{{3d gravity from Virasoro TQFT:
  Holography, wormholes and knots}},
  \href{https://doi.org/10.21468/SciPostPhys.17.5.134}{\emph{SciPost Phys.}
  {\bfseries 17} (2024) 134}
  [\href{https://arxiv.org/abs/2401.13900}{{\ttfamily 2401.13900}}].

\bibitem{Gaiotto:2024tpl}
D.~Gaiotto and J.~Teschner, \emph{{Quantum Analytic Langlands Correspondence}},
   \href{https://arxiv.org/abs/2402.00494}{{\ttfamily 2402.00494}}.

\bibitem{Chen:2020tes}
Y.~Chen, V.~Gorbenko and J.~Maldacena, \emph{{Bra-ket wormholes in
  gravitationally prepared states}},
  \href{https://doi.org/10.1007/JHEP02(2021)009}{\emph{JHEP} {\bfseries 02}
  (2021) 009} [\href{https://arxiv.org/abs/2007.16091}{{\ttfamily
  2007.16091}}].

\bibitem{Ribault:2014hia}
S.~Ribault, \emph{{Conformal field theory on the plane}},
  \href{https://arxiv.org/abs/1406.4290}{{\ttfamily 1406.4290}}.

\bibitem{Collier:2017shs}
S.~Collier, P.~Kravchuk, Y.-H. Lin and X.~Yin, \emph{{Bootstrapping the
  Spectral Function: On the Uniqueness of Liouville and the Universality of
  BTZ}}, \href{https://doi.org/10.1007/JHEP09(2018)150}{\emph{JHEP} {\bfseries
  09} (2018) 150} [\href{https://arxiv.org/abs/1702.00423}{{\ttfamily
  1702.00423}}].

\bibitem{Collier:2019weq}
S.~Collier, A.~Maloney, H.~Maxfield and I.~Tsiares, \emph{{Universal dynamics
  of heavy operators in CFT$_{2}$}},
  \href{https://doi.org/10.1007/JHEP07(2020)074}{\emph{JHEP} {\bfseries 07}
  (2020) 074} [\href{https://arxiv.org/abs/1912.00222}{{\ttfamily
  1912.00222}}].

\bibitem{Saad:2019lba}
P.~Saad, S.~H. Shenker and D.~Stanford, \emph{{JT gravity as a matrix
  integral}},  \href{https://arxiv.org/abs/1903.11115}{{\ttfamily 1903.11115}}.

\bibitem{Coleman:1980aw}
S.~R. Coleman and F.~De~Luccia, \emph{{Gravitational Effects on and of Vacuum
  Decay}}, \href{https://doi.org/10.1103/PhysRevD.21.3305}{\emph{Phys. Rev. D}
  {\bfseries 21} (1980) 3305}.

\bibitem{Baumann:2021fxj}
D.~Baumann, W.-M. Chen, C.~Duaso~Pueyo, A.~Joyce, H.~Lee and G.~L. Pimentel,
  \emph{{Linking the singularities of cosmological correlators}},
  \href{https://doi.org/10.1007/JHEP09(2022)010}{\emph{JHEP} {\bfseries 09}
  (2022) 010} [\href{https://arxiv.org/abs/2106.05294}{{\ttfamily
  2106.05294}}].

\bibitem{Collier:2018exn}
S.~Collier, Y.~Gobeil, H.~Maxfield and E.~Perlmutter, \emph{{Quantum Regge
  Trajectories and the Virasoro Analytic Bootstrap}},
  \href{https://doi.org/10.1007/JHEP05(2019)212}{\emph{JHEP} {\bfseries 05}
  (2019) 212} [\href{https://arxiv.org/abs/1811.05710}{{\ttfamily
  1811.05710}}].

\bibitem{Melville:2021lst}
S.~Melville and E.~Pajer, \emph{{Cosmological Cutting Rules}},
  \href{https://doi.org/10.1007/JHEP05(2021)249}{\emph{JHEP} {\bfseries 05}
  (2021) 249} [\href{https://arxiv.org/abs/2103.09832}{{\ttfamily
  2103.09832}}].

\bibitem{Goodhew:2021oqg}
H.~Goodhew, S.~Jazayeri, M.~H.~G. Lee and E.~Pajer, \emph{{Cutting cosmological
  correlators}},
  \href{https://doi.org/10.1088/1475-7516/2021/08/003}{\emph{JCAP} {\bfseries
  08} (2021) 003} [\href{https://arxiv.org/abs/2104.06587}{{\ttfamily
  2104.06587}}].

\bibitem{Goodhew:2020hob}
H.~Goodhew, S.~Jazayeri and E.~Pajer, \emph{{The Cosmological Optical
  Theorem}}, \href{https://doi.org/10.1088/1475-7516/2021/04/021}{\emph{JCAP}
  {\bfseries 04} (2021) 021}
  [\href{https://arxiv.org/abs/2009.02898}{{\ttfamily 2009.02898}}].

\bibitem{Banerjee:2013mca}
S.~Banerjee, A.~Belin, S.~Hellerman, A.~Lepage-Jutier, A.~Maloney, D.~Radicevic
  et~al., \emph{{Topology of Future Infinity in dS/CFT}},
  \href{https://doi.org/10.1007/JHEP11(2013)026}{\emph{JHEP} {\bfseries 11}
  (2013) 026} [\href{https://arxiv.org/abs/1306.6629}{{\ttfamily 1306.6629}}].

\bibitem{Cotler:2016fpe}
J.~S. Cotler, G.~Gur-Ari, M.~Hanada, J.~Polchinski, P.~Saad, S.~H. Shenker
  et~al., \emph{{Black Holes and Random Matrices}},
  \href{https://doi.org/10.1007/JHEP05(2017)118}{\emph{JHEP} {\bfseries 05}
  (2017) 118} [\href{https://arxiv.org/abs/1611.04650}{{\ttfamily
  1611.04650}}].

\bibitem{Law:2020cpj}
Y.~T.~A. Law, \emph{{A compendium of sphere path integrals}},
  \href{https://doi.org/10.1007/JHEP12(2021)213}{\emph{JHEP} {\bfseries 12}
  (2021) 213} [\href{https://arxiv.org/abs/2012.06345}{{\ttfamily
  2012.06345}}].

\bibitem{Anninos:2011af}
D.~Anninos, S.~A. Hartnoll and D.~M. Hofman, \emph{{Static Patch Solipsism:
  Conformal Symmetry of the de Sitter Worldline}},
  \href{https://doi.org/10.1088/0264-9381/29/7/075002}{\emph{Class. Quant.
  Grav.} {\bfseries 29} (2012) 075002}
  [\href{https://arxiv.org/abs/1109.4942}{{\ttfamily 1109.4942}}].

\bibitem{Loganayagam:2023pfb}
R.~Loganayagam and O.~Shetye, \emph{{Influence phase of a dS observer. Part I.
  Scalar exchange}}, \href{https://doi.org/10.1007/JHEP01(2024)138}{\emph{JHEP}
  {\bfseries 01} (2024) 138}
  [\href{https://arxiv.org/abs/2309.07290}{{\ttfamily 2309.07290}}].

\bibitem{Belin:2020hea}
A.~Belin and J.~de~Boer, \emph{{Random statistics of OPE coefficients and
  Euclidean wormholes}},
  \href{https://doi.org/10.1088/1361-6382/ac1082}{\emph{Class. Quant. Grav.}
  {\bfseries 38} (2021) 164001}
  [\href{https://arxiv.org/abs/2006.05499}{{\ttfamily 2006.05499}}].

\bibitem{Chandra:2022bqq}
J.~Chandra, S.~Collier, T.~Hartman and A.~Maloney, \emph{{Semiclassical 3D
  gravity as an average of large-c CFTs}},
  \href{https://doi.org/10.1007/JHEP12(2022)069}{\emph{JHEP} {\bfseries 12}
  (2022) 069} [\href{https://arxiv.org/abs/2203.06511}{{\ttfamily
  2203.06511}}].

\bibitem{deBoer:2024mqg}
J.~de~Boer, D.~Liska and B.~Post, \emph{{Multiboundary wormholes and OPE
  statistics}}, \href{https://doi.org/10.1007/JHEP10(2024)207}{\emph{JHEP}
  {\bfseries 10} (2024) 207}
  [\href{https://arxiv.org/abs/2405.13111}{{\ttfamily 2405.13111}}].

\bibitem{Belin:2023efa}
A.~Belin, J.~de~Boer, D.~L. Jafferis, P.~Nayak and J.~Sonner,
  \emph{{Approximate CFTs and random tensor models}},
  \href{https://doi.org/10.1007/JHEP09(2024)163}{\emph{JHEP} {\bfseries 09}
  (2024) 163} [\href{https://arxiv.org/abs/2308.03829}{{\ttfamily
  2308.03829}}].

\bibitem{Jafferis:2024jkb}
D.~L. Jafferis, L.~Rozenberg and G.~Wong, \emph{{3d Gravity as a random
  ensemble}},  \href{https://arxiv.org/abs/2407.02649}{{\ttfamily 2407.02649}}.

\bibitem{Caron-Huot:2017vep}
S.~Caron-Huot, \emph{{Analyticity in Spin in Conformal Theories}},
  \href{https://doi.org/10.1007/JHEP09(2017)078}{\emph{JHEP} {\bfseries 09}
  (2017) 078} [\href{https://arxiv.org/abs/1703.00278}{{\ttfamily
  1703.00278}}].

\bibitem{Karateev:2018oml}
D.~Karateev, P.~Kravchuk and D.~Simmons-Duffin, \emph{{Harmonic Analysis and
  Mean Field Theory}},
  \href{https://doi.org/10.1007/JHEP10(2019)217}{\emph{JHEP} {\bfseries 10}
  (2019) 217} [\href{https://arxiv.org/abs/1809.05111}{{\ttfamily
  1809.05111}}].

\bibitem{Dorn:1994xn}
H.~Dorn and H.~J. Otto, \emph{{Two and three point functions in Liouville
  theory}}, \href{https://doi.org/10.1016/0550-3213(94)00352-1}{\emph{Nucl.
  Phys. B} {\bfseries 429} (1994) 375}
  [\href{https://arxiv.org/abs/hep-th/9403141}{{\ttfamily hep-th/9403141}}].

\bibitem{Zamolodchikov:1995aa}
A.~B. Zamolodchikov and A.~B. Zamolodchikov, \emph{{Structure constants and
  conformal bootstrap in Liouville field theory}},
  \href{https://doi.org/10.1016/0550-3213(96)00351-3}{\emph{Nucl. Phys. B}
  {\bfseries 477} (1996) 577}
  [\href{https://arxiv.org/abs/hep-th/9506136}{{\ttfamily hep-th/9506136}}].

\bibitem{Anninos:2022ujl}
D.~Anninos, D.~A. Galante and B.~M\"uhlmann, \emph{{Finite features of quantum
  de Sitter space}},
  \href{https://doi.org/10.1088/1361-6382/acaba5}{\emph{Class. Quant. Grav.}
  {\bfseries 40} (2023) 025009}
  [\href{https://arxiv.org/abs/2206.14146}{{\ttfamily 2206.14146}}].

\bibitem{Volkov:2000ih}
M.~S. Volkov and A.~Wipf, \emph{{Black hole pair creation in de Sitter space: A
  Complete one loop analysis}},
  \href{https://doi.org/10.1016/S0550-3213(00)00287-X}{\emph{Nucl. Phys. B}
  {\bfseries 582} (2000) 313}
  [\href{https://arxiv.org/abs/hep-th/0003081}{{\ttfamily hep-th/0003081}}].

\bibitem{Castro:2011xb}
A.~Castro, N.~Lashkari and A.~Maloney, \emph{{A de Sitter Farey Tail}},
  \href{https://doi.org/10.1103/PhysRevD.83.124027}{\emph{Phys. Rev. D}
  {\bfseries 83} (2011) 124027}
  [\href{https://arxiv.org/abs/1103.4620}{{\ttfamily 1103.4620}}].

\bibitem{AtiyahBott}
M.~F. Atiyah and R.~Bott, \emph{{The Yang-Mills Equations over Riemann
  Surfaces}}, \href{https://doi.org/10.1098/rsta.1983.0017}{\emph{Philosophical
  Transactions of the Royal Society of London. Series A, Mathematical and
  Physical Sciences} {\bfseries 308} (1983) 523}
  [\href{https://arxiv.org/abs/1306.6629}{{\ttfamily 1306.6629}}].

\bibitem{Hitchin:1986vp}
N.~J. Hitchin, \emph{{The Selfduality equations on a Riemann surface}},
  \href{https://doi.org/10.1112/plms/s3-55.1.59}{\emph{Proc. Lond. Math. Soc.}
  {\bfseries 55} (1987) 59}.

\bibitem{Chakraborty:2023yed}
T.~Chakraborty, J.~Chakravarty, V.~Godet, P.~Paul and S.~Raju, \emph{{The
  Hilbert space of de Sitter quantum gravity}},
  \href{https://doi.org/10.1007/JHEP01(2024)132}{\emph{JHEP} {\bfseries 01}
  (2024) 132} [\href{https://arxiv.org/abs/2303.16315}{{\ttfamily
  2303.16315}}].

\bibitem{Cotler:2024xzz}
J.~Cotler and K.~Jensen, \emph{{Non-perturbative de Sitter Jackiw-Teitelboim
  gravity}}, \href{https://doi.org/10.1007/JHEP12(2024)016}{\emph{JHEP}
  {\bfseries 12} (2024) 016}
  [\href{https://arxiv.org/abs/2401.01925}{{\ttfamily 2401.01925}}].

\bibitem{Friedan:1986ua}
D.~Friedan and S.~H. Shenker, \emph{{The Analytic Geometry of Two-Dimensional
  Conformal Field Theory}},
  \href{https://doi.org/10.1016/0550-3213(87)90418-4}{\emph{Nucl. Phys. B}
  {\bfseries 281} (1987) 509}.

\bibitem{Witten:1988hf}
E.~Witten, \emph{{Quantum Field Theory and the Jones Polynomial}},
  \href{https://doi.org/10.1007/BF01217730}{\emph{Commun. Math. Phys.}
  {\bfseries 121} (1989) 351}.

\bibitem{Axelrod:1989xt}
S.~Axelrod, S.~Della~Pietra and E.~Witten, \emph{{Geometric quantization of
  Chern-Simons gauge theory}}, {\emph{J. Diff. Geom.} {\bfseries 33} (1991)
  787}.

\bibitem{Drinfeld:1984qv}
V.~G. Drinfeld and V.~V. Sokolov, \emph{{Lie algebras and equations of
  Korteweg-de Vries type}}, \href{https://doi.org/10.1007/BF02105860}{\emph{J.
  Sov. Math.} {\bfseries 30} (1984) 1975}.

\bibitem{Elitzur:1989nr}
S.~Elitzur, G.~W. Moore, A.~Schwimmer and N.~Seiberg, \emph{{Remarks on the
  Canonical Quantization of the Chern-Simons-Witten Theory}},
  \href{https://doi.org/10.1016/0550-3213(89)90436-7}{\emph{Nucl. Phys. B}
  {\bfseries 326} (1989) 108}.

\bibitem{Lempert}
L.~Lempert, \emph{{The problem of complexifying a Lie group}},
  {\emph{Contemporary Mathematics} {\bfseries 205} (1997) 169}.

\bibitem{Hitchin_1979}
N.~J. Hitchin, \emph{Polygons and gravitons},
  \href{https://doi.org/10.1017/S0305004100055924}{\emph{Mathematical
  Proceedings of the Cambridge Philosophical Society} {\bfseries 85} (1979)
  465–476}.

\bibitem{Kronheimer1990AHS}
P.~B. Kronheimer, \emph{{A Hyper‐K\"ahlerian Structure on Coadjoint Orbits of
  a Semisimple Complex Group}}, {\emph{Journal of The London Mathematical
  Society-second Series} (1990) 193}.

\bibitem{Gawedzki:1990qt}
K.~Gawedzki and A.~Kupiainen, \emph{{$\mathrm{SU}(2)$ Chern-Simons theory at
  genus zero}}, \href{https://doi.org/10.1007/BF02104120}{\emph{Commun. Math.
  Phys.} {\bfseries 135} (1991) 531}.

\bibitem{Witten:2013tpa}
E.~Witten, \emph{{Notes On Holomorphic String And Superstring Theory Measures
  Of Low Genus}},  \href{https://arxiv.org/abs/1306.3621}{{\ttfamily
  1306.3621}}.

\end{thebibliography}\endgroup
\end{document}